\documentclass[11pt,a4paper,english]{article}
\usepackage{graphicx}
\usepackage{amssymb,latexsym}
\usepackage{amsmath}
\usepackage{epsf}
\usepackage{pdfsync}
\usepackage{cite}
\usepackage{epsfig}
\usepackage[parfill]{parskip}
\usepackage[matrix,arrow,color]{xy}
\usepackage[usenames]{color}
\usepackage{hyperref}
\usepackage{mathrsfs}
\usepackage[font={small}]{caption}
\usepackage{subcaption}
\usepackage[export]{adjustbox}
\usepackage{float}
\usepackage{wrapfig}
\usepackage{microtype}
\usepackage{slashed}
\allowdisplaybreaks

\input{xy}
\xyoption{all}

\input{epsf}

\makeatletter

\usepackage{setspace}

\usepackage{lscape}

\catcode`@=11
\renewcommand{\section}
{\@startsection{section}{1}{0pt}{\medskipamount}{\medskipamount}{\large\bf}}
\makeatletter\renewcommand{\subsection}
{\@startsection{subsection}{2}{\z@}{-3.25ex plus -1ex minus -.2ex}
{1.5ex plus .2ex}{\it }}

\numberwithin{equation}{section}
\catcode`@=12

\newcommand{\ban}{\begin{eqnarray}}
\newcommand{\ean}{\end{eqnarray}}

\newcommand{\Tr}{{\rm Tr}}

\newcommand{\cW}{{\cal W}}
\newcommand{\cN}{{\cal N}}
\newcommand{\cM}{{\cal M}}
\newcommand{\cS}{{\cal S}}
\newcommand{\cB}{{\cal B}}

\newcommand{\cC}{{\cal C}}
\newcommand{\cQ}{{\cal Q}}

\newcommand{\re}{{\rm Re \,}}
\newcommand{\im}{{\rm Im \,}}

\newcommand{\mbf}[1]{{\boldsymbol {#1} }}
\newcommand{\complex}{{\mathbb C}} 
\newcommand{\zed}{{\mathbb Z}} 
\newcommand{\real}{{\mathbb R}} 
\newcommand{\torus}{{\mathbb T}}

\def\e{{\,\rm e}\,}

\def\ii{{\,{\rm i}\,}}
\def\dd{{\rm d}}

\newcommand{\Hom}{\mathrm{Hom}}

\def\beq{\begin{equation}}
\def\bee{\begin{equation}}
\def\eeq{\end{equation}}
\def\bea{\begin{eqnarray}}
\def\eea{\end{eqnarray}}
\def\bd{\begin{displaymath}}
\def\ed{\end{displaymath}}

\newcommand{\Cint}{\int\kern-10.5pt-\kern7pt}

\newcommand{\be}{\begin{equation}}
\newcommand{\ee}{\end{equation}}
\newcommand{\bal}{\begin{align}}
\newcommand{\eal}{\end{align}}

\newcommand\fverbit{\egroup\item[\fbox{\unhbox\pippobox}]}
\newbox\pippobox

\def\a{\alpha}

{
\def\d{\delta}

\def\be{\begin{equation}}
\def\ee{\end{equation}}
\def\bea{\begin{eqnarray}}
\def\eea{\end{eqnarray}}

\begin{document}

\begin{titlepage}
\setcounter{page}{1}

\vskip 5cm

\begin{center}

\vspace*{3cm}

{\Huge Quivers, Line Defects \\[10pt] and Framed BPS Invariants}

\vspace{15mm}

{\large\bf Michele Cirafici}
\\[6mm]
\noindent{\em Center for Mathematical Analysis, Geometry and Dynamical Systems,\\
Instituto Superior T\'ecnico, Universidade de Lisboa, \\
Av. Rovisco Pais, 1049-001 Lisboa, Portugal}\\ Email: \ {\tt michelecirafici@gmail.com}

\vspace{15mm}

\begin{abstract}
\noindent

A large class of $\mathcal{N}=2$ quantum field theories admits a BPS quiver description and the study of their BPS spectra is then reduced to a representation theory problem. In such theories the coupling to a line defect can be modelled by framed quivers. The associated spectral problem characterises the line defect completely. Framed BPS states can be thought of as BPS particles bound to the defect. We identify the framed BPS degeneracies with certain enumerative invariants associated with the moduli spaces of stable quiver representations. We develop a formalism based on equivariant localization to compute explicitly such BPS invariants, for a particular choice of stability condition. Our framework gives a purely combinatorial solution of this problem. We detail our formalism with several explicit examples.

\end{abstract}

\vspace{15mm}

\today

\end{center}
\end{titlepage}


\tableofcontents

\section{Introduction}

The problem of counting BPS states of $\mathcal{N}=2$ quantum field theory has often an algebraic reformulation in terms of quivers. Quivers can be studied using powerful techniques from representation theory and provide conceptually clear categorical tools to address the physical problem. In many cases the theory of quiver mutations and the associated quantum cluster algebra structures give an elegant formalism to study the BPS spectral problem and the wall-crossing phenomena. The relation between wall-crossing phenomena, BPS spectra and non-perturbative effects is at the core of the Seiberg-Witten solution for the Wilsonian effective action \cite{Seiberg:1994rs}.

The information about the spectrum of BPS states in a given chamber can be elegantly encoded in the quantum monodromy, or Kontsevich-Soibelman operator. Its invariance upon crossing walls of marginal stability is a way to formulate the Kontsevich-Soibelman wall-crossing formula for generalized Donaldson-Thomas invariants \cite{KS1}. When a given chamber in the Coulomb moduli space admits a discrete $R$-symmetry the model admits finer fractional quantum monodromy operators, whose iteration reproduce the full quantum monodromy. Within the context of quantum field theory this program was initiated in \cite{Fiol:2000wx,Fiol:2000pd,Cecotti:2010fi,Cecotti:2011rv,Alim:2011ae,Alim:2011kw} and successfully applied to a variety of cases \cite{Cecotti:2011gu,DelZotto:2011an,Xie:2012Dw,Cecotti:2012va,Cecotti:2012sf,Xie:2012jd,Xie:2012gd,Cecotti:2013lda,Cecotti:2013sza,Cordova:2014oxa,Cordova:2015vma,Cecotti:2015qha,Caorsi:2016ebt}. Similar concepts have been pursued in the study of D-branes in Calabi-Yau varieties \cite{Douglas:1996sw,Douglas:2000ah,Douglas:2000qw,Denef:2002ru}. There are by now several approaches to determine the BPS degeneracies; for example spectral networks \cite{Gaiotto:2009hg,Gaiotto:2012rg,Gaiotto:2012Db,Galakhov:2013oja,Maruyoshi:2013fwa,Galakhov:2014xba,Longhi:2016rjt,Hollands:2016kgm,Longhi:2016wtv,Longhi:2016bte}, the MPS wall-crossing formula \cite{Manschot:2010qz,Manschot:2011xc,Sen:2011aa,Manschot:2012rx,Manschot:2013sya,Manschot:2013dua,Manschot:2014fua}, or a direct localization approach  \cite{Hori:2014tda,Kim:2015fba,Cordova:2014oxa}.

In this paper, as well as in the companion \cite{BPSlinesCluster}, we take a step to extend this program to line defects in theories of class $\mathcal{S}[A_k]$. This is part of a project initiated in \cite{Cirafici:2013bha} for the case of $\mathcal{S}[A_1]$ theories. In the deep infra-red these defects look like a bound state of an infinitely massive dyonic particle with a halo of ordinary BPS states. These bound states are called framed BPS states \cite{Gaiotto:2010be} and can be described algebraically in terms of framed quivers \cite{Cordova:2013bza}. Framed BPS states pose a new BPS spectral problem in quantum field theory. In the UV line defects can be identified with certain paths on a curve $\mathcal{C}$, whose covering is the Seiberg-Witten curve $\Sigma$ \cite{Gaiotto:2010be,Drukker:2009tz,Alday:2009fs,Ito:2011ea}. Line defects and the corresponding framed BPS spectra have been studied using semiclassical methods \cite{Ito:2011ea,Lee:2011ph,Moore:2015szp,Moore:2015qyu,Moore:2014gua,Moore:2014jfa,Brennan:2016znk} or spectral networks \cite{Hollands:2013qza,Gabella:2016zxu,Longhi:2016wtv,Xie:2013lca,Gaiotto:2014bza}. The work \cite{BPSlinesCluster} establishes a connection between line defects, quantum discrete integrable systems and cluster algebras. The latter aspect is also discussed in \cite{Cirafici:2013bha,Williams:2014efa,Allegretti:2015nxa}.

When we rephrase the problem in terms of BPS quivers, studying the spectrum of stable BPS states becomes equivalent to classifying stable quiver representations up to isomorphisms. Similarly the wall-crossing formula is associated with certain quantum dilogarithm identities corresponding to sequences of quiver mutations. A BPS quiver is constructed directly from the data of the $\cN=2$ quantum field theory: the nodes of the quiver correspond to a basis of the lattice of charges while the arrow structure is dictated by the Dirac-Schwinger-Zwanziger pairing between charges \cite{Cecotti:2011rv,Alim:2011ae,Alim:2011kw}. In many cases BPS quivers can be engineered by studying D-branes on Calabi-Yau threefolds \cite{Douglas:1996sw,Douglas:2000ah,Douglas:2000qw,Denef:2002ru}, where the quiver nodes corresponds to generators of the compactly supported $K$-theory. Alternatively they can be obtained via the $4d/2d$ correspondence \cite{Cecotti:2010fi}. Mathematically the BPS spectra and the wall-crossing formula arise from the generalized Donaldson-Thomas theory associated with the quiver \cite{KS1,keller,nagao}.

In this paper we aim to develop a systematic framework to determine the framed BPS degeneracies for several classes of BPS quivers. We will identify the framed BPS degeneracies with BPS invariants of Donaldson-Thomas type associated with the moduli spaces of cyclic modules of framed quivers. The main tool we will use is equivariant localization with respect to a natural toric action which rescales all the morphisms associated to the arrows of the quivers. This formalism is naturally rooted in the analog problem of counting supersymmetric bound states of a gas of D0 and D2 branes with a single D6 brane wrapping a local Calabi-Yau \cite{szendroi,reineke,Ooguri:2008yb,Cirafici:2008sn,Cirafici:2010bd,Chuang:2013wt}. The D0 and D2 branes are supported on compact cycles and are the analog of the unframed BPS states, while the non-compact brane is naturally associated with a line defect. Indeed in many cases this analogy can be made concrete by directly taking a certain scaling limit \cite{Chuang:2013wt}. Very closely related to this paper is the result of \cite{szendroi} which provides a combinatorial solution for the BPS invariants in the case of the noncommutative crepant resolution of the conifold; such a solution can be expressed in terms of pyramid partitions, certain combinatorial arrangements. We extend such a formalism to a variety of quivers and line defects, and for each case we provide a simple combinatorial solution. These techniques can be easily extended to any framed quiver with superpotential, upon choosing appropriate stability conditions.

This paper is organized as follows. Section \ref{classS} contains a brief review of theories of class $\mathcal{S}$ and their line defects. Section \ref{BPSquivers} contains a discussion of the relation between BPS states and the representation theory of quivers, and introduces framed quivers to model line defects. In Section \ref{eqloc} we set up a formalism to compute the framed BPS degeneracies using equivariant localization, which we then apply in Sections \ref{SU2loca}, \ref{SU3loca} and \ref{SO8loca}. These sections contain several results for SU(2), SU(3) and SO(8) line defects: some of them are checks that our formalism reproduces the results available in the literature, others are new. We end with the Conclusions.

\section{Theories of class $\mathcal{S}$ and line defects} \label{classS}

\subsection{Theories of class $\mathcal{S}$}

Theories of class $\mathcal{S}[A_n, \mathcal{C} , D]$ are four dimensional $\mathcal{N}=2$ supersymmetric field theories which arise from the compactification of the six dimensional $\mathcal{N}= (0,2)$ superconformal theory on a Riemann surface $\mathcal{C}$ (the ``UV curve") with punctures $s_n$ and some extra data $D_n$ at the punctures. We will denote with $\mathcal{B}$ their Coulomb branches. A generic point in $\mathcal{B}$ is a tuple of meromorphic $k$-differentials $u = (\varphi_1 , \dots ,  \varphi_r)$, where $\varphi_k$ is a sections of the $k$-th power of the canonical bundle $\mathcal{K}_{\mathcal{C}}^{\otimes k}$ with prescribed residues at $s_n$. The low energy Wilsonian effective action is completely determined in terms of a family of ``IR curves" $\Sigma_u$. The Seiberg-Witten curve $\Sigma_u$ is a $r$-fold branched covering of $\mathcal{C}$, defined by
\begin{equation}
\lambda^r + \lambda^{r-1} \, \varphi_1 + \cdots + \varphi_r = 0 \, .
\end{equation}
At a generic point $u \in \mathcal{B}$, the gauge group is spontaneously broken down to its maximal torus $U(1)^r$. The lattice of electric and magnetic charges $\Gamma_g$ is identified with a quotient of $\Gamma = H_1 (\Sigma_u , \mathbb{Z})$ and endowed with an antisymmetric integral pairing $\langle \ , \ \rangle$. Locally $\Gamma = \Gamma_g \oplus \Gamma_f$, where the lattice of flavor charges $\Gamma_f$ is the annihilator of the pairing. Due to supersymmetry the central charge operator is represented by an holomorphic function on $\mathcal{B}$
\begin{equation}
Z_{\gamma} (u) = \frac{1}{\pi} \, \int_\gamma \ \lambda \, ,
\end{equation}
written in terms of the periods of the Seiberg-Witten differential \cite{Seiberg:1994rs}.

If we compactify the theory on a circle $S^1_R$ with radius $R$ and periodic boundary conditions for the fermions, the theory reduces to a three dimensional sigma model with $\mathcal{N}=4$ supersymmetry \cite{Seiberg:1996nz}. Due to supersymmetry, the target of the sigma model $\mathcal{M}_H$ is a smooth hyperK\"ahler manifold of $\mathrm{dim}_{\mathbb{C}} \, \mathcal{M}_H = \frac12 \, \mathrm{dim}_{\mathbb{C}} \, \mathcal{B}$. The space $\mathcal{M}_H$ carries a family of complex structures $J_\zeta$ and symplectic forms $\omega_\zeta$ (holomorphic with respect to $J_\zeta$) parametrized by $\zeta \in \mathbb{P}^1$. Geometrically $\mathcal{M}_H$ is the Hitchin moduli space, which parametrizes harmonic bundles on $\mathcal{C}$ \cite{Gaiotto:2009hg}, that is solutions of the Hitchin equations
\begin{align}
F_A + R \, [ \varphi , \overline{\varphi} ] = & 0 \, ,  \cr
\overline{\partial}_A \, \phi = & 0 \, , \cr
\partial_A \, \overline{\phi} = & 0 \, ,
\end{align}
for a unitary connection $A$ of a rank $r$ Hermitian vector bundle $E$ on $\mathcal{C}$ and a section $\varphi$ of $\mathrm{End} (E) \otimes \mathcal{K}_{\mathcal{C}}$, with prescribed singularities, and up to gauge equivalence. We refer the reader to \cite{Neitzke:2014cja} for a more detailed discussion.

In complex structure $J_0$, $\mathcal{M}_H$ coincides with the moduli space of Higgs bundles $(E_h , \phi)$, where $E_h$ is the holomorphic bundle defined by $\overline{\partial}_A$ and $\phi$ the $(1,0)$ part of $\varphi$. In this complex structure the Hitchin fibration realizes $\mathcal{M}_H$ as a bundle over $\mathcal{B}$ whose fibers are compact complex tori. The exact metric on $\mathcal{M}_H$ is smooth after taking into account the quantum corrections. As a first approximation it arises from the naive dimensional reduction of the four dimensional lagrangian field theory, where the $r$ complex scalars parametrize the base $\mathcal{B}$. The scalars parametrizing the torus fibers come from the holonomies of the four dimensional gauge fields along $S^1_R$, as well as from dualizing the three dimensional gauge fields into periodic scalars. The smoothness of the quantum corrected metric is equivalent to the condition that the BPS degeneracies enjoy the wall-crossing formula \cite{Gaiotto:2008cd}.

The moduli space $\mathcal{M}_H$ has a canonical set of Darboux coordinates $\{ X_{\gamma} (u , \zeta ) \}$ for $\gamma \in \Gamma$ and $(u , \zeta) \in \mathcal{B} \times \mathbb{P}^1$ (we will usually suppress the dependence from $(u ; \zeta)$). They satisfy the twisted group algebra
\begin{equation}
X_{\gamma} \, X_{\gamma'} = (-1)^{\langle \gamma , \gamma' \rangle} \ X_{\gamma+ \gamma'} \, ,
\end{equation}
and are piecewise holomorphic on $\mathcal{M}_H$ in the sense that at fixed $\zeta \in \mathbb{C}^*$, $X_{\gamma} (u , \zeta )$ the dependence on $u$ is holomorphic with respect to $J_{\zeta}$. They satisfy the Poisson bracket relation
\begin{equation}
\{ X_{\gamma}  ,  X_{\gamma'} \} = \langle \gamma , \gamma^\prime \rangle \, X_{\gamma + \gamma'}
\end{equation}
induced by the symplectic structure. The coordinates $\{ X_{\gamma} (u , \zeta ) \}$ jump at real codimension one walls in $\mathcal{B} \times \mathbb{C}^*$. These jumps occur at BPS walls, or walls of second kind;  loci where
\begin{equation} 
Z(\gamma')/\zeta \in \mathbb{R}_- \, ,
\end{equation}
and which can be thought of as rays $\ell_{\gamma'}$ in the $\zeta$-plane. The effect of crossing the wall is captured by the transformation
\begin{equation} \label{jumpX}
X_{\gamma} \longrightarrow \mathcal{K}_{\gamma'}^{\Omega (\gamma' , u)} (X_{\gamma})
\end{equation}
expressed in terms of the Kontsevich-Soibelman symplectomorphism
\begin{equation}
\mathcal{K}_{\gamma'} (X_{\gamma}) = X_{\gamma} \, \left( 1 - X_{\gamma'} \right)^{\langle \gamma, \gamma' \rangle} \ .
\end{equation}

\subsection{Line defects}

We will discuss line defects which are straight lines in $\mathbb{R}^{1,3}$, located at the spatial origin and extended along the timelike direction. We will follow \cite{Gaiotto:2010be}   
 in the presentation. We will require that such defects preserve a subalgebra of the $\mathcal{N}=2$ supersymmetry algebra labeled by a phase $\zeta \in \mathbb{C}^*$, along with $\frak{so} (3)$ rotations around the insertion point of the defect, time translations and the $\frak{su} (2)_R$ R-symmetry group. For theories which have a lagrangian description in a certain region of the moduli space, these line defects admit a set of UV labels which specify them uniquely. For a given gauge group $G$, these labels are a pair of weights $ \alpha = (\lambda_e , \lambda_m) \in \Lambda_w \times \Lambda_{mw} / \mathfrak{w}$, elements of the weight lattice of $\frak g$ and the weight lattice of the Langlands dual algebra $\frak g^*$ respectively, modulo the action of the Weyl group $\mathfrak{w}$. For example supersymmetric Wilson lines along the path $\ell$ have the familiar form
\begin{equation}
W ( \mathbf{R}) = \Tr_{\mathbf{R}} \  \mathrm{P}\exp \int_\ell \dd^3 x \left( \frac{\phi}{2 \zeta} - \ii A - \zeta \frac{\overline{\phi}}{2} \right) \, ,
\end{equation}
where $\mathbf{R}$ is an irreducible representation of $G$, and in this formula we have denoted by $(\phi, \overline{\phi} , A)$ the bosonic fields in the $4d$ $\mathcal{N}=2$ vector multiplet. Similarly for a 't Hooft operator, the boundary conditions consist in defining a G-bundle over a small linking sphere $S^2$, which are classified by magnetic weights. For general dyonic charges, the allowed set of labels $(\lambda_e , \lambda_m)$ is further restricted by a Dirac-like quantization condition: for any pair $(\lambda_e , \lambda_m)$ and $(\lambda'_e , \lambda'_m)$ of line defect charges, we must have \cite{Gaiotto:2010be,Aharony:2013hda}
\begin{equation}
\langle (\lambda_e , \lambda_m) , (\lambda'_e , \lambda'_m) \rangle \in \mathbb{Z} \, .
\end{equation}
For more general field theories, one can assume that such a discrete labeling exists, and takes value in an appropriate lattice; we will denote by $\mathcal{L}_{UV}$ the lattice of UV labels of a given theory. 

In the presence of a defect $L_{\zeta , \alpha}$ the Hilbert space of states is modified to $\mathcal{H}_{L_{\zeta , \alpha}}$. Line defects form interesting algebraic structures. To begin with, they can be endowed with an obvious addition operation: the line defect $L_1+L_2$ is defined as the defect whose correlators are simply the sum of the correlators of $L_1$ and of $L_2$. More formally the Hilbert space of the theory in the presence of the sum of two defects is the direct sum $\mathcal{H}_{L_1+L_2} = \mathcal{H}_{L_1} \oplus \mathcal{H}_{L_2} $. This allows us to define a \textit{simple} line operator, as a line operator which is not the sum of other line operators.

Similarly the product structure is defined by inserting two line defects $L_1$ and $L_2$ in the functional integral. Supersymmetry guarantees that the correlators are independent on the relative distance between the defects. Therefore by locality, letting them approach each other gives a new, composite, defect. The latter can be expressed in terms of simple line defects. More formally at the level of the Hilbert spaces, this procedure implies that
\begin{equation} \label{HilbOPE}
\mathcal{H}^{BPS}_{L_i} \otimes \mathcal{H}^{BPS}_{L_j} = \bigoplus_k \, \mathcal{N}_{ij}^k \otimes \mathcal{H}^{BPS}_{L_k} \, .
\end{equation}
The presence of the vector spaces $\mathcal{N}_{ij}^k$ follows from quantization of the electromagnetic field sourced by the defects, seen as infinitely heavy dyonic particles inserted in the functional integral. 

\subsection{Framed BPS degeneracies}

As we have mentioned the presence of a line defect modifies the Hilbert space of the theory. To be more precise, $\mathcal{H}_{L,u}$ will depend explicitly on the defect, as well as on a point of the Coulomb branch $u \in \mathcal{B}$. The Hilbert space is graded by the electromagnetic charge as measured at infinity
\begin{equation}
\mathcal{H}_{L,u} = \bigoplus_{\gamma \in \Gamma_L} \, \mathcal{H}_{L , u , \gamma} \ .
\end{equation}
Here $\Gamma_L$ denotes the lattice of charges in the presence of the defect $L$. This lattice has generically the form of a torsor for $\Gamma$, that is $\Gamma_L = \Gamma + \gamma_L$, where $\langle \gamma_L , \gamma \rangle \in \zed$ for all $\gamma \in \Gamma$. The charge $\gamma_L$ does not have to be an element of $\Gamma$. Physically it can be interpreted as an IR label for the line defect. The charge $\gamma_L$ has the form of a core charge plus a dummy flavor charge; the role of the latter is to introduce a mass parameter in the central charge associated with $\gamma_L$. This mass is then send to infinity to obtain a modified BPS bound. After such a procedure the new BPS bound is 
\begin{equation}
E = \re \left( Z_{\gamma} (u) / \zeta \right) \, ,
\end{equation}
and the quantum states which saturate this bound are called framed BPS states \cite{Gaiotto:2010be}. As in the case without the defect, we can introduce the framed protected spin character as a trace over the single-particle BPS Hilbert space 
\begin{equation} \label{psc}
\underline{\overline{\Omega}} \left( u , L , \gamma ; q \right) := \Tr_{\mathcal{H}_{L,u,\gamma}^{BPS}} \ q^{2 J_3} (-q)^{2 I_3} \ ,
\end{equation}
defined in terms of an $\frak{so} (3)$ generator $J_3$ and an $\frak{su}_R (2)$ generator $I_3$.  
By taking the $q \longrightarrow 1$ limit of (\ref{psc}), one finds
\begin{equation} \label{Omegaplus}
\underline{\overline{\Omega}} \left( u , L , \gamma ; q = 1\right) := \Tr_{\mathcal{H}_{L,u,\gamma}^{BPS}} \  (-1)^{2 I_3} \ .
\end{equation}
The \textit{no-exotic conjecture} states that the protected spin characters receive contributions only from states with trivial $\frak{su}_R (2)$ quantum numbers \cite{Gaiotto:2010be}. In the absence of exotic states, $I_3 = 0$ identically and $\underline{\overline{\Omega}} \left( u , L , \gamma ; q = 1\right) = \dim \mathcal{H}_{L , u , \gamma}^{BPS}$ is a non negative integer. Such a conjecture has been by now proven in many cases \cite{Chuang:2013wt,DelZotto:2014bga}. On the other hand the $q \longrightarrow -1$ limit yields
\begin{equation} \label{Omegaminus}
\underline{\overline{\Omega}} \left( u , L , \gamma ; q = -1 \right) := \Tr_{\mathcal{H}_{L,u,\gamma}^{BPS}} \ (-1)^{2 J_3} \ ,
\end{equation}
which is an ordinary Witten index and therefore counts the net number of ground states.

Framed BPS states can be roughly pictured as particles bound to the defect, separated by a non vanishing energy gap from the continuum of unbound states. As the Coulomb branch parameters vary, the gap might close and a particle halo is free to join the continuum. As a result the protected spin character jumps at BPS walls, the loci where $Z_\gamma (u) / \zeta \in \mathbb{R}_-$.

One can use the Protected Spin Character to compute the OPE coefficients of the algebra of line defects from \eqref{HilbOPE}
\begin{equation} \label{opeCq}
c_{ij}^k (q) = \Tr_{\mathcal{N}_{ij}^k} \ q^{2 J_3} (-q)^{2 I_3} \, .
\end{equation}
If we assume that no exotics are present, then the $\frak{su}_R (2)$ generator $I_3$ in (\ref{opeCq}) acts trivially. In particular this implies that in the $q \longrightarrow +1$ limit
\begin{equation}
c_{ij}^k (q = 1) = \Tr_{\mathcal{N}_{ij}^k} \ (-1)^{2 I_3} = \dim \mathcal{N}_{ij}^k
\end{equation}
and in particular is manifestly positive. Therefore the absence of exotics implies that the coefficients of the OPE are non-negative integers in the $q \longrightarrow + 1$ limit.

It is useful to adopt a more geometrical perspective and consider the framed BPS degeneracies as enumerative invariants of Donaldson-Thomas type, as in \cite{Chuang:2013wt}. To do so we model the Hilbert space of states on the cohomology of an appropriate moduli space of BPS states
\begin{equation}
\mathcal{H}_{L, u , \gamma} = \bigoplus_{p,q} \, H^{p,q} (\mathcal{M}^{\tt{BPS}} (L , \gamma ; u)) \, .
\end{equation}
This formula assumes that $\mathcal{M}^{\tt{BPS}} (\gamma ; u)$ can be defined as a smooth variety. When this is not the case, we assume that analog quantities can be defined. We can think of $\mathcal{M}^{\tt{BPS}} (L , \gamma ; u)$ as parametrizing stable objects in a certain category of quiver representations. Continuing with this analogy, the Protected Spin Character has the form of a refined Donaldson-Thomas invariant \cite{Chuang:2013wt}
\begin{equation}
{\tt DT}^{ref} (L , u , \gamma , q ) = \underline{\overline{\Omega}}  (L , u , \gamma , q )  = \sum_{p,q, \in \mathbb{Z}} \, (-1)^{p-q} \, q^{2p-m} \, h^{p,q} (\mathcal{M}^{\tt{BPS}} (L , \gamma ; u))
\end{equation}
where the dependence on $m = \mathrm{dim}_\mathbb{C} \, \mathcal{M}^{\tt{BPS}} (L , \gamma ; u)$ comes from the Lefschetz action on cohomology, identified with the action of $SU(2)_{spin}$. In particular the Protected Spin Character can be written as 
\begin{equation} 
{\tt DT}^{ref} (L , u , \gamma , q ) = \underline{\overline{\Omega}} (u,L,\gamma ; q) = q^{-m} \ \chi_y (\mathcal{M}^{\tt{BPS}} (L , \gamma ; u)) \vert_{y=q^2} \ ,
\end{equation}
in terms of the $\chi_y$-genus
\begin{equation}
\chi_y (\mathcal{M}^{\tt{BPS}} (L , \gamma ; u)) = \sum_{p,q \in \mathbb{Z}} \, (-1)^{p+q} \, y^p \, h^{p,q} (\mathcal{M}^{\tt{BPS}} (L , \gamma ; u)) \, .
\end{equation}
We can also provide a geometrical interpretation of the $q \longrightarrow \pm 1$ limits. In particular when $q \longrightarrow +1$ the Protected Spin Character specializes to the Euler characteristic 
\begin{equation} 
 \underline{\overline{\Omega}} (u,L,\gamma ; q=+1) =  \chi (\mathcal{M}^{\tt{BPS}} (L , \gamma ; u))  \ .
\end{equation}
In the $q \longrightarrow -1$ limit the framed BPS degeneracies coincide with the numerical Donaldson-Thomas invariants
\begin{equation} 
 \underline{\overline{\Omega}} (u,L,\gamma ; q=-1) =  {\tt DT} (\mathcal{M}^{\tt{BPS}} (L , \gamma ; u))  \ .
\end{equation}
These are the quantities that we will discuss and compute in this paper. We will often use the notations ${\tt DT} (L , u , \gamma )$, or more simply ${\tt DT} (L)$. 

In general $\mathcal{M}^{\tt{BPS}} (L , \gamma ; u)$ is not a smooth manifold and all of the above quantities have to be defined appropriately: for example numerical Donaldson-Thomas invariants can be defined as weighted Euler characteristics \cite{behrend} or via localization \cite{szendroi}. However the above relations are still expected to hold. 

\subsection{IR vevs and core charges}

When we define the theory on $\mathbb{R}^3 \times S^1_R$, a line defect becomes a local operator. Since a defect $L_{\zeta , \alpha}$ preserves a supersymmetry sub-algebra parametrized by the phase $\zeta$, its vacuum expectation value defines a function on the moduli space $\mathcal{M}_H$ which is holomorphic in complex structure $J_\zeta$. In particular there exists a distinguished set of $J_\zeta$-holomorphic functions on $\mathcal{M}_H$ which correspond to simple line defects. When we compactify the direction where the line defect is stretched into an $S^1_R$, the path integral turns into a trace
\begin{equation} \label{vevindex}
\langle L_{\zeta , \alpha} \rangle_{q,u} = \Tr_{\mathcal{H}_{L , u}} \, (-1)^F \, \e^{- 2 \pi R H} (-q)^{2 J_3 + 2 I_3} \, \e^{\ii \theta \cdot \mathcal{Q}} \, \sigma (\mathcal{Q})
\end{equation}
where $H$ is the Hamiltonian and the factor $\e^{\ii \theta \cdot \mathcal{Q}} \, \sigma (\mathcal{Q})$ is required to properly define boundary conditions for both electric and magnetic defects, as discussed in \cite{Gaiotto:2010be}.

The $q \longrightarrow -1$ limit of \eqref{vevindex} reduces to \cite{Gaiotto:2010be,Neitzke:2014cja}
\begin{equation} \label{vevLX}
\langle L_{\zeta , \alpha} \rangle_{q=-1 , u} = \sum_{\gamma \in \Gamma_L} \underline{\overline{\Omega}} ( u , L , \gamma ; q = -1) \ X_{\gamma} \, .
\end{equation}
This equation gives a direct meaning to the Darboux coordinates $X_\gamma$ as vevs of IR line operators, $X_\gamma (u , \zeta) \equiv \langle L_{\zeta , \gamma}^{IR} \rangle_{q=-1,u}$. These functions are not simple line defect of an abelian theory, they receive an infinite series of non-perturbative corrections from the four dimensional BPS states running around the $S^1_R$. The functions $X_\gamma$ are discontinuous across BPS walls, with jumps given by \eqref{jumpX}. On the other hand $\langle L_{\zeta , \alpha} \rangle_{q=-1 , u}$ is a continuous function of $(u,\zeta)$, as no phase transition is present in the UV. Therefore consistency requires that the framed degeneracies $\underline{\overline{\Omega}} ( u , L , \gamma ; q = -1)$ have discontinuities which precisely cancel those of the $X_\gamma$. Physically such jumps describe a process in which a framed BPS bound state forms or decays, by capturing or emitting a vanilla BPS particle.

The IR line operators $X_\gamma$ are $J_\zeta$-holomorphic functions on $\cM_H$ and in particular have the following asymptotic behavior for $\zeta \longrightarrow 0$
\begin{equation} \label{leadingX}
X_{\gamma} \sim c_\gamma \e^{\pi R \frac{Z_\gamma}{\zeta} }
\end{equation}
where $c_\gamma$ is a constant independent of $\zeta$. Therefore a line defect vev as in \eqref{vevLX} will have a similar expansion. The charge $\gamma_c$ determined by the smallest term defines the core charge of the defect. Physically it can be interpreted as the ground state of the defect. Note that as we move in the $(\zeta,u)$ space the core charge will jump, according to the discontinuities of the functions $X_\gamma$ and of the framed degeneracies $\underline{\overline{\Omega}} ( u , L , \gamma ; q = -1)$.

It is also useful to introduce untwisted coordinates $Y_{\gamma}$, so that
\begin{equation}
Y_{\gamma} \, Y_{\gamma'} = Y_{\gamma+ \gamma'} \ .
\end{equation}
These coordinates describe locally a space $\widetilde{\mathcal{M}}_H$ and up to a quadratic refinement can be identified with the Darboux coordinates $X$ on $\mathcal{M}_H$, establishing a conjectural isomorphism between $\mathcal{M}_H$ and $\widetilde{\mathcal{M}}_H$. We will assume that indeed these spaces are isomorphic. The isomorphism between the two sets of coordinates is given by a quadratic refinement, a map
\begin{equation}
\sigma \, : \, \Gamma \longrightarrow \{ \pm 1 \}
\end{equation}
such that
\begin{equation}
\sigma (\gamma) \sigma (\gamma') = (-1)^{\langle \gamma , \gamma' \rangle} \, \sigma (\gamma + \gamma') \, .
\end{equation}
In the following we will identify $Y_\gamma$ with $\sigma (\gamma) X_\gamma$ with the choice $\sigma (\gamma) = (-1)^{J_3 + I_3}$ \cite{Gaiotto:2010be}. With these choices, the transformation law for the coordinates $Y_\gamma$ upon crossing the BPS wall for a single hypermultiplet with charge $\gamma'$ becomes a cluster transformation
\begin{equation}
Y_\gamma \longrightarrow Y_\gamma (1 + Y_{\gamma'})^{\langle \gamma , \gamma' \rangle} \ .
\end{equation}
Such a transformation endows $\mathcal{M}_H$ locally with the structure of a cluster variety. 

The coordinates $X_\gamma$ and $Y_\gamma$ on the moduli space obey the TBA-like equations \cite{Gaiotto:2008cd}
\begin{equation}
\log Y_\gamma (\zeta) = \frac{R}{\zeta} Z_\gamma + \ii \theta_\gamma + R \zeta \overline{Z}_\gamma + \sum_{\gamma'} \, \Omega (\gamma') \frac{\langle \gamma , \gamma' \rangle}{4 \pi \ii} \ \int_{\ell_{\gamma'}} \, \frac{d \zeta'}{\zeta'} \frac{\zeta' + \zeta}{\zeta'- \zeta} \, \log \left( 1 - \sigma (\gamma') Y_{\gamma'}  (\zeta') \right)
\end{equation}
in terms of the degeneracies of stable BPS particles $\Omega (\gamma)$. Above $\theta$ parametrizes the holonomies of the gauge field along $S^1_R$. The discontinuities of these coordinates at the BPS rays are the reason certain dynamical systems play a role in \cite{BPSlinesCluster}. 

The reason why we are discussing both $X$ and $Y$ coordinates is that in this paper we will compute the vevs $\langle L_{\zeta , \alpha} \rangle_{q=-1 , u}$  while in \cite{BPSlinesCluster} the uses of cluster algebra techniques naturally led to results for the $q \longrightarrow +1$ limit of the same quantity. To compare the result of this paper with \cite{BPSlinesCluster} we simply need to pass to the untwisted coordinates $Y_\gamma$. The twisted vevs $\langle L_{\zeta , \alpha} \rangle_{q=-1 , u}$ contain more information since the minus signs in the BPS invariants $\underline{\overline{\Omega}} ( u , L , \gamma ; q = -1)$ keep track of the spin of the bound state, and can for example distinguish a vector multiplet for which $\underline{\overline{\Omega}} ( u , L , \gamma ; q = -1) = -2$ from two hypermultiplets, for which $\underline{\overline{\Omega}} ( u , L , \gamma ; q = -1) = +2$.

To summarize, in the IR the defect splits as a sum of elementary line defects with coefficients given by the framed degeneracies. At a certain point $(u,\zeta) \in \cB \times \complex^*$ the state with the smallest energy $\re ( Z_{\gamma} / \zeta )$ can be seen as the defect ground state. Physically the defect appears to an IR observer as an infinitely massive dyon, surrounded by a cloud of (in general mutually non-local) halos. The ground state charge is the core charge $\gamma_c$ which plays the role of IR label for the defect. As the central charge depends explicitly on the Coulomb branch parameters, the core charge can jump as the ground state becomes degenerate. The condition for this to happen is that another state in the framed spectrum has its energy lowered to that of the core charge. This can happen at loci where $\re (Z_{\gamma_1} / \zeta ) = \re (Z_{\gamma_2} / \zeta )$, or equivalently when $ Z_{\gamma} / \zeta \in - \ii \real_+ $ with $\gamma = \gamma_1 - \gamma_2$. The latter condition defines \textit{anti-walls}. Crossing an anti-wall corresponding to a charge $\gamma$, the core charge $\gamma_c$ transforms as
\begin{equation}
\gamma_{c'} = \gamma_c + [ \langle \gamma , \gamma_c \rangle]_+ \, \gamma
\end{equation}
where $\gamma_{c'}$ is the ground state charge at the other side of the wall \cite{Gaiotto:2010be}.

Therefore to any defect, labeled in the UV by a $\alpha  \in \Lambda_w \times \Lambda_{mw} / \mathfrak{w}$, we can associate an IR label $\gamma_c$, uniquely defined modulo wall crossings at the anti-walls. This map 
\begin{equation}
\mathbf{RG} (\cdot , \zeta , u) : \mathcal{L}_{UV}  \longrightarrow \Gamma   
\end{equation}
was called defect renormalization group flow in \cite{Cordova:2013bza}. This map was studied for complete theories, where it appears to be invertible. In other words, a line defect in the UV can be completely identified by its IR decomposition into elementary line defects. It is likely that this property holds in general; possibly at the price of restricting the variables $(u , \zeta) \in \cB \times \complex^*$ to the physically accessible region if the theory is not complete. In the following we will also label defects by their core charge, i.e. as $L_{\zeta , \gamma_c}$, whenever we want to emphasize it .

\section{BPS quivers and representation theory} \label{BPSquivers}

For a large class of four dimensional $\mathcal{N}=2$ supersymmetric field theories, the spectrum of BPS states can be described in terms of quivers, at least in certain chambers of their quantum moduli space. In this case we say that the theory has the quiver property \cite{Cecotti:2010fi,Cecotti:2011rv,Alim:2011ae,Alim:2011kw}.

Recall that a quiver is a finite directed graph, consisting in the quadrupole $Q = (Q_0 ,Q_1 ,t,s)$. Here $Q_0$ and $Q_1$ are two finite sets which represents the nodes and the arrows of the quiver, respectively. The two linear maps $s , t \, : \, Q_1 \longrightarrow Q_0$ specify the starting node $s(a) \in Q_0$ and the ending node $t(a) \in Q_0$ of every arrow. For every quiver $Q$ we can define its algebra of paths $\complex Q$ as the algebra whose elements are the arrows and where multiplication is given by the concatenation of paths whenever possible. To a quiver we can associate a superpotential $\cW \, : \, Q_1 \longrightarrow \complex Q$ which has the form of a sum of cyclic monomials. On such a function we can define an formal derivative $\partial_a$ for each $a \in Q_1$, which cyclically permutes the elements of a monomial until $a$ is at the first position and then deletes it, or gives zero if $a$ does not appear in the monomial. The Jacobian algebra $\mathscr{J}_\cW = \complex Q / \langle  \partial \, \cW  \rangle$ is the quotient of $\complex Q$ by the two sided ideal of relations $\mathsf{R} = \langle \partial_a \, \cW \, \vert \, a \in Q_1 \rangle$.

When a theory has the quiver property, one can find a basis $\{ e_i \}$ of the lattice of charges $\Gamma$ corresponding to stable hypermultiplets, and whose central charges $Z_{e_i} (u)$ lie in the upper half plane $\mathfrak{h}_\theta = \mathrm{e}^{- \ii \theta} \mathfrak{h}$. Furthermore we require this basis to be positive, that is for every BPS state of charge $\gamma$, we can write $\gamma = \sum_i n_i e_i$ where all the $n_i$ are positive (or negative) integers. These conditions are typically met at a point $p$ of the parameter space $\mathscr{P}$ and explicitly depend on $(p, \theta)$. Out of this basis we construct the BPS quiver $Q_{p, \theta}$ by labelling the vertices with the basis elements $\{ e_i \}$ and connecting two vertices $e_i$ and $e_j$ by a (signed) number of arrows given by the Dirac-Schwinger-Zwanziger pairing $\langle e_i , e_j \rangle$. We call $B_{ij} = \langle e_i , e_j \rangle$ the adjacency matrix of the quiver. At the same point $p \in \mathscr{P}$ two different basis $\{ e_i \}$ and $\{ e' \}$ which obey these properties are \textsc{pct} equivalent. Quivers $Q_{p,\theta}$ associated to $\textsc{pct}$ equivalent basis are related by sequences of quiver mutations. An elementary quiver mutation $\mu_{\hat{e}}^{\pm}$ is defined as 
\be \label{mutpm}
e^\prime \equiv \mu^{\pm}_{\hat{e}}(e) \equiv \begin{cases} -e &\text{if } e = \hat{e}\\ e + \text{max}(0,\pm\langle \hat{e} ,\, e \rangle) \hat{e} &\text{otherwise}\end{cases} \, ,
\ee
or equivalently in terms of the adjacency matrix
\begin{equation} \label{mutB}
B'_{ij} = \left\{ \begin{matrix} - B_{ij} & \text{if} \ i=k  \ \text{or} \ j=k \\ B_{ij} + \text{sgn} (B_{ik}) [ B_{ik} \, B_{kj} ]_+ & \text{otherwise} \end{matrix} \right. \ .
\end{equation}

Consider a theory with the quiver property. Then the low energy dynamics of a particle can be described by an effective matrix quantum mechanics based on the quiver \cite{Cecotti:2010fi,Cecotti:2011rv,Alim:2011ae,Alim:2011kw}. Such a model has four supercharges, so that stable BPS states correspond to its supersymmetric ground states. Such ground states can be elegantly described in terms of stable quiver representations\footnote{More precisely one trades the F-term and D-term equations of the matrix quantum mechanics, modulo gauge transformations, for the F-terms equations modulo complexified gauge transformations with an extra stability condition.}. Quiver representations are defined by the assignment of a vector space $V_i$ to each node $i \in Q_0$ and morphisms $B_a \, : \, V_{s (a)} \longrightarrow V_{t (a)}$, for each arrow $a \in Q_1$. For example the aforementioned basis $E_{p,\theta}$ of stable hypermultiplets correspond to simple quiver representations: an element $e_i$ of the basis is the representation where all the maps are set to zero and only one one-dimensional vector space $V_i$ is assigned to a single node $i$. We will denote by $\mathsf{rep} (Q)$ the category of representations of the quiver $Q$. When $Q$ has a superpotential $\cW$, we require the representation morphisms to be compatible with the relations $\partial \, \cW = 0$, and denote the category of representation of the quiver with superpotential $(Q , \cW)$ by $\mathsf{rep} (Q , \cW)$. In many occasions, as it will be the case in this paper, it is convenient to switch from the language of quiver representations to the language of left modules over the algebra $\mathscr{J}_\cW$. The two perspectives are equivalent and the category of left $\mathscr{J}_\cW$-modules $\mathscr{J}_\cW-\mathsf{mod}$ is equivalent to $\mathsf{rep} (Q , \cW)$. Finally in physical applications we are always interested in isomorphism classes of representations under the action of $\prod_{i \in Q_0} \, \mathrm{GL} (\dim_\complex V_i , \complex)$. 

A state with charge $\gamma = \sum_i d_i \, e_i$ with all $d_i \in \mathbb{Z}_+$ corresponds to a representation with dimension vector $\mathbf{d}$ with components  $d_i = \dim_{\mathbb{C}} V_i$. 

Finally the stability condition is determined by the central charge. At fixed $u$ the action of the central charge $Z (u)$ on the lattice of charges $\Gamma$ induces an action on the topological K-theory group $Z (u) \, : \, K (\mathsf{rep}(Q , \cW)) \longrightarrow \complex$. We say that a BPS state with charge $\gamma_r$, corresponding to a representation $\mathsf{R} \in \mathsf{rep}(Q , \cW)$ is stable (semi-stable) if for every proper sub-representation $\mathsf{S} \in \mathsf{rep}(Q , \cW)$, associated with a BPS state with charge $\gamma_s$, one has $\arg \, Z_{\gamma_s} (u) < \arg \, Z_{\gamma_r} (u)$ ($\arg \, Z_{\gamma_s} \le \arg \, Z_{\gamma_r}$).

When the spectrum of BPS states of a model can be described in terms of quiver representation theory, the same is true for framed BPS states \cite{Chuang:2013wt,Cirafici:2013bha,Cordova:2013bza}. When a model is coupled to a line defect $L_{\zeta,\alpha}$ the low energy description of framed BPS states is captured by an effective quantum mechanics which describes the low energy dynamics of BPS states coupled to an infinitely massive dyonic particle, at a certain point in the Coulomb branch. The dyonic particle has core charge $\gamma_c$, which however depends on the point in $\cB$. The core charge is determined by the renormalization group map $\mathbf{RG}$. At the level of the quiver, this coupling is described by the framing. The framed BPS quiver is defined as follows: first we extend the lattice of charges $\Gamma$ to the torsor $\Gamma_L = \Gamma + \gamma_L$, where $\langle \gamma_L , \gamma \rangle \in \zed$ for all $\gamma \in \Gamma$. Then we add to the quiver $Q_{\theta,p}$ an extra node $e_f$ where the charge $\gamma_f$ is the core charge of the defect at $p \in \mathscr{P}$, and connect it to the rest of the quiver via the Dirac pairing. We denote by $Q_{\theta,p} [f]$ or simply $Q[f]$ the resulting framed quiver. The central charge function is extended to $Q [f]$ by linearity. The superpotential for $Q [f]$ has now two terms
\be
\cW \equiv \cW_Q + \cW_{L} \, ,
\ee
where $\cW_Q$ is the superpotential of $Q$ and $\cW_L$ contains arrows that are connected to the framing node $f$. Note that in general $\cW$ will not be the same superpotential obtained by considering $Q[f]$ as an unframed quiver and then sending the mass of a BPS particle to infinity. The reason for this is that when deriving a BPS quiver one has already taken a Wilsonian limit. After such a limit is taken, heavy degrees of freedom have been integrated out and it is not anymore possible to send the mass of any state to infinity. In general $\cW_L$ has to be determined by other methods. A correct microscopic procedure would be first to engineer the model with a collection of D-branes on a local threefold; then to take the size of a cycle corresponding to the core charge to be very large, in an appropriate scaling limit; and only as the final step take the Wilsonian limit to derive the low energy effective quantum mechanics \cite{Chuang:2013wt}. In this paper we will determine $\cW_L$ indirectly. Furthermore since we will employ localization techniques within the context of topological models, we will always have the freedom to change the relevant actions by BRST-exact terms in order to choose a more convenient form.

Consider now the special case of \eqref{vevLX} when the line defect is a Wilson line $W_{\zeta , \mathbf{k}}$ in the representation $\mathbf{k}$ of an asymptotically free theory based on a gauge group $G$. In the Coulomb branch $G$ is broken to its maximal torus and the representation spaces of $\mathbf{k}$ will decompose into their weight spaces. We can therefore always pick one of the weight of the representation $\mathbf{k}$ as IR label. Indeed in this situation the core charge can be identified with the highest weight of the representation $\mathbf{k}$ \cite{Gaiotto:2010be,Moore:2015szp}. Gauge invariance guarantees that all weights should then appear in the expansion \eqref{vevLX}, as color states of the core charge:
\begin{equation} \label{Wexpweights}
\langle W_{\zeta , \mathbf{k}} \rangle  = \sum_{\mathbf{w} \, \text{weight}} \ \underline{\overline{\Omega}} (u,W,\gamma; q=-1) \ X_{\mathbf{e} \cdot B^{-1} \cdot \mathbf{w}} + \text{quantum effects} \, .
\end{equation}
Here $\mathbf{e}$ represents the standard basis of $\real^{\dim \mathbf{k}}$ and the sums runs over all the weight vectors $\mathbf{w}$. The combination $\mathbf{e} \cdot B^{-1} \cdot \mathbf{w}$ expresses elements of the charge lattice $\Gamma$ in terms of the weights of the representation $\mathbf{k}$ and the BPS quiver adjacency matrix $B$. A more detailed description is in \cite{BPSlinesCluster}. The extra terms represent quantum effects which are not visible from the perturbative limit. Physically they correspond to the fact that the line operators in the Coulomb branch are not simply those of an abelian theory, but receive an infinite series of quantum corrections from ordinary BPS particles.

\section{Framed BPS degeneracies from equivariant localization} \label{eqloc}

In this Section we will set up equivariant localization techniques to compute the framed BPS degeneracies $\underline{\overline{\Omega}} (L , u , \gamma ; q=-1)$ corresponding to line defects. In the next Sections we will see explicitly how these techniques can be used in practice in various cases. We will use cyclic and co-cyclic stability conditions to define appropriate moduli spaces with a natural toric action and show how the framed BPS degeneracies can be computed directly using localization with respect to this toric action. These techniques are rooted in the analysis of \cite{Moore:1997dj,Moore:1998et} which pioneered localization in the context of topological quantum mechanics, and have been widely used in the mathematics \cite{szendroi,reineke} and physics \cite{Ooguri:2008yb,Cirafici:2008sn,Cirafici:2010bd} literature to compute enumerative invariants of Donaldson-Thomas type on noncommutative resolutions of Calabi-Yau singularities. These invariants are typically formulated in terms of framed quivers associated with the singularities. In that setup the framing nodes represent infinitely massive $D6$ branes wrapping the local threefold and the enumerative invariants count bound states with lower dimensional light branes on compactly supported cycles. This is similar to the situation we have with line defects: the framing node correspond to an infinitely heavy particle coupled to lighter BPS states. Indeed our analysis will support the conclusions of \cite{Chuang:2013wt} that framed BPS degeneracies can be identified with the noncommutative Donaldson-Thomas invariants of \cite{szendroi}. 

In this Section we will discuss our formalism for the particular case of an SU(2) model. We will however set up the discussion in more general terms, such that it can be extended straightforwardly to more general framed quivers.

\subsection{Generalities}

We begin by discussing some general qualitative ideas about the localization computation, which we will make more precise in the remaining of this Section. Techniques of localization are by now commonly used in quantum field theory after the seminal works \cite{Witten:1988ze,Nekrasov:2002qd,Pestun:2007rz}. We refer the reader to the reviews \cite{Szabo:2000fs,Cordes:1994fc,Pestun:2014mja,Pestun:2016qko} for a more in depth discussion. The expert reader can safely skip this Subsection.

Our task is to study the ground states of a certain supersymmetric quiver quantum mechanics associated with a framed BPS quiver. This in practice entails studying the moduli space of solutions of the D-term and F-term equations modulo gauge transformations, and its cohomology. In these situations it is usually convenient to study a closely related moduli space, obtained by only imposing the F-term equations and taking the quotient respect to complexified gauge transformations. Upon imposing a suitable stability condition, the moduli space obtained in this way coincides with the moduli space of physical vacua. 

This approach is particularly convenient when dealing with quiver quantum mechanics, since after imposing the F-terms the relevant moduli space is the moduli space of quiver representations, or modules over the quiver Jacobian algebra, a well studied object. For the problem at hand, there are two particularly natural stability conditions, which correspond to cyclic and co-cyclic modules \cite{Cordova:2013bza}. We will usually choose cyclic stability conditions, for which the moduli space of vacua will be represented by the moduli space $\mathcal{M}_{\mathbf{d}} (Q , W ; v)$ of cyclic modules over the Jacobian algebra $\mathbb{C} Q / \partial \, \mathcal{W}$, generated by a framing vector $v$. 

Note that since we are only interested in the quantum mechanics ground states, one can equivalently perform a topological twist and study the partition function of the resulting topological quantum mechanics \cite{Moore:1997dj,Moore:1998et}. This problem can be studied using techniques of equivariant localization. In this paper we will use localization techniques to compute BPS invariants directly; it is however useful to have this quantum mechanics as a concrete physical model in mind. We are interested in counting four dimensional BPS states which correspond to ground states of a supersymmetric quiver quantum mechanics with four supercharges. The problem of counting ground states can be solved by going to the topologically twisted sector of the quantum mechanics; equivalently one can construct directly a topological quiver quantum mechanics whose partition function compute the relevant index of bound states, using the formalism of \cite{Moore:1997dj,Moore:1998et}. Here we review this formalism, following \cite{Cirafici:2008sn,Cirafici:2010bd} in the exposition. Since we will not use this formalism we will be rather schematic. One starts with a set of F-term and D-term equations $\vec{\mathcal{E}}= \vec0$. Such equations will depend on complex fields, which we denote collectively by $X_a$, associated to the arrows of the quivers. One forms multiplets $(X_a , \Psi_a)$ with BRST transformations
\begin{equation} \cQ \, X_a = \Psi_a \, , \qquad \mbox{and}
\qquad \, \cQ \, \Psi_a = [\phi_k , X_a] \ .
\end{equation}
Geometrically we can think of the $\Psi_a$ as differentials on the moduli space parametrized by the $X_a$. In the quiver setting the fields $X_a$ are really morphisms $X_a \in \mathrm{Hom} (V_{s (a)} , V_{t (a)})$ associated with the arrows while the gauge parameters $\phi_k$ are $GL (V_k , \complex)$-valued and associated with the nodes. The linearized gauge transformation $[\phi_k , X_a] $ should be properly written as $ \phi_{s (a)} \, X_a - X_a \, \phi_{t (a)} $. We will avoid spelling out these details and simply write $[\phi_k , X_a] $ for the gauge transformations, confident that no confusion can arise. To this set of fields we add the Fermi multiplet $(\vec{\chi} , \vec{H}\,)$ of antighosts and auxiliary fields with the same quantum numbers as the equations $\vec{\mathcal{E}}$. In case some of the equations are overdetermined, additional multiplets with opposite statistics have to be added to avoid overcounting of degrees of freedom. In this case there exists an additional set $\mathcal{\tilde{E}}$ of ``relations between the relations'' and the corresponding multiplets. By abuse of notation we will still denote the set of all these equations and multiplets by $\vec{\mathcal{E}}$ and  $(\vec{\chi} , \vec{H}\,)$. Finally the gauge multiplets $(\phi_k ,\overline{\phi}_k ,  \eta_k)$ has to be added in order to close the BRST algebra.

We will work equivariantly with respect to a natural torus $\torus$ generated by parameters $\epsilon_a$, which rescales the fields $X_a \rightarrow X_a ~\e^{\ii \epsilon_a}$. This modifies the BRST differential to an equivariant differential
\begin{equation}
\cQ_\epsilon \, X_a = \Psi_a \, ,  \qquad \mbox{and}
\qquad \, \cQ_\epsilon \, \Psi_a = [\phi , X_a] + \epsilon_a\,X_a \ .
\end{equation}
These transformations can be extended to the Fermi multiplets by keeping track of the transformations of the equations $\vec{\mathcal{E}}$ under $\torus$, while the gauge multiplet transformations are unchanged.  In particular let us denote by $\epsilon_{c,i}$ the toric weights associated with the transformation of the F-term equations, which are just linear combinations of the parameters $\epsilon_a$. The quantum mechanics constructed in this fashion is cohomological and its action
\begin{equation}
S = \cQ_\epsilon \Tr\left( \eta\, \big[\phi \,,\, \overline{\phi}~\big] -
\vec{\chi} \cdot \vec{\mathcal{E}} + g\, \vec{\chi} \cdot \vec{H} +
\Psi_i\,\big[X_i \,,\, \overline{\phi}~\big] \right)
\end{equation}
a BRST variation. Note that this action can be modified by adding $\cQ_\epsilon$-exact terms. For example we can chose 
\begin{equation}
t_1 \, \cQ_\epsilon  \sum_r \Tr\,\chi_r \,\overline{\phi} + t_2 \, \cQ_\epsilon \Tr\big(X_i\, \Psi_i^{\dagger} - X_i^\dag\,\Psi_i\big) \ .
\end{equation}
where $\chi_r$ are the antighosts associated with the D-term equations, one for each node of the quiver. Since the theory is cohomological, it does not depend on the three parameters $g$, $t_1$ and $t_2$ which only enter in $\cQ_\epsilon$-exact terms. To evaluate the path integral one firstly diagonalizes all the gauge parameters $\phi_k$ producing a set of Vandermonde determinants. Then can now evaluate the path integral in three steps, taking first the $t_1 \rightarrow \infty$ limit, then the limit $g \rightarrow \infty$ and finally sending $t_2 \rightarrow \infty$. These limits allow to integrate out fields and produce ratios of determinants, depending on the bosonic or fermionic statistics of the field which is integrated out. The result is a contour integral over the Cartan subalgebras of the gauge parameters $\phi_k$, which has the structure
\begin{equation}
\int_\cC \prod_k \dd \phi_k \ \frac{\prod_{\rm g} \left( \mathrm{ad} \ \phi \right)_{\rm g}  \prod_{\rm r} \left(  \mathrm{ad} \  \phi + \epsilon \right)_{r}  }{ \prod_{\rm f} \left(  \mathrm{ad} \ \phi + \epsilon \right)_{\rm f}  \prod_{\rm rr} \left(  \mathrm{ad} \ \phi + \epsilon \right)_{\rm rr} }
\end{equation}
In this notation the products are as follows: g over the gauge parameters associated with the nodes of the quiver, f over the fields associated over the arrows of the quiver, e over the relations of the quiver and rr over the doubly determined relations, if any. Furthermore each determinant is constructed from the gauge and equivariant transformations of the field which was integrated out. For example for a field $X_a \in \mathrm{Hom} (V_{s (a)} , V_{t (a)})$ which transforms as $X_a \rightarrow X_a ~\e^{\ii \epsilon_a}$, the notation $\left(  \mathrm{ad} \ \phi + \epsilon \right)_{\rm f} $ stands for the operator which acts on $X_a$ as $ \phi_{s (a)} \, X_a - X_a \, \phi_{t (a)}  + \epsilon_a \, X_a $. Similarly for all the other determinants. Finally the contour has to be chosen appropriately to select the fixed points of the $\cQ_\epsilon$ differential. This derivation has been very sketchy and we refer the reader to \cite{Cirafici:2008sn,Cirafici:2010bd} for a lengthier discussion.

In the rest of the paper we will skip the cohomological quiver quantum mechanics formalism and evaluate equivariant integrals directly. Indeed what the cohomological formalism construct is a (Mathai-Quillen) representative of a certain equivariant class over the moduli space, see \cite{Pestun:2016qko} for a review. This class is a characteristic class of the obstruction bundle, whose sections are the F-term relations. Roughly speaking equivariant localization reduces the computations of integrals over a moduli space upon which an algebraic group acts, to a simpler integration over sub-varieties which are fixed by the action of said group. In the simplest and most useful case, the fixed locus consists of isolated points. In our applications the group acting on our moduli spaces will always be a torus $\mathbb{T}$. The Atiyah-Bott localization formula then states that
\begin{equation} \label{ABformula}
\int_{\mathcal{M}} \, \alpha = \sum_{f \in \mathcal{M}^{\mathbb{T}}} \frac{\alpha \vert_f}{\mathrm{eul}_{\mathbb{T}}  \left( T_f \mathcal{M} \right)}
\end{equation} 
for any equivariant cohomology class $\alpha$. The integral of $\alpha$ over a smooth variety $\mathcal{M}$ is expressed as a sum over the fixed points $f$ in the fixed point locus $\mathcal{M}^{\mathbb{T}}$. Each fixed point contributes with the class $\alpha$ evaluated at the fixed point, and each contribution is weighted by the Euler class of the tangent space at the fixed point $\mathrm{eul}_{\mathbb{T}}  \left( T_f \mathcal{M} \right)$. The usefulness of the formula is that, no matter how complicated are the physical configurations parametrized by $\mathcal{M}$, only certain configurations fixed by the toric action will contribute to the equivariant integral. Therefore integrals over a moduli space $\mathcal{M}$ are in principle computable once the fixed point configurations are classified, and the local structure of the moduli space around each fixed point $T_f \mathcal{M}$ is known. This can be done by constructing a local model of the tangent space around each fixed point by linearizing the moduli space equations. The information about the contribution of each fixed point is then elegantly encoded in a deformation complex, adopting the approach of \cite{nakajima}. Later on we will see how this formalism applies to the computation of BPS indices associated with framed quivers. The computation of BPS indices will be greatly simplified and reduced to a simple combinatorial problem.

In many cases physical moduli spaces $\mathcal{M}$ are not smooth varieties; however if they are smooth ``generically" and have a well defined virtual tangent space, a stronger version of \eqref{ABformula} holds, the virtual localization formula \cite{graber}. This is the version of equivariant integration that we will use to compute BPS invariants. The main reason is that the corresponding virtual counts are what typically enters in physics as BPS indices. In simple terms our formalism will be a simple extension of the well known localization computations for BPS invariants of the Hilbert scheme of points $\mathrm{Hilb}^n (\complex^3)$. We will now discuss in more detail our formalism. The reader who wishes to see more clearly the physical counterpart of the virtual formalism is encouraged to think in terms of topological quiver quantum mechanics.

\subsection{Moduli spaces and framed degeneracies}

In the following we will freely switch between representations of the quiver $Q$ and finitely generated left $\mathscr{J}_\mathcal{W}$-modules. To begin with, our quiver $Q$ will be the Kronecker quiver, framed by an extra node $f_{\mathbf{n}}$ connected to the quiver with $n-1$ arrows for each node:
\begin{equation} \label{KroneckerSU2}
\xymatrix@C=8mm{
&  \bullet  \ar@{..>}@<-0.5ex>[dl]_{B_1, \cdots , B_{n-1}}  \\
f_{\mathbf{n}}  \ar@{..>}@<-0.5ex>[dr]_{C_1,  \cdots , C_{n-1}}   & \\
& \circ \ar@<-0.5ex>[uu]_{\tilde{A}}  \ar@<0.5ex>[uu]^{A} 
}
\end{equation}
We will denote this quiver as $Q[f_{\mathbf{n}}]$ or just $Q[\mathbf{n}]$, and its superpotential by $\mathcal{W}$. Later on we will consider more general quivers. Introduce the representation space
\begin{align}
\mathsf{Rep} (Q [f_{\mathbf{n}}]) & =  \bigoplus_{a \in Q_1 [f_{\mathbf{n}}]} \, \mathrm{Hom} (V_{s (a)} , V_{t (a)}) 
\cr
&  =  \bigoplus_{a \in Q_1 } \, \mathrm{Hom} (V_{s (a)} , V_{t (a)})  \oplus \Hom (V_\bullet , V_f)^{\oplus n-1} \oplus \Hom (V_f , V_\circ)^{\oplus n-1}
\end{align}
where $V_f = \mathbb{C}$ will conventionally denote the representation space supported at the framing node, which will always be one dimensional. We have introduced the notation $V_\bullet$ and $V_\circ$ to denote the representation spaces based at the two nodes $\bullet , \circ$ of the quiver. We will employ this notation in the rest of the paper.

Assume we are given a superpotential $\cW$. We define $\mathsf{Rep} (Q [f_{\mathbf{n}}] , \mathcal{W})$ as the subspace obtained by imposing the relations $\partial \, \mathcal{W} = 0$. The moduli space of quiver representations of dimension $\mathbf{d}$ is the smooth Artin stack
\begin{equation}
\mathcal{M}_{\mathbf{d}} (Q [f_{\mathbf{n}}]) = \left[ 
\mathsf{Rep}_{\mathbf{d}} (Q [f_{\mathbf{n}}] , \mathcal{W}) / GL (d_{\circ} , \mathbb{C}) \times GL (d_{\bullet} , \mathbb{C})
\right]
\end{equation}
where $\mathsf{Rep}_{\mathbf{d}} (Q [f_{\mathbf{n}}] , \mathcal{W})$ denotes the representation space with fixed dimension vector and $d_{\circ} = \dim_\mathbb{C}  V_\circ$ and $d_\bullet = \dim_\mathbb{C} V_\bullet$.  The gauge group $GL (d_{\circ} , \mathbb{C}) \times GL (d_{\bullet} , \mathbb{C})$ acts by basis rotation at each node, and the superpotential $\mathcal{W}$ is gauge invariant. We are interested in extracting Donaldson-Thomas type invariants from these moduli spaces. To do so we need to impose a stability condition which selects the stable framed BPS states. Note that mathematically we expect framed moduli spaces to be better behaved that their unframed counterparts, since a generic enough framing condition will not be preserved by the automorphisms of the original moduli space.

In the following we will impose cyclic and co-cyclic stability conditions\footnote{Recall that a representation is \textit{cyclic} if it admits a cyclic vector. Namely, let $V$ be a representation for an algebra $\mathsf{A}$. Then a vector $v \in V$ is called cyclic if $V = \mathsf{A} \, v$, and $V$ is called a cyclic representation.  Cyclic representations are naturally associated with ideals. Indeed a representation $V$ is cyclic if and only if it is of the form $V = \mathsf{A} / \mathsf{I}$ where $\mathsf{I}$ is a left ideal of $\mathsf{A}$. Similar arguments hold for co-cyclic representations. We refer the reader to the text \cite{wisba} for more details.}. These two choices are particularly natural and correspond to the fact that the phase of the central charge of the defect is much bigger, or much smaller, than that of all the other particles, while keeping the mass of the defect much bigger than all the other masses \cite{Cordova:2013bza}. These two choices greatly simplify the stability analysis. A representation $R \in \mathsf{rep} (Q [f_{\mathbf{n}}] , \mathcal{W})$ is cyclic if the only non trivial sub-representation which has non vanishing support at the framing node $V_f$, is $R$ itself. Similarly a representation  $R \in \mathsf{rep} (Q [f_{\mathbf{n}}] , \mathcal{W})$ is co-cyclic if all the non trivial sub-representations of $R$ have non vanishing support at the framing node $V_f$. Note that the two notions are interchanged upon opposing the quiver. 

Equivalently we can talk about cyclic and co-cyclic modules of $\mathscr{J}_\mathcal{W}$. In this case a module $M$ is a cyclic left $\mathscr{J}_\mathcal{W}$-module if it is generated by a vector $v \in M$. We will be interested in the case where the cyclic module is based at a certain vertex of the quiver, that is $v \in \mathsf{e}_i \, M$, with $\mathsf{e}_i$ an idempotent of the Jacobian algebra $\mathscr{J}_\mathcal{W}$ of $Q [f_{\mathbf{n}}]$. In this case the relevant representation space is $\mathsf{Rep} (Q [f_{\mathbf{n}}] ; v_k)$, defined as the sub-scheme of  $\mathsf{Rep} (Q [f_{\mathbf{n}}])$ which consists of modules generated by the vector $v_k \in V_k$, with $k \in Q_0 [f_{\mathbf{n}}]$. In the same way $\mathsf{Rep} (Q [f_{\mathbf{n}}]  , \mathcal{W} ; v_k)$ denotes the sub-scheme of $\mathsf{Rep} (Q [f_{\mathbf{n}}] ; v_k)$ cut out by the F-term equations $\partial \, \mathcal{W} = 0$. Similarly a co-cyclic module can be characterized by the property that there exists a simple submodule $N$ of $M$ which is contained in every non zero submodule of $M$ (for example see \cite[Thm 14.8]{wisba}). We will adopt a more practical view of co-cyclic modules and simply regard them as cyclic modules of the opposite quiver $(Q [f_{\mathbf{n}}])^{\textsf{op}}$.

Note that in general we will have more than one cyclic vector. For example if we pick a vector $v_f \in V_f$, all vectors of the form $C_i \, v_f = v^{(i)} \in V_\circ$ can be used to generate cyclic modules, as in \cite{nagaonakajima}. Note that here $C_i$ are \textit{fixed} framing morphisms. To be precise we should therefore speak of cyclic modules generated by a collection of fixed distinct vectors $\{ v^{(i)} \}$ determined by the framing. To avoid cluttering the notation, we will loosely speak of cyclic modules generated by $v_f \in V_f$, hoping that this will cause no confusion.

The relevant moduli space is now the moduli space of cyclic modules $\mathcal{M}_{\mathbf{\delta}}^{(c)} (Q [f_{\mathbf{n}}], v) \subset \mathcal{M}_{\mathbf{\delta}} (Q [f_{\mathbf{n}}])$, defined by the condition that each module is generated by a vector $v \in V_f$ with fixed framing morphisms, or the moduli space of co-cyclic modules $\mathcal{M}_{\mathbf{\delta}}^{(cc)} (Q [f_{\mathbf{n}}] , v) \subset \mathcal{M}_{\mathbf{\delta}} (Q [f_{\mathbf{n}}])$. The dimension vector $\mathbf{\delta}$ is of the form $\mathbf{\delta} = (1 , \mathbf{d})$, which corresponds to the condition that the vector space $V_f$ at the framing node is always one-dimensional. Physically this corresponds to the fact that the ground state of the defect can be thought of as an infinitely massive dyon, in a certain region of the moduli space.

We will also use the short-hand notation $\mathcal{M}_{\mathbf{d}}^{(c)}$ and $\mathcal{M}_{\mathbf{d}}^{(cc)}$ for these moduli spaces. Note that this is slightly different from the definition used in the quiver literature \cite{szendroi,reineke}, for which cyclic modules are modules of the \textit{unframed} quiver $Q$ together with a morphism from the framing node to a node of $Q$. We will argue later on that these two definitions are equivalent, thanks to the constraint that $\dim V_f = 1$ for a line defect. From our perspective fixing a vector within the framing node or its image in the unframed quiver is merely a convenient notational choice. The reason we prefer this notation is that often in our problems the arrows starting from and ending at the framing node will be constrained by superpotential terms.

We will interpret the degeneracies of framed BPS states as noncommutative Donaldson-Thomas invariants associated with these moduli spaces, for example defined as weighted topological Euler characteristics \cite{behrend} of these moduli spaces. This is physically natural and expected from the properties of D-brane bound states on local toric Calabi-Yaus, where in the noncommutative crepant resolution chamber the infinitely massive D6 brane wrapping the threefold is modeled as a framing node \cite{Ooguri:2008yb,Cirafici:2008sn,Cirafici:2010bd,Cirafici:2011cd,Cirafici:2012qc,Cirafici:2013tna,Chuang:2013wt,Chuang:2008aw}. For example we can define BPS invariants by integration over the moduli space of cyclic modules as 
\begin{align} \label{DTinv}
\underline{\overline{\Omega}} (u , L_{\zeta , \alpha} , \gamma , q=-1) &= {\tt DT}_{\mathbf{d}}^{(c)} (L_{\zeta , \alpha} ) = \chi (\mathcal{M}_{\mathbf{d}}^{(c)}  (Q [f_{\mathbf{n}}] , v)  , \nu_\mathsf{A}) 
\nonumber \\[4pt]  & = (-1)^{\dim T \mathcal{M}_{\mathbf{d}}^{(c)}  (Q [f_{\mathbf{n}}] , v)} \chi (\mathcal{M}_{\mathbf{d}}^{(c)}  (Q [f_{\mathbf{n}}] , v))
\end{align}
where $\gamma = \sum_{i \in Q_0 [f_{\mathbf{n}}]} e_i \, \d_i = e_f +  \sum_{i \in Q_0}  e_i \, d_i$ and $e_f$ is the core charge of the defect. Similarly for co-cyclic modules. Here $\nu_\mathsf{A}$ is a canonical constructible function \cite{behrend}. Note that the indices \eqref{DTinv} contain much more information than just the Euler characteristics of the moduli spaces, which is at the origin of their intricate wall-crossing behavior. Out of these enumerative invariants, we will construct the generating functions
\begin{equation}
\langle L_{\zeta , \alpha} \rangle_{q=-1} = \sum_{\mathbf{d}=(d_\circ , d_\bullet)}  {\tt DT}_{\mathbf{d}}^{(c)} (L_{\zeta , \alpha} ) \ X_{e_f + d_\circ \, e_\circ  + d_\bullet \, e_\bullet}
\end{equation}
We will compute the framed degeneracies using equivariant virtual localization to integrate over the moduli spaces (see \cite[Section 3.5]{Szabo:2009vw} for a review of this approach). This is a two step procedure: firstly we have to define an appropriate toric action on the moduli space and classify the fixed points; secondly by studying the local structure of the moduli space around each fixed point, we can use the localization formula to compute (\ref{DTinv}).

\subsection{Toric action}

As a working model, we will use the point of view of the supersymmetric quantum mechanics associated with the line defect, appropriately twisted so that it localizes onto its ground states \cite{Moore:1997dj,Moore:1998et}. From the quantum mechanics perspective there is a natural toric action, which rescales each field in such a way that the F-term relations are preserved. In other words we have a flavor torus $\mathbb{T}_F = (\mathbb{C}^*)^{|Q_1 [f_\mathbf{n}]|}$ which rescales the arrows, and a subtorus $\mathbb{T}_{F , \, \partial \mathcal{W}} \subset \mathbb{T}_F$ which preserves the $\partial \, \mathcal{W} = 0$ relations. Note that this torus acts on the moduli spaces $\cM_{\mbf d}^{c}$ since these are constructed using the F-term relations. Part of this toric action is however induced by gauge transformations; the gauge group $GL (d_0 , \mathbb{C}) \times GL (d_1 , \mathbb{C})$ contains a  gauge torus $\mathbb{T}_G = (\mathbb{C}^*)^2$. An element $(\mu_1 , \mu_2) \in  (\mathbb{C}^*)^2$ will act on a map $X_a \, : \, V_{t(a)} \longrightarrow V_{h(a)}$ as $\mu_1 \, X_a \, \mu_2^{-1} $. The induced action is however just $\mathbb{C}^* = (\mathbb{C}^*)^2 / \mathbb{C}^*$, since diagonal elements act trivially. We will take the torus action to be $\mathbb{T}_{\mathcal{W}} = \mathbb{T}_{F , \, \partial \mathcal{W}} / \mathbb{C}^*$.

To be definite assume a cyclic stability condition. As a working example it is useful to keep in mind the quiver $Q [f_{\mathbf{2}}]$ which describe a Wilson line in the fundamental representation of $SU(2)$. We take as superpotential $\mathcal{W} = B A C - B \tilde{A} C$. The $\partial \, \mathcal{W} = 0$ relations are
\begin{eqnarray} \label{eomSU2-2}
r_A \ , \ r_{\tilde{A}} &:& CB=0 \cr
r_B &:& AC - \tilde{A} C = 0 \cr
r_C &:& BA-B \tilde{A} = 0
\end{eqnarray}
Also note the following relations between the relations (\ref{eomSU2-2})
\begin{eqnarray} \label{rrSU2-2}
rr_\bullet &:& - A \, r_A + \tilde{A} \, r_{\tilde{A}}  + r_B \, B = 0 \cr
rr_\circ &:& - r_A \, A + r_{\tilde{A}} \tilde{A} + C \, r_C = 0
\end{eqnarray}
The torus $\mathbb{T}_{F , \partial \, \mathcal{W}}$ acts on a quantum mechanics field $X$ as
\begin{equation}
X \longrightarrow \e^{\ii \epsilon_X} \, X
\end{equation}
where compatibility with the F-term equations (\ref{eomSU2-2}) requires $\epsilon_A = \epsilon_{\tilde{A}}$. A fixed point is a field configuration such that the toric action can be compensated by a gauge transformation (a change of basis in the representation spaces). As we have remarked before part of the toric action is induced by diagonal gauge transformations. For example such a transformation would rescale all the arrows between $\circ$ and $\bullet$ by a factor $\lambda_1 = g'_\circ (g'_\bullet)^{-1}$, where both $g'_\circ$ and $g'_\bullet$ are diagonal and distinct, and the arrows connecting the framing node by $(g'_\circ)^{-1}$ or $g'_\bullet$. For convenience we can ``partially gauge fix", by imposing an additional condition on the toric weights, without altering the fixed points classification. The toric action of $\mathbb{T}_{\mathcal{W}}$ is obtained by imposing this additional condition; we can choose for example to impose $\epsilon_{A} + \epsilon_B + \epsilon_C = \epsilon_{\tilde{A}} + \epsilon_B + \epsilon_C = 0$ which leaves the superpotential invariant. 
In our case, due to the cyclic stability condition the arrow labeled by $B$ can be effectively removed from the analysis. 
In the following we will give a general argument why this is the case for the quivers under consideration, based on the physical requirement that $\mathrm{dim} V_f =1$, which constraints the allowed toric fixed points.

\subsection{Fixed points and pyramid partitions}

The main use of localization techniques is that they reduce the problem of computing enumerative invariants to a simple combinatorial problem. Enumerative invariants associated with moduli spaces of cyclic modules are well studied in the literature using pyramid partitions \cite{szendroi}, dimer models \cite{Ooguri:2008yb}, plane partitions \cite{Cirafici:2008sn,Cirafici:2010bd,Cirafici:2009ga} and techniques from algebraic topology \cite{Cirafici:2015sdg}. In the following we will adopt the prescription of \cite{szendroi} and use pyramid partitions to classify toric fixed points. While we will focus on our example given by \eqref{KroneckerSU2}, we will discuss how the classification of fixed points works in quite some generality, so that we can apply it word by word to more general quivers.

Cyclic modules over $\mathscr{J}_{\mathcal{W}} = \mathbb{C} Q / \partial \, \mathcal{W}$ are generated by a vector $v \in V_{f}$. Denote by $v_k \in V_k$ its image in $Q$. In our example at hand $v_k$ is based at $V_\circ$ and has the form $C_i \, v$ for some morphism $C_i$. Since $\dim V_f = 1$, we can pick a basis of $V_f$ consisting only of the vector $v$.  Stable states correspond to cyclic modules and therefore to set up the localization formalism we need to characterize the fixed point set $\mathcal{M}_{\mathbf{d}}^c (Q [f] , v)^{\mathbb{T}_\mathcal{W}}$, which corresponds to toric invariant cyclic modules. Equivalently we can reason in terms of ideals, following \cite{szendroi}. For a cyclic module $M$, we have a canonical map $\mathscr{J}_\mathcal{W} \longrightarrow M$ which sends an element $a \in \mathscr{J}_\mathcal{W}$ to $a \, v$. We will denote this map by $\overline{v}$. Then to the module $M$ we can associate the annihilator of the cyclic generator, $I_M = \ker \overline{v}$. This is an ideal, and is generally of the form $I_M = I_f \oplus \bigoplus_{j \neq k} P_j$ where the $P_j$ are projective modules based at any node but $f$ and $k$, and $I_f$ is an ideal based at the framing node $f$ (in our example $I_f =  I_\circ v$ where $I_\circ$ is an ideal based on $\circ$). The projective modules $P_j$ will play only a passive role in the following and will be generically omitted. Since cyclic modules $M$ are in one to one correspondence with ideals $I_M$, we can classify toric fixed points in $\mathcal{M}_{\mathbf{d}}^{c}$ by looking at toric fixed ideals.

The generators of $I_f$ naturally split into path algebra elements which have the same starting point and the same endpoint. On all these paths the gauge torus will act by multiplication of the same constant diagonal matrix. Therefore the covering torus $\mathbb{T}_{F , \partial \mathcal{W}}$ also acts on the ideals and is enough to classify fixed points \cite{szendroi}. In other words there is a one to one correspondence between $\mathbb{T}_\mathcal{W}$-fixed points in $\mathcal{M}_{\mathbf{d}}^{c}$ and $\mathbb{T}_{F , \partial \mathcal{W}}$-fixed ideals $I_f$. In the following we will switch freely between the two. Generators of a $\mathbb{T}_{F , \partial \mathcal{W}}$-fixed ideal must necessarily have a definite weight under the $\mathbb{T}_{F , \partial \mathcal{W}}$ action. Furthermore a $\mathbb{T}_{F , \partial \mathcal{W}}$-fixed ideal $I_f$ is generated by monomials, eigenvectors of $\mathbb{T}_{F , \partial \mathcal{W}}$. Recall that fixed points are such that the toric action can be compensated by a gauge transformation. Therefore a $\mathbb{T}_{F , \partial \mathcal{W}}$-fixed ideal $I_f$ will be generated by linear combinations of path algebra monomials with the same toric weight. For the quivers at hand this can be verified explicitly by direct inspection, listing all the path algebra monomials with the same $\mathbb{T}_{F , \partial \mathcal{W}}$ weight. The latter give the set of vectors
\begin{equation}
\{Êv , C \, v , A C \, v , \tilde{A} C \, v , \cdots \} \in M
\end{equation}
which span the toric fixed module $M$. However not all these vectors will be linearly independent, but some of them will be identified (or set to zero) by the Jacobian algebra relations. Since $I_f$ is a monomial ideal, after imposing these relations we are left with a set of linearly independent vectors (for example by picking one representative for each equivalent class determined by $\partial \, \mathcal{W} = 0$) and form a finite dimensional basis for $M$. Note that in general this will not be true for ideals which are not $\mathbb{T}_{F , \partial \mathcal{W}}$-fixed.

\textbf{Remark:} Note that in principle we could consider vectors of the form $B \, A \, C \, v \in V_f$. By the above argument this can be a vector in a toric fixed module only if it is linearly independent from the vector $v \in V_f$, since they have different $\mathbb{T}_{F , \partial \mathcal{W}}$ weights\footnote{
In general the vector $v$ has trivial toric weight while $B \, A \, C \, v $ has not. We could however try to remove this weight by using the residual gauge transformation to gauge away its phase. However this is a global transformation and will reintroduce the same phase, now multiplying $v$. This is just the statement that fixed points are classified by the $\mathbb{T}_{F , \partial \mathcal{W}}$ action. Later on we will impose gauge conditions which remove the toric weight from vectors of the form $B \, A \, C \, v$, but as we have just stated this does not affect the classification of $\mathbb{T}_{F , \partial \mathcal{W}}$-fixed points
.}. In other words, while modules where $B \, A \, C \, v $ and $v$ are two parallel vectors will generically be present within the moduli space, they drop out of the localization formula and don't contribute to the index. Since in a toric fixed module  $B \, A \, C \, v $ and $v$ are linearly independent, this means that $V_f$ is at least two-dimensional. Considering more complicated cycles we would conclude that the vector space $V_f$ could have arbitrary dimension. This is however in contradiction with our condition: since the framing vector represents a line defect, obtained by sending to infinity the mass of a stable particle, its vector space $V_f$ is strictly one dimensional. This extra condition implies that, for what concerns counting toric fixed cyclic modules associated with line defects, the Jacobian algebra of the quiver is effectively restricted only to the unframed quiver. We see now that our definition of moduli spaces is indeed equivalent to the standard definition \cite{szendroi,reineke}: since the arrow $B$ can be effectively removed (keeping however its equation of motion as relations in $\mathscr{J}_\cW$), we can simply dispose of the framing node altogether and consider cyclic modules generated by $C \, v \in V_\circ$.

So far we have provided a classification of $\mathbb{T}_{F , \partial \mathcal{W}}$-fixed ideals. However to count BPS invariants associated to framed quivers it is customary in the literature to use an elegant reformulation in terms of certain combinatorial arrangements, known as \textit{pyramid partitions} \cite{szendroi,Ooguri:2008yb}. Pyramid partitions are certain configurations of colored stones, where each stone is associated with a node of the unframed quiver $Q$ and different colors corresponds to different nodes\footnote{The simplest example of a pyramid partition is a Young tableau, corresponding to a framed quiver with a single node and two arrows from that node to itself \cite{nakajima}. Cyclic modules of this quiver correspond to point-like instantons and play a fundamental role in Seiberg-Witten theory \cite{Nekrasov:2002qd}}. This construction is a simplified version of the construction of \cite{szendroi}, Proposition 2.5.1, which we follow closely. To define a pyramid partition one starts from a \textit{pyramid arrangement}. We arrange the stones in layers, and the color of the first layer is determined by which node is associated with the vector $C \, v$. Conventionally, and for practical drawing reasons, we will not associate any stone to the zeroth layer associated with the framing vector $v \in V_f$.

\begin{figure}[H]
    \centering
    \includegraphics[width=0.20\textwidth]{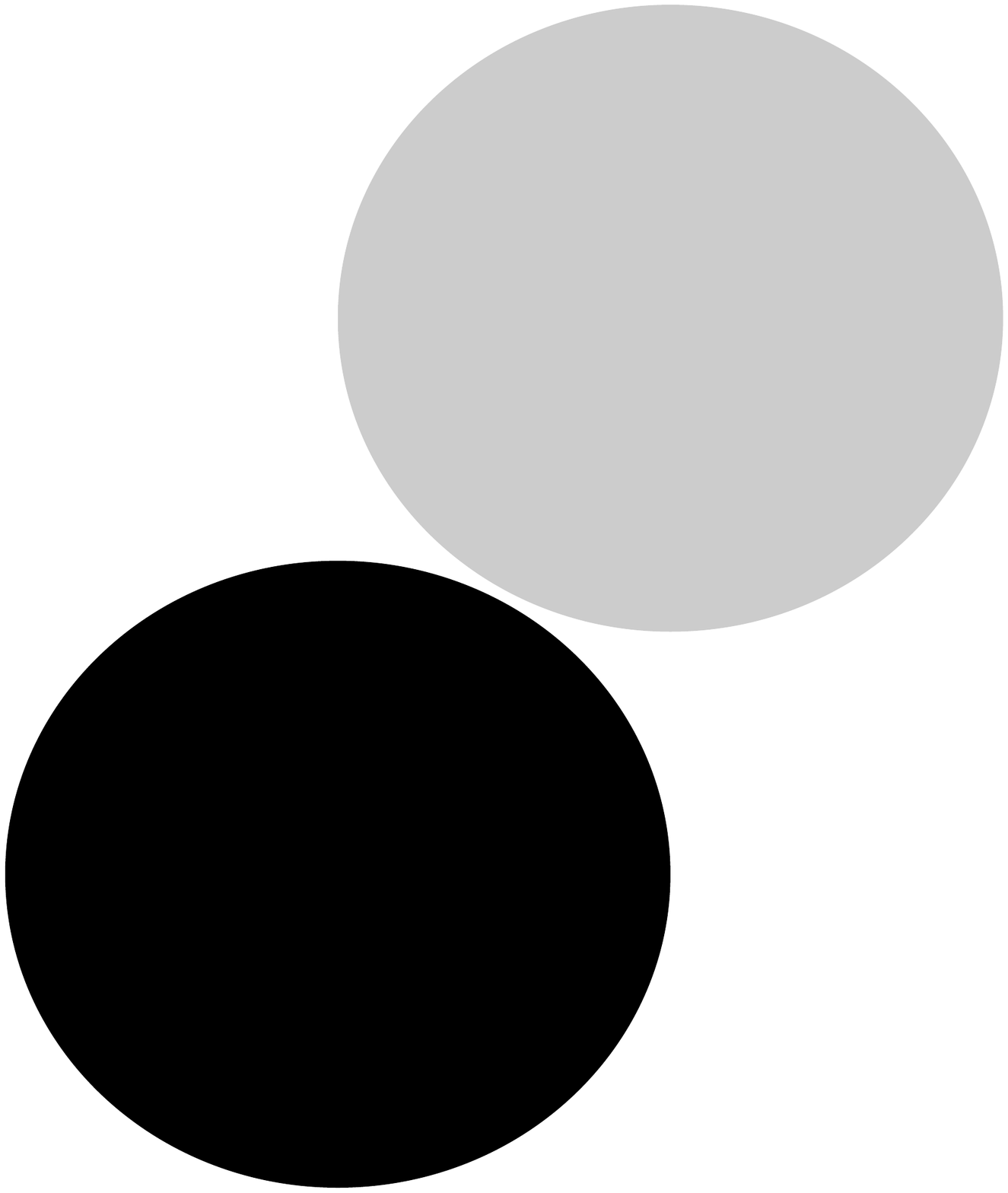}
    \caption{Pyramid arrangement for $SU(2)$ with a fundamental Wilson line}
    \label{initialSU2-2}
\end{figure}

\begin{figure}[H]
    \centering
    \includegraphics[width=0.20\textwidth]{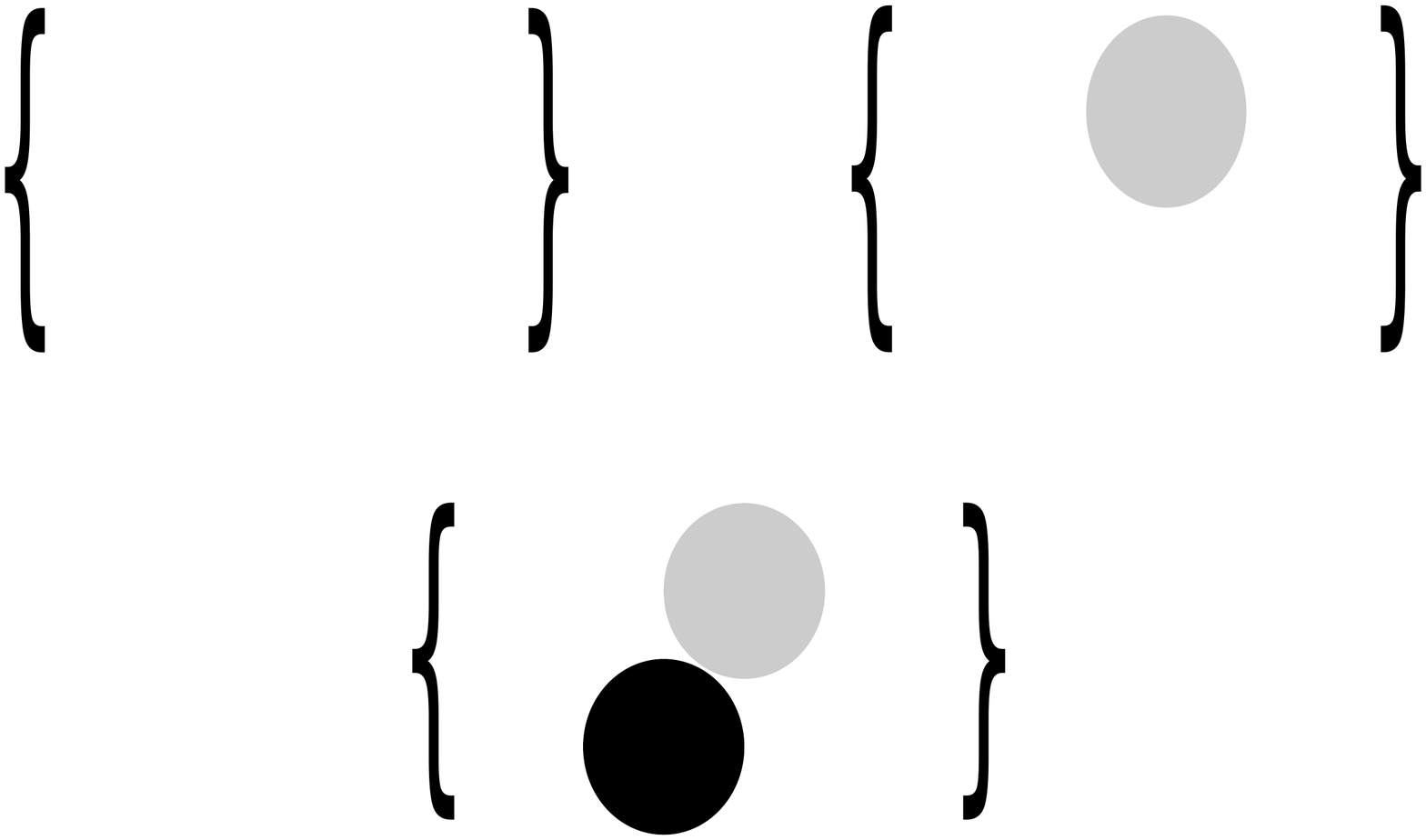}
    \caption{Pyramid partitions for $SU(2)$ with a fundamental Wilson line}
    \label{fixedSU2-2}
\end{figure}

In the case of $Q [f_{\mathbf{2}}]$ the first layer contains only one stone, since there is only one cyclic vector $C \, v$. This is generic for defects in the fundamental or anti-fundamental representation, but will change for other representations. The second layer consists of stones corresponding to nodes which can be reached by an arrow $a \in \mathscr{J}_\mathcal{W} = \mathbb{C} \, Q / \partial \, \mathcal{W}$ from the first layer, the third layer consists of stones corresponding to nodes which can be reached by an arrow from the second layer, and so on. The relations $\partial \, \mathcal{W} = 0$ determine the shape of the pyramid arrangement; in other words the pyramid arrangement is just a combinatorial representation of the Jacobian algebra based at the framing node. In our case the pyramid arrangement is rather simple: from the framing vector there is only one possibility of reaching the node $\bullet$, since the F-term relations identify $A C = \tilde{A} C$, as in Figure \ref{initialSU2-2}.
A pyramid partition is a configuration $\pi$ of stones such that for each stone in $\pi$, the stones immediately above it (of a different color) are in $\pi$ as well. For example, in the case of Figure \ref{initialSU2-2} there are just three pyramid partitions, listed in Figure \ref{fixedSU2-2}.

Note that, by definition of a pyramid partition $\pi$, if a configuration of stones of the form $a \, p$ is part of $\pi$ for some $a \in \mathscr{J}_\mathcal{W}$, then also the configuration of stones $p$ must be part of $\pi$. On the other hand consider the complement $I$ of the vector space spanned by the elements of $\pi$ in $\mathscr{J}_\mathcal{W} \cdot v$. Then by definition, if any $q \in I$ then it must be $a \, q \in I$ for all $a \in \mathscr{J}_\mathcal{W}$ (or equivalently, if $q \notin \pi$, then $a \, q \notin \pi$ for every $a \in \mathscr{J}_\mathcal{W}$). In other words $I$ is an ideal of $\mathscr{J}_\mathcal{W}$. Indeed $I$ is the annihilator of the cyclic module described by $\pi$. 

Recall that a cyclic module $M$ is generated by applying all the arrows to a cyclic vector, and then imposing the F-term relations. The vectors obtained in this way form a basis for the module $M$. In our particular case $M = V_f \oplus V_\circ \oplus V_\bullet$. Each of this basis vectors corresponds to a stone in a pyramid partition. After removing a pyramid partition, the remaining configuration in the pyramid arrangement corresponds to a $\mathbb{T}_{F , \partial \mathcal{W}}$-fixed ideal, generated by monomials. Therefore $\mathbb{T}_\mathcal{W}$-fixed cyclic modules are in one to one correspondence with pyramid partitions.

On the other hand, given a pyramid partition $\pi$ we can construct a cyclic module explicitly. We simply assign to each stone in $\pi$ a basis vector $v_a$ where the labels $a$ runs over the number of stones in $\pi$, and  construct the vector spaces $V_i$ spanned by the basis vectors of the same color, with $i = \circ, \bullet$. In our specific case these vector spaces are one dimensional, given the allowed configurations in Figure \ref{fixedSU2-2}, but in general will be of arbitrary dimension. The arrows of the quiver induce maps between basis vectors, and these obey the F-term relations by construction. Therefore a pyramid partition gives an $\mathscr{J}_\cW$-module $M = V_f \oplus V_{\circ} \oplus V_{\bullet}$. Furthermore this module is cyclic and generated by the vector $v$ by construction. Finally these modules are in the torus fixed locus, since the action of $\mathbb{T}_{\mathcal{W}}$ can be compensated by a change of basis. The toric fixed ideals corresponding to the pyramid partitions in Figure \ref{fixedSU2-2} are respectively $\langle v \rangle$ $\langle C \, v\rangle$, and $\langle A \, C \, v=\tilde{A} \, C \, v \rangle$, where we only write down the generator based at $v$ for simplicity.

Therefore the problem of counting the number of toric fixed ideals which correspond to cyclic modules with fixed dimension vector, can be more easily dealt with by counting pyramid partitions with a prescribed number of stones for each color. Note that for $Q [f_{\mathbf{2}}]$ this combinatorial construction is not necessary as the Jacobian algebra is rather simple; on the other hand it will simplify the computations in the case of more general framings. We will see plenty of examples later on.

When we have more than one framing arrow, say $n$ arrows, we can generate cyclic modules by acting with each one of them on the vector $v$ based at the framing node. As a consequence the corresponding pyramid arrangements will have $n$ stones in the first layer. We will use this fact in the next Sections.

So far we have only discussed cyclic modules, but everything we have said also holds for co-cyclic modules. Since co-cyclic modules can be obtained from cyclic modules by opposing the quiver, the modifications to our construction are completely straightforward, and only amount in switching black and white stones in the pyramid partitions. By partial abuse of language we will denote co-cyclic modules by multiplication on the right, such as $v \, B$, $v \, B \, A$, and so on.

\subsection{Localization}

The localization formula reduces equivariant integration over the moduli space to a sum over contributions coming from toric fixed points. The contribution of each fixed point is determined by the local structure of the moduli space around that fixed point. We have reduced the problem of classifying torus fixed points to a combinatorial problem. To compute the framed BPS degeneracies we have to determine the contribution of each fixed point. We will now show that each fixed point just contributes with a sign. We are going to compute the degeneracies using the virtual localization formula \cite{graber}, which generalizes the Atiyah-Bott localization formula,
\begin{equation}
\int_{[\mathcal{M}_{\mathbf{d}}^{c} (Q [f])]^{vir}} \ 1 = \sum_{\pi \in \mathcal{M}_{\mathbf{d}}^{c} (Q [f])^{\mathbb{T}_{\mathcal{W}}}} \frac{1}{\mathrm{eul} (T_{\pi}^{vir} \mathcal{M}_{\mathbf{d}}^{c} (Q [f]) )} \, ,
\end{equation}
to integrate over the moduli spaces (as reviewed for example in \cite[Section 3.5]{Szabo:2009vw}). By doing so we will bypass the question if our moduli spaces are smooth manifolds or not. This formula expresses the invariants as a sum over the fixed points of the toric action with weights determined by the local structure of the moduli space around each fixed point. Here $T^{vir}_\pi = \rm{Def} - \rm{Obs}$ is the virtual tangent space at the fixed point $\pi$, which has the canonical form of deformations minus obstructions. It is precisely this structure which allows the localization approach to reconstruct the full integral from only a finite number of fixed points. The approach we are using is basically the one outlined in \cite[Section 4]{MNOP}, although in a much more simpler setting: in our cases the fixed points of interest are a finite number and therefore we can verify many statements directly. We have already classified the fixed points in terms of the combinatorics of pyramid partitions; what is left to do is compute the equivariant Euler class of the virtual tangent space at a fixed point. Here the virtual fundamental class $[\mathcal{M}_{\mathbf{d}}^{c} (Q [f])]^{vir}$ is simply the one associated with the vanishing locus of the F-term relations \cite[Remark 3.12]{behrend}.

To do so we write down a local model for the moduli space in the form of a deformation complex at a point $\pi$
\begin{equation} \label{defcomplex}
\xymatrix@C=8mm{  0 \ar[r] & \mathsf{S}^0_{\pi} \ar[r]^{\delta_0} & \mathsf{S}^1_\pi \ar[r]^{\delta_1} & \mathsf{S}^2_\pi \ar[r]^{\delta_2} & \mathsf{S}^3_\pi \ar[r] & 0
} \, ,
\end{equation}
similarly as to what is done in \cite[Appendix E]{Chuang:2013wt}, where
\begin{eqnarray}
\mathsf{S}^0_\pi &=&\mathrm{Hom}_{\mathbb{C}} (V_{\circ , \pi} , V_{\circ , \pi}) \oplus \mathrm{Hom}_{\mathbb{C}} (V_{\bullet , \pi} , V_{\bullet , \pi}) \, , \cr
\mathsf{S}^1_\pi &=& \mathrm{Hom}_{\mathbb{C}} (V_{\circ , \pi} , V_{\bullet , \pi}) \otimes (t_A + t_{\tilde{A}}) \oplus \mathrm{Hom}_{\mathbb{C}} (V_{\bullet , \pi} , V_{f , \pi}) \otimes t_B \oplus \mathrm{Hom}_{\mathbb{C}} (V_{f , \pi} , V_{\circ , \pi}) \otimes t_{C} \, , \cr
\mathsf{S}^2_\pi &=& \mathrm{Hom}_{\mathbb{C}} (V_{\bullet , \pi} , V_{\circ , \pi}) \otimes (t_A ^{-1}+ t_{\tilde{A}}^{-1}) \oplus \mathrm{Hom}_{\mathbb{C}} ( V_{f , \pi} , V_{\bullet , \pi} ) \otimes t_B^{-1} \oplus \mathrm{Hom}_{\mathbb{C}} ( V_{\circ , \pi} , V_{f , \pi} ) \otimes t_{C}^{-1} \, , \cr
\mathsf{S}^3_\pi &=&\mathrm{Hom}_{\mathbb{C}} (V_{\circ , \pi} , V_{\circ , \pi}) \oplus \mathrm{Hom}_{\mathbb{C}} (V_{\bullet , \pi} , V_{\bullet , \pi}) \, .
\end{eqnarray}
With $t_X$ we denote the one dimensional $\mathbb{T}_\mathcal{W}$-module generated by $\e^{\ii \epsilon_X}$. Recall that at each fixed point we can use the gauge torus $\mathbb{C}^*$ to impose a condition on the toric weights. We have chosen the condition $\epsilon_A + \epsilon_B + \epsilon_C = 0$ so that the superpotential is invariant. This is not necessary but will simplify the computations. We stress again that this is a gauge choice and does not affect the fixed point classification. In particular due to this condition we have that $\mathsf{S}^0 \simeq \mathsf{S}^{3,*}$ and $\mathsf{S}^1 \simeq \mathsf{S}^{2,*}$ and the complex is self-dual. At each fixed point $\pi$, each vector space $V_i$, with $i=\bullet,\circ,f$ decomposes into $\mathbb{T}_{\mathcal{W}}$ modules. Let us describe the deformation complex more explicitly. The map $\delta_0$ is the linearized gauge transformation
\begin{equation} \label{lingauge}
\delta_0 \left( \begin{matrix} \phi_\circ \\ \phi_\bullet \end{matrix} \right) = \left(
\begin{matrix}
\phi_\bullet \, A - A  \, \phi_\circ \\
\phi_\bullet \, \tilde{A} - \tilde{A} \, \phi_\circ \\
 - B \, \phi_\bullet \\
\phi_\circ \, C
\end{matrix}
\right) \ ,
\end{equation}
the map $\delta_1$ is the linearization of the F-term relations (\ref{eomSU2-2})  
\begin{equation}
\delta_1 \left( 
\begin{matrix} \eta_A \\ \eta_{\tilde{A}} \\ \eta_B \\ \eta_C \end{matrix}
\right) = \left(
\begin{matrix}
C \, \eta_B + \eta_C \, B \\
C \, \eta_B + \eta_C \, B \\
A \, \eta_C + \eta_A \, C - \tilde{A} \, \eta_C - \eta_{\tilde{A}} \, C \\
B  \, \eta_A + \eta_B \, A - B \, \eta_{\tilde{A}} - \eta_B \, \tilde{A}
\end{matrix}
\right) \ ,
\end{equation} 
and finally $\delta_2$ corresponds to the linearized relations between the relations (\ref{rrSU2-2})
\begin{equation}
\delta_2 \left(
\begin{matrix} \sigma_A \\ \sigma_{\tilde{A}} \\ \sigma_B \\ \sigma_C \end{matrix}
\right)
= \left( \begin{matrix}
- A \, \sigma_A + \tilde{A} \, \sigma_{\tilde{A}} + \sigma_B \, B \\
- \sigma_A \, A + \sigma_{\tilde{A}} \, \tilde{A} + C \, \sigma_{C}
\end{matrix}\right) \ .
\end{equation}
The differentials are linearizations around the fixed point labelled by $\pi$, which therefore corresponds to a configuration $(A , \tilde{A} , B , C)$ which is a solution of the F-term equations. One can see directly that indeed $\delta_1 \circ \delta_0 = 0$ and $\delta_2 \circ \delta_1 = 0$. Indeed we have
\begin{align}
\delta_1 \circ \delta_0 \, \left( \begin{matrix} \phi_\circ \\ \phi_\bullet \end{matrix} \right) & = \delta_1 \, \left(
\begin{matrix}
\phi_\bullet \, A - A \, \phi_\circ \\ \phi_\bullet \, \tilde{A} - \tilde{A} \, \phi_\circ \\ - B \, \phi_\bullet \\ \phi_\circ \, C
\end{matrix} 
\right)
\\ \nonumber & = \left(
\begin{matrix}
- C \, B \, \phi_\bullet + \phi_\circ \, C \, B = 0 \\
- C \, B \, \phi_\bullet + \phi_\circ \, C \, B = 0 \\
A \, \phi_\circ \, C + (\phi_\bullet \, A - A \, \phi_\circ) \, C - \tilde{A} \, (\phi_\circ \, C) -(\phi_\bullet \, \tilde{A} - \tilde{A} \phi_\circ) C = \phi_\bullet (A \, C - \tilde{A} \, C ) = 0 \\
B (\phi_\bullet \, A - A \, \phi_\circ) - B \, \phi_\bullet \, A - B (\phi_\bullet \, \tilde{A} - \tilde{A} \, \phi_\circ) + B \, \phi_\bullet \, \tilde{A} = (- B \, A + B \, \tilde{A}) \phi_\circ = 0
\end{matrix}
\right) \, ,
\end{align}
where we have used the F-term equations \eqref{eomSU2-2}. Similarly
\begin{equation}
\delta_2 \circ \delta_1 \, \left( 
\begin{matrix} \eta_A \\ \eta_{\tilde{A}} \\ \eta_B \\ \eta_C \end{matrix}
\right)  = \delta_2 \, 
\left(
\begin{matrix}
C \, \eta_B + \eta_C \, B \\ C \, \eta_B + \eta_C \, B \\ A \, \eta_C + \eta_A \, C - \tilde{A} \, \eta_C - \eta_{\tilde{A}} \, C \\ B \, \eta_A + \eta_B \, A  - B \, \eta_{\tilde{A}} - \eta_B \, \tilde{A}
\end{matrix}
\right) \, .
\end{equation}
Vanishing of this expression is equivalent to the conditions
\begin{align}
- A \,  (C \, \eta_B + \eta_C \, B) + \tilde{A} (C \, \eta_B + \eta_C \, B) + (A \, \eta_C + \eta_A \, C - \tilde{A} \, \eta_C - \eta_{\tilde{A}} \, C) \, B 
\cr
= (- A \, C + \tilde{A} \, C) \, \eta_B + \eta_A \, C \, B - \eta_{\tilde{A}} \, C \, B = 0  \, ,
\cr
- (C \, \eta_B + \eta_C \, B) A + (C \, \eta_B + \eta_C \, B) \tilde{A} + C (B \, \eta_A + \eta_B \, A - B \, \eta_{\tilde{A}} - \eta_B \, \tilde{A}) 
\cr
= - \eta_C (B \, A - B \, \tilde{A}) + C \, B \, \eta_C - C \, B \, \eta_{\tilde{A}} = 0 \, ,
\end{align}
which are indeed fulfilled due to the F-term equations  \eqref{eomSU2-2}. We conclude that $\delta_1 \circ \delta_0 = 0$ and $\delta_2 \circ \delta_1 = 0$ as claimed.

We will assume that the complex (\ref{defcomplex}) has trivial cohomology at the first position, that is $\ker \delta_0 = 0$. This is equivalent to consider only irreducible representations. Indeed consider an irreducible representation $X$: then by (\ref{lingauge}) the maps $\phi_\circ$ and $\phi_\bullet$ commute with $A$ and $\tilde{A}$ and take value in $\mathrm{End}(X)$. By Schur's lemma, both maps are therefore proportional to the identity. However the kernel equations $-B \, \phi_\bullet = 0$ and $\phi_\circ \, C = 0$ ensure that this proportionality constant vanishes. Therefore for irreducible representations the kernel is empty. Similarly the cohomology $\ker \, \delta_1 / \im \, \delta_0$ parametrizes infinitesimal displacements at a fixed point, up to gauge transformations, and is therefore a local model for the tangent space $ T_{\pi} ( \mathcal{M}_{\mathbf{d}}^{c} (Q [f_{\mathbf{2}}]))$. The linearized deformations at a fixed points can however be obstructed, and this is measured by the cohomology at the third position $\mathcal{N}_\pi$ (the ``normal bundle" or ``obstruction bundle"). Finally, since the complex is self-dual, we will also assume that cohomology at the fourth position is trivial. In this case the alternating sum of the cohomologies of the complex is precisely the virtual tangent space $T_{\pi}^{vir} \mathcal{M}_{\mathbf{d}}^{c} (Q [f_{\mathbf{2}}])  =T_{\pi} \mathcal{M}_{\mathbf{d}}^{c} (Q [f_{\mathbf{2}}]) \ominus \mathcal{N}_\pi$.

The Euler class $\mathrm{eul} (T_{\pi}^{vir} \mathcal{M}_{\mathbf{d}}^{c} (Q [f_{\mathbf{2}}] )$ can be reconstructed from the equivariant character of the complex. The latter is given by the cohomology of the complex, which can be expressed as the alternating sum of the weights of the $\mathbb{T}_\mathcal{W}$ representations
\begin{equation} \label{Char}
\mathrm{ch}_{\mathbb{T_\mathcal{W}}} (T_{\pi}^{vir} \mathcal{M}_{\mathbf{d}}^{c} (Q [f_{\mathbf{2}}])) = \mathsf{S}^0_\pi -  \mathsf{S}^1_\pi + \mathsf{S}^2_\pi - \mathsf{S}^3_\pi \ ,
\end{equation}
in the representation ring of $\mathbb{T}_\mathcal{W}$. The character (\ref{Char}) is the generalization, within the present context, of \cite[Proposition 5.7]{nakajima}.

Note that by decomposing each term as a $\mathbb{T}_\mathcal{W}$ module, this character can be schematically written as $\sum_i \e^{w_i} - \sum_j \e^{\tilde{w}_j}$, which in the language of localization parametrizes deformations minus obstructions. Therefore from the character we can read directly the Euler class as
\begin{equation} \label{eulweights}
\mathrm{eul} (T_{\pi}^{vir} \mathcal{M}_{\mathbf{d}}^{c} (Q [f_{\mathbf{2}}]) ) = \frac{\prod_i \, w_i}{\prod_j \, \tilde{w}_j} \ .
\end{equation}
Since the complex is self dual, the toric weights $w_i$ and $\tilde{w}_j$ are exactly paired up, and $\tilde{w}_i = - w_i$.  
In other words, numerator and denominator of (\ref{eulweights}) cancel up to an overall sign given by $\dim \mathsf{S}^0_\pi + \dim \mathsf{S}^2_\pi = \dim \mathsf{S}^1_\pi + \dim \mathsf{S}^3_\pi $. In particular the dependence on the toric weights drops out. Comparing with (\ref{DTinv}) we see
\begin{equation}
(-1)^{\dim T_\pi \mathcal{M}_{\mathbf{d}}^{c} (Q [f_{\mathbf{2}}])} = (-1)^{ d_\circ^2 + d_\bullet^2 + d_\circ + d_\bullet + 2 d_\circ d_\bullet } \, .
\end{equation}
Putting everything together we arrive at a compact expression for the framed BPS degeneracies
\begin{equation} 
{\tt DT}^c_{\mathbf{d}}  (W_{\zeta , \mathbf{2}})= \sum_{\pi \in \mathcal{M}_{\mathbf{d}}^{c} (Q [f_{\mathbf{2}}])^{\mathbb{T}_\mathcal{W}}} \ (-1)^{ d_\circ^2 + d_\bullet^2 + d_\circ + d_\bullet + 2 d_\circ d_\bullet } \, .
\end{equation}
We can think of this as a supersymmetric quiver quantum mechanics derivation of the result of \cite[Thm.~3.4]{fantechi} (also compare with \cite[Def.~7.21]{joycesong}).

Finally we conclude that the Donaldson-Thomas invariants are
\begin{eqnarray}
{\tt DT}^{c}_{(d_\circ = 0,d_\bullet = 0)} (W_{\zeta , \mathbf{2}}) &=& +1 \, , \cr
{\tt DT}^{c}_{(d_\circ=1,d_\bullet =0)} (W_{\zeta , \mathbf{2}}) &=& +1\, , \cr
{\tt DT}^{c}_{(d_\circ=1,d_\bullet =1)} (W_{\zeta , \mathbf{2}}) &=& +1 \, .
\end{eqnarray}
A similar computation for co-cyclic stability conditions (which can be equivalently obtained by opposing the quiver) yields
\begin{eqnarray}
{\tt DT}^{cc}_{(d_\bullet = 0,d_\circ = 0)} (W_{\zeta , \mathbf{2}}) &=& +1 \, , \cr
{\tt DT}^{cc}_{(d_\bullet=1,d_\circ =0)} (W_{\zeta , \mathbf{2}}) &=& +1 \, , \cr
{\tt DT}^{cc}_{(d_\bullet=1,d_\circ =1)} (W_{\zeta , \mathbf{2}}) &=& +1 \, .
\end{eqnarray}
For cyclic boundary conditions we can write down the framed quantum mechanics partition function as
\begin{eqnarray}
\langle W_{\zeta , \mathbf{2}} \rangle_{q=-1} &=& \sum_{\mathbf{d} = (d_\circ , d_\bullet)} \ {\tt DT}^{c}_{\mathbf{d}} (W_{\zeta , \mathbf{2}})  \ X_{e_\mathbf{2} + d_\circ e_{\circ} + d_{\bullet} e_{\bullet}} \cr
&=& X_{-\frac12 (e_\circ + e_{\bullet})} + X_{-\frac12 e_\circ + \frac12 e_\bullet} + X_{+\frac12 (e_\circ + e_\bullet)} \, ,
\end{eqnarray}
with $e_\mathbf{2} = -\frac12 (e_{\circ} + e_{\bullet})$. This function is precisely reproduces the known framed BPS spectrum in this case, see for example \cite{Gaiotto:2010be,Cirafici:2013bha,Cordova:2013bza,BPSlinesCluster}.

\subsection{$SU(2)$ quivers: general structure}

We have so far discussed quite in detail the most simple example of an SU(2) model coupled to a fundamental Wilson line. We have however set up the formalism in such a way that it can be extended almost word by words to more general quivers. We will now discuss the case of a general framed $SU(2)$ quiver $Q [f_\mathbf{n+1}]$. In this case there are $n$ arrows connecting the framing node $f_\mathbf{n+1}$ with the node $\circ$, and $n$ arrows connecting the node $\bullet$ with $f_\mathbf{n+1}$. We consider the superpotential
\begin{equation}\label{generalWSU2}
\mathcal{W} = A C_1 B_1 + \sum_{i \ \text{odd}}^{n-1} B_{i+1} ( \tilde{A}  C_{i+1}-  \tilde{A}  C_i ) + \sum_{i \ \text{even}}^{n-2} B_{i+1} (A C_{i+1}  - A C_i) \, .
\end{equation}
This superpotential is a generalization of the superpotential we have used in the previous case. We do not have a microscopic derivation of \eqref{generalWSU2}. This form of the superpotential reproduces the results available in the literature, for example in \cite{BPSlinesCluster}.

\textbf{Remark}. Note that such a superpotential should be regarded as a datum of the \textit{topological} quantum mechanics, which computed BPS indices associated with the framed quiver. In particular we are free to add any BRST exact term in the action of the topological model. Virtual counting for BPS states are robust under such deformations. For example such a superpotential differs from the one used in \cite{BPSlinesCluster} by a field redefinition. Mathematically such a redefinition is necessary to kill certain residual automorphisms of the quiver moduli space. From the perspective of virtual counts the two superpotentials are therefore equivalent and produce the same result, if one deals properly with the automorphisms. We should however stress that we are \textit{not} claiming that the two corresponding physical (i.e. non topological) quantum mechanics models are equivalent or that the relevant representation theories are the same; as far as we know the models are equivalent only for the computation of the BPS index.

\begin{figure}[h]
    \centering
    \includegraphics[width=0.20\textwidth]{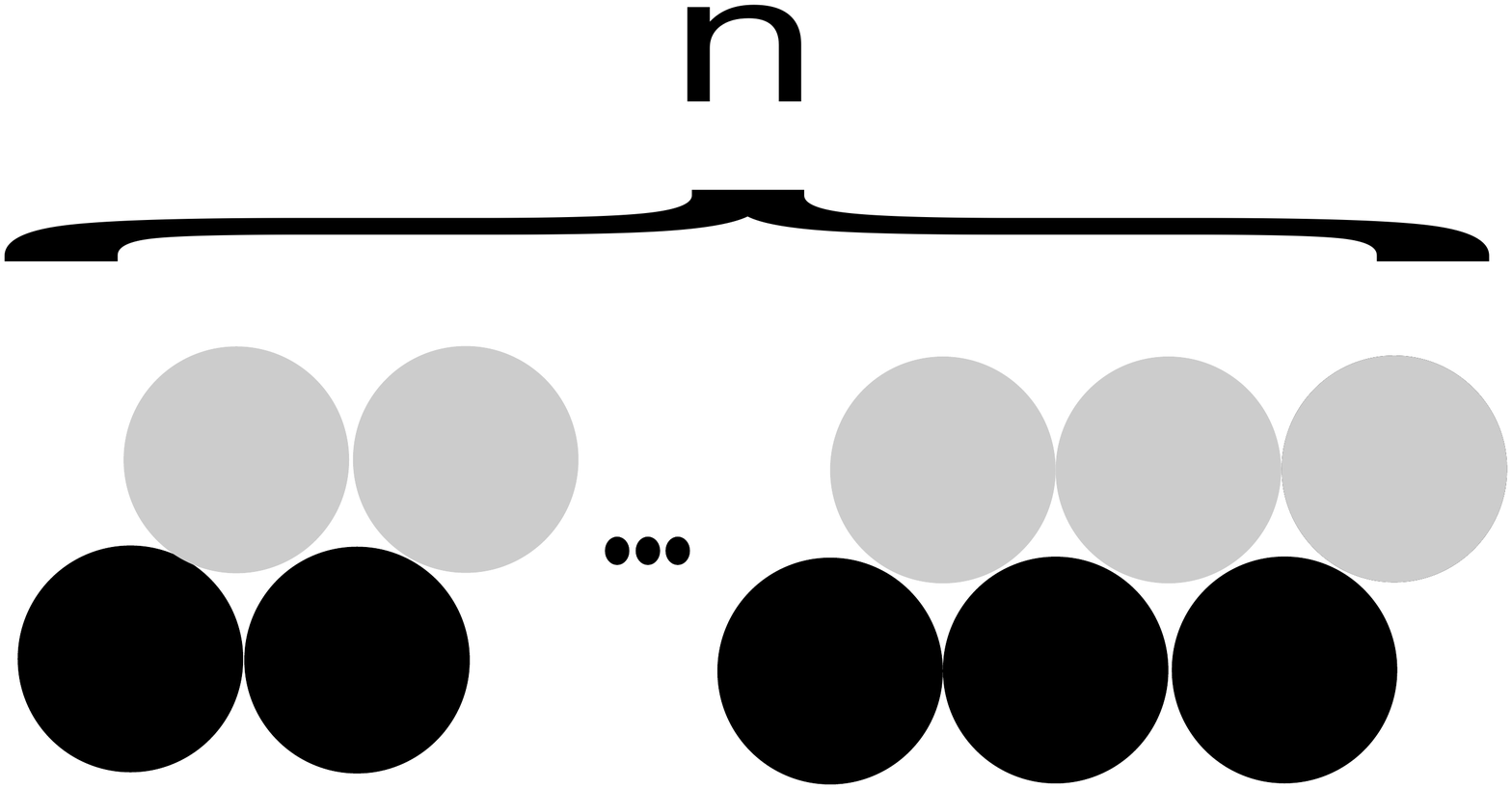}
    \caption{Initial configuration}
    \label{initialSU2-n}
\end{figure}

The torus $\mathbb{T}_\mathcal{W}$ acts by rescaling all the fields in the quantum mechanics, with the constraint that its action preserves the F-term equations $\partial \, \mathcal{W} = 0$, and modulo the action of the gauge subtorus. Consider for example cyclic stability conditions, where cyclic modules are generated by the vector $v$; in this case the pyramid arrangement has $n$ white stones in the first layer and $n$ black stones in the second, and its shape is given by the Jacobian algebra obtained from the superpotential (\ref{generalWSU2}), as shown in Figure \ref{initialSU2-n}. 
Fixed points $\pi$ of the toric action correspond to pyramid partitions which can be removed from this pyramid arrangement. We will discuss several explicit examples in the next Section.

We write down the following quiver deformation complex around the fixed point $\pi$
\begin{equation}
\xymatrix@C=8mm{  0 \ar[r] & \mathsf{S}^0_\pi \ar[r]^{\delta_0} & \mathsf{S}^1_\pi \ar[r]^{\delta_1} & \mathsf{S}^2_\pi \ar[r]^{\delta_2} & \mathsf{S}^3_\pi \ar[r] & 0
} \, ,
\end{equation}
where
\begin{eqnarray}
\mathsf{S}^0_\pi &=& \bigoplus_{i \in Q_0} \, \mathrm{Hom}_\mathbb{C} (V_{i, \pi} , V_{i, \pi}) \, , \\
\mathsf{S}^1_\pi &=& \bigoplus_{a \in Q_1 [f_{\mathbf{n}}]} \, \mathrm{Hom}_\mathbb{C}  (V_{s (a), \pi} , V_{t (a), \pi}) \otimes t_a \, ,\\
\mathsf{S}^2_\pi &=& \bigoplus_{r \in \mathsf{R}} \, \mathrm{Hom}_\mathbb{C}  (V_{s (r), \pi} , V_{t (r), \pi}) \otimes t_r \, , \\
\mathsf{S}^3_\pi &=& \bigoplus_{rr \in \mathsf{R} \mathsf{R}} \mathrm{Hom}_\mathbb{C}  (V_{s (rr), \pi} , V_{t (rr), \pi}) \, ,
\end{eqnarray}
where $t_r$ denotes the one dimensional $\mathbb{T}_\mathcal{W}$-module generated by the toric weight of the relation $r \in \mathsf{R}$ (recall that by definition of $\mathbb{T}_\mathcal{W}$ all the monomials in a relation have the same weight). The map $\delta_0$ is the linearization of the gauge transformations, the map $\delta_1$ is the linearization of the F-term relations $\partial \, \mathcal{W} = 0$ and finally $\delta_2$ corresponds to the linearized relations between the relations. We will write down the associated set of equations explicitly in a few cases in Section \ref{SU2loca}.

The parity of the tangent space can be computed directly
\begin{equation} \label{paritySU2-n}
(-1)^{\dim T_\pi \, \mathcal{M}_{\mathbf{d}}^{c} (Q [f_\mathbf{n+1}])} = (-1)^{ d_\circ^2 + d_\bullet^2 + n d_\circ + n d_\bullet + 2 d_\circ d_\bullet } \, ,
\end{equation}
and the framed BPS degeneracies are given by
\begin{equation}
{\tt DT}^{c}_{\mathbf{d}} (W_{\zeta , \mathbf{n+1}}) = \sum_{\pi \in \mathcal{M}_{\mathbf{d}}^{c} (Q [f_\mathbf{n+1}])^{\mathbb{T}_\mathcal{W}}} \ (-1)^{ d_\circ^2 + d_\bullet^2 + n d_\circ + n d_\bullet + 2 d_\circ d_\bullet } \, .
\end{equation}
The generating functions of framed BPS states now have the generic form
\begin{equation}
\langle W_{\zeta , \mathbf{n+1}} \rangle_{q=-1} = \sum_{\mathbf{d} = (d_\circ , d_\bullet)} \, {\tt DT}^{c}_{\mathbf{d}} (W_{\zeta , \mathbf{n+1}}) \, X_{e_t + d_\circ \, e_\circ + d_\bullet \, e_\bullet} \ .
\end{equation} 
We will compute them explicitly in the next Section for a variety of cases.

\subsection{Arbitrary quivers}

Now we will argue that these ideas extend to arbitrary quivers with framing, not just of the Kronecker type, under some generic assumptions. The details of the argument will depend on the specific quiver, framing vector and superpotential. However the general structure continues to hold, eventually involving arbitrarily complicated combinatorial arrangements. Sections \ref{SU3loca} and \ref{SO8loca} will contain more detailed examples. We streamline the argument as follows:

\underline{Quiver}. Consider a generic quiver $Q$ and add a framing node $f$. We will denote the framed quiver by $Q [f]$. To simplify the notation, we pick a superpotential which is a sum of cubic terms, but our arguments hold more generically. We write
\begin{equation}
\mathcal{W} = \sum_{i,j,k} X_{ij} \, X_{jk} \, X_{ki}
\end{equation}  
where $X_{ij}$ denotes and element of $\mathrm{Hom} (V_j , V_i)$ and as usual the product is defined to be zero if the arrows do not concatenate. The sum includes the framing node and can be over a subset of $Q[f]_0$ only. The equations of motions have the form
\begin{equation} \label{eomgen}
\frac{\partial \mathcal{W}}{\partial X_{lm}} = \sum_k^\prime \, X_{mk} \, X_{kl} = 0
\end{equation}
where the $\prime$ over the sum is a reminder that the vertices $k$, $l$ and $m$ form a closed triangle in the quiver. Finally cycles starting and ending at the same node $m$ obey
\begin{equation} \label{rrgen}
\sum_l^\prime \left( \frac{\partial \mathcal{W}}{\partial X_{lm}} X_{lm} - X_{ml} \frac{\partial \mathcal{W}}{\partial X_{ml}} \right) = 0
\end{equation}
for every $m \in Q_0$.

\underline{Moduli spaces}. We will consider cyclic stability conditions. Therefore the relevant moduli space is the moduli space of cyclic modules. As before this is obtained by looking at a quotient of the representation space $\mathsf{Rep}_{\mathbf{d}} \left( Q [f] ,\mathcal{W}  ; v  \right)$, the subscheme of the space of representations generated by $v \in V_f$ defined by the equations $\partial \, \mathcal{W} = 0$, by the gauge group
\begin{equation}
G_{\mathbf{d}} = \prod_{i \in Q_0} \, GL (d_i , \mathbb{C}) \ .
\end{equation}
The resulting moduli space $\mathcal{M}^{c}_{\mathbf{d}} (Q [f] ; v)$ parametrizes cyclic $\mathscr{J}_\mathcal{W}$-modules generated by the vector $v \in V_f$.

\underline{Toric action}. This moduli space carries a natural toric action generated by rescaling the arrows of the quiver $Q [f]$. Denote the fields of the quantum mechanics by $X_a$ for each $a \in Q[f]_1$. We let the torus $\mathbb{T}_F = (\mathbb{C}^*)^{|Q_1[f] |}$ act by rescaling $X_a \longrightarrow X_a \e^{\ii \epsilon_a}$, and define the sub-torus $\mathbb{T}_{F , \partial W}$ by the condition that it preserves the F-term equations. The gauge group $G_{\mathbf{d}}$ has a sub-torus $\mathbb{T}_G = (\mathbb{C}^*)^{|Q_0|-1}$ which has to be mod out. The overall ``-1" in the exponent is due to diagonal gauge transformations which act trivially. Therefore the full torus is $\mathbb{T}_{\mathcal{W}} = \mathbb{T}_{F , \partial \mathcal{W}} / \mathbb{T}_G$. This is practically obtained by imposing $|Q_0|-1$ conditions on the toric parameters of $\mathbb{T}_{F , \partial \mathcal{W}}$. We will assume for simplicity that such conditions can be chosen in such a way as to leave the superpotential $\mathcal{W}$ invariant (although this is not necessary).

\underline{Fixed points}. In this case we cannot make a general argument for the structure of the pyramid arrangement, which will depend sensitively on the Jacobian algebra $\mathscr{J}_\mathcal{W}$, and in particular on the superpotential $\mathcal{W}$. However the construction of the pyramid arrangement is completely algorithmic: one draws colored stones for each node, connected by paths in the Jacobian algebra, eventually identified by the F-term relations. The first layer of the pyramid arrangement will contain as many stones as arrows from the framing node $f$ to the unframed quiver $Q$. For arbitrary quivers the combinatorial structure of the pyramid arrangement might become rather intricate. However fixed points are still counted by removing stone configurations in the form of a pyramid partition; pyramid partitions by construction correspond to $\mathbb{T}_{F , \partial \mathcal{W}}$ ideals in the Jacobian algebra. Each stone in a pyramid partition identifies a basis vector in the  corresponding vector space, and the collection of these vectors together with the induced maps between them, reproduces a toric fixed cyclic module (or a co-cyclic module for the opposite quiver). Given a quiver, the classification of pyramid partitions is a completely algorithmic combinatorial problem.

\underline{Localization}. We assume that the deformation complex is self-dual. This holds for example if the toric action can be chosen in such a way as to leave the superpotential invariant. In case the deformation complex is not self-dual, the same construction would nevertheless apply but the contribution of each point have to be computed one by one.  Within this assumption the local structure of the moduli space around a fixed point labelled by a pyramid partition $\pi$ is given by the deformation complex
\begin{equation}
\xymatrix@C=8mm{  0 \ar[r] & \mathsf{S}^0_\pi \ar[r]^{\delta_0} & \mathsf{S}^1_\pi \ar[r]^{\delta_1} & \mathsf{S}^2_\pi \ar[r]^{\delta_2} & \mathsf{S}^3_\pi \ar[r] & 0 \, ,
}
\end{equation}
where
\begin{eqnarray}
\mathsf{S}^0_\pi &=& \bigoplus_{i \in Q_0} \, \mathrm{Hom}_\mathbb{C} (V_{i, \pi} , V_{i, \pi})  \simeq (\mathsf{S}_\pi^{3})^* \, , \\
\mathsf{S}^1_\pi &=& \bigoplus_{a \in Q_1 [f]} \, \mathrm{Hom}_\mathbb{C}  (V_{s (a), \pi} , V_{t (a), \pi}) \otimes t_a \simeq (\mathsf{S}_\pi^{2})^* \, .
\end{eqnarray}
The differentials are defined as follows. For simplicity we forget about the $\mathbb{T}$-equivariant structure. The map $\delta_0$ correspond to an infinitesimal gauge transformation. If we conventionally denote with $\phi_i$, with $i \in Q_0$ the linearized gauge parameter, then the image of $\delta_0$ has the form
\begin{equation}
( \phi_i \, X_{ij} - X_{ij} \, \phi_j) \in \mathrm{Hom} (V_j , V_i)  \ \, \qquad \forall \ i,j \in Q[f]_0 \, ,
\end{equation}
where $\phi_f = 0$ if $f$ is the framing node. We use the same set of indices both for $Q_0$ and $Q[f]_0$, hoping that this won't cause confusion. The differential $\delta_1$ is the linearization of the equations of motions, and acts as
\begin{equation}
\sum_{k}^\prime \left( \eta_{mk} \, X_{kl} + X_{mk} \, \eta_{kl} \right) \, ,
\end{equation}
on $ \eta_{ij} \in \mathrm{Hom} (V_j , V_i)$. Similarly the differential $\delta_2$ acts as
\begin{equation}
\sum_{l}^\prime \left( \sigma_{lm} \, X_{lm} - X_{ml} \, \sigma_{ml}  \right) \, , 
\end{equation}
on $\sigma_{ij} \in \mathrm{Hom} (V_j , V_i)$. The $\prime$ in all the sums is a reminder that all the maps involved must concatenate. 

We can now check explicitly that the differentials are nilpotent around a solution of the equations of motions. Indeed computing $\delta_1 \circ \delta_0$ we see
\begin{align}
\sum_k^\prime \left( 
(\phi_m \, X_{mk} - X_{mk} \, \phi_k) X_{kl} + X_{mk} (\phi_k \, X_{kl} - X_{kl} \phi_l)
\right)
\cr
= \phi_m \left(  \sum_k^\prime X_{mk} X_{kl} \right) - \left( \sum_k^\prime X_{mk} X_{kl} \right) \, \phi_l = 0 \, ,
\end{align}
where we have used the equations of motion \eqref{eomgen}. Similarly evaluation of $\delta_2 \circ \delta_1$ leads to
\begin{align}
\sum_l^\prime \sum_k^\prime \left( \eta_{mk} X_{kl} + X_{mk} \eta_{kl}  \right) X_{lm} - \sum_l^\prime X_{ml} \sum_k^\prime \left( \eta_{lk} X_{km} - X_{lk} \, \eta_{km} \right)
\cr
= \sum_{l,k}^\prime \left( \eta_{mk} X_{kl} \, X_{lm} - X_{ml} \, X_{lk} \, \eta_{km} \right) = 0 \, ,
\end{align}
using again \eqref{eomgen}.

Since the equivariant complex is self-dual, when computing the framed BPS degeneracies the toric weights cancel out and 
\begin{equation}
{\tt DT}^{c}_{\mathbf{d}}  (Q[f]) = \sum_{\pi \in \mathcal{M}_{\mathbf{d}}^{c} (Q [f] , v )^{\mathbb{T}_\mathcal{W}}} \ (-1)^{q_{Q [f]}  (\mathbf{d}) - 1} \, ,
\end{equation}
where 
\begin{equation}
q_{Q[f]} (\mathbf{d}) = \sum_{i \in Q_0 [f] } d_i^2 - \sum_{i,j \in Q_0 [f]} B_{ij}^f d_i d_j \, ,
\end{equation}
is the Tits form of the quiver $Q [f]$ (the minus one removes the quadratic contribution from the framing node) and $B_{ij}^f$ the adjacency matrix of the framed quiver. These results are a natural generalization of \cite[Chapter 7]{joycesong}.

\section{$SU(2)$ super Yang-Mills} \label{SU2loca}

We will now put the formalism described in the previous Section to practice and compute the framed BPS spectra of several Wilson lines in $SU(2)$ super Yang-Mills. These correspond to defects with core charges $\mathbf{RG} [W_{\zeta , \mathbf{n}}] = e_{\mathbf{n}} = -\frac{n-1}{2} (e_\circ + e_\bullet)$ and can be modeled by a framed quiver $Q [f_\mathbf{n}]$, where the framing node is connected to the quiver by $n-1$ arrows. We will go through the examples in some detail.

\subsection{$SU(2)$ with an adjoint Wilson line}

We begin by considering the following quiver
\begin{equation}
\xymatrix@C=8mm{
&  \bullet  \ar@{..>}@<-0.5ex>[dl]_{\tilde{B}}  \ar@{..>}@<0.5ex>[dl]^{B} \\
f_{\mathbf{3}}  \ar@{..>}@<-0.5ex>[dr]_{\tilde{C}}  \ar@{..>}@<0.5ex>[dr]^{C}  & \\
& \circ \ar@<-0.5ex>[uu]_{\tilde{A}}  \ar@<0.5ex>[uu]^{A} 
}
\end{equation}
We model the coupling of $SU(2)$ Yang-Mills to a Wilson line in the adjoint representation with the following superpotential in the topological quiver quantum mechanics
\begin{equation} \label{SU2adjW}
\mathcal{W} = A C B + \tilde{B} \left( \tilde{A} \tilde{C} - \tilde{A} C \right) \, .
\end{equation}
We will now compute explicitly the corresponding vev. The relations $\partial \, \cW = 0$ derived from \eqref{SU2adjW} are
\begin{align}
r_A \, : \, CB  = 0 \, , \ & \qquad  r_{\tilde{A}}  \, : \, \tilde{C} \tilde{B} = C \tilde{B} \, , \\
r_{B} \, : \, AC  =0 \, , \ & \qquad  r_{\tilde{B}}  \, : \,  \tilde{A} \tilde{C} =  \tilde{A} C  \, ,\\
r_C \, : \, BA = \tilde{B} \tilde{A}  \, , \ & \qquad  r_{\tilde{C}}  \, : \, \tilde{B} \tilde{A} = 0 \, .
\end{align}
Note the following relations between the relations 
\begin{align}
rr_{\bullet} \, : \qquad  A \, r_A - r_B B - r_{\tilde{B}} \tilde{B} + \tilde{A} r_{\tilde{A}} &= 0 \, , \\
rr_{\circ} \, : \qquad  C r_C - r_A A -r_{\tilde{A}} \tilde{A} + \tilde{C} r_C &= 0 \, . 
\end{align}
The gauge group $GL (V_\circ) \times GL (V_\bullet)$ of the quantum mechanics acts as
\begin{align}
(A,\tilde{A}) &\longrightarrow  ( g_\bullet \, A \, g_\circ^{-1},  g_\bullet  \, \tilde{A} \, g_\circ^{-1} ) \, ,\\
(B,\tilde{B}) &\longrightarrow  ( B \, g_\bullet^{-1} , \tilde{B} \, g_\bullet^{-1})  \, ,\\
(C,\tilde{C}) &\longrightarrow   ( g_\circ \, C , g_\circ \, \tilde{C} ) \, .
\end{align}
We pick a toric action $(\mathbb{C}^*)^6$ which acts as the rescaling by a phase $X \longrightarrow \e^{\ii \epsilon_X} X$ on each quantum mechanics field $X$. For this action to be compatible with the relations $\partial \, \mathcal{W} = 0$ we must have
\begin{equation}
\epsilon_C = \epsilon_{\tilde{C}} \, , \qquad \epsilon_A + \epsilon_B = \epsilon_{\tilde{A}} + \epsilon_{\tilde{B}} \, .
\end{equation}

\begin{figure}[H]
    \centering
    \includegraphics[width=0.5\textwidth]{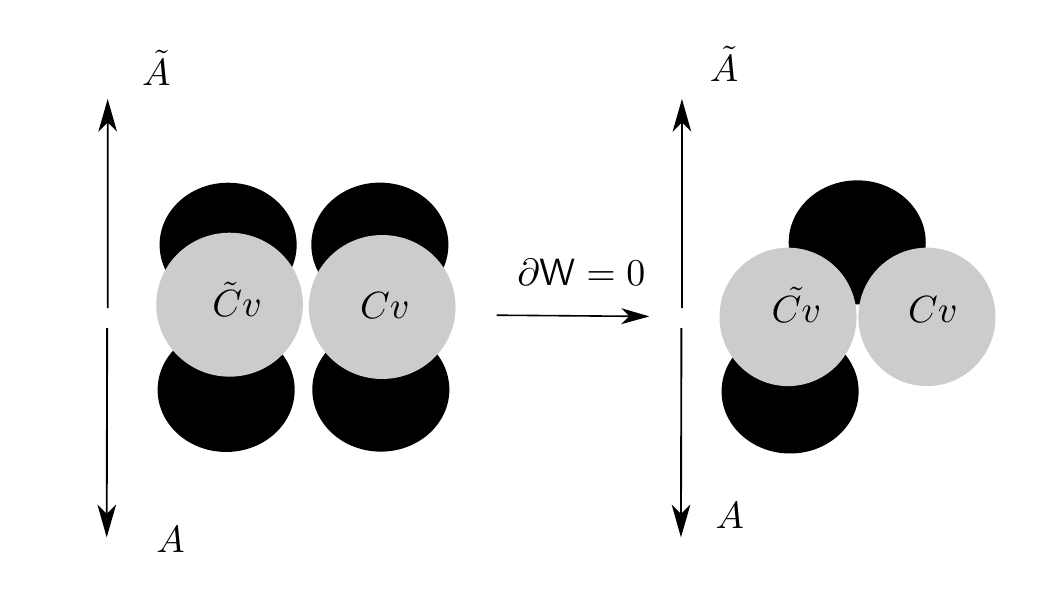}
    \caption{How a pyramid arrangement arises after imposing the F-terms relations}
    \label{pyramid}
\end{figure}

\begin{figure}[h]
    \centering
    \includegraphics[width=0.20\textwidth]{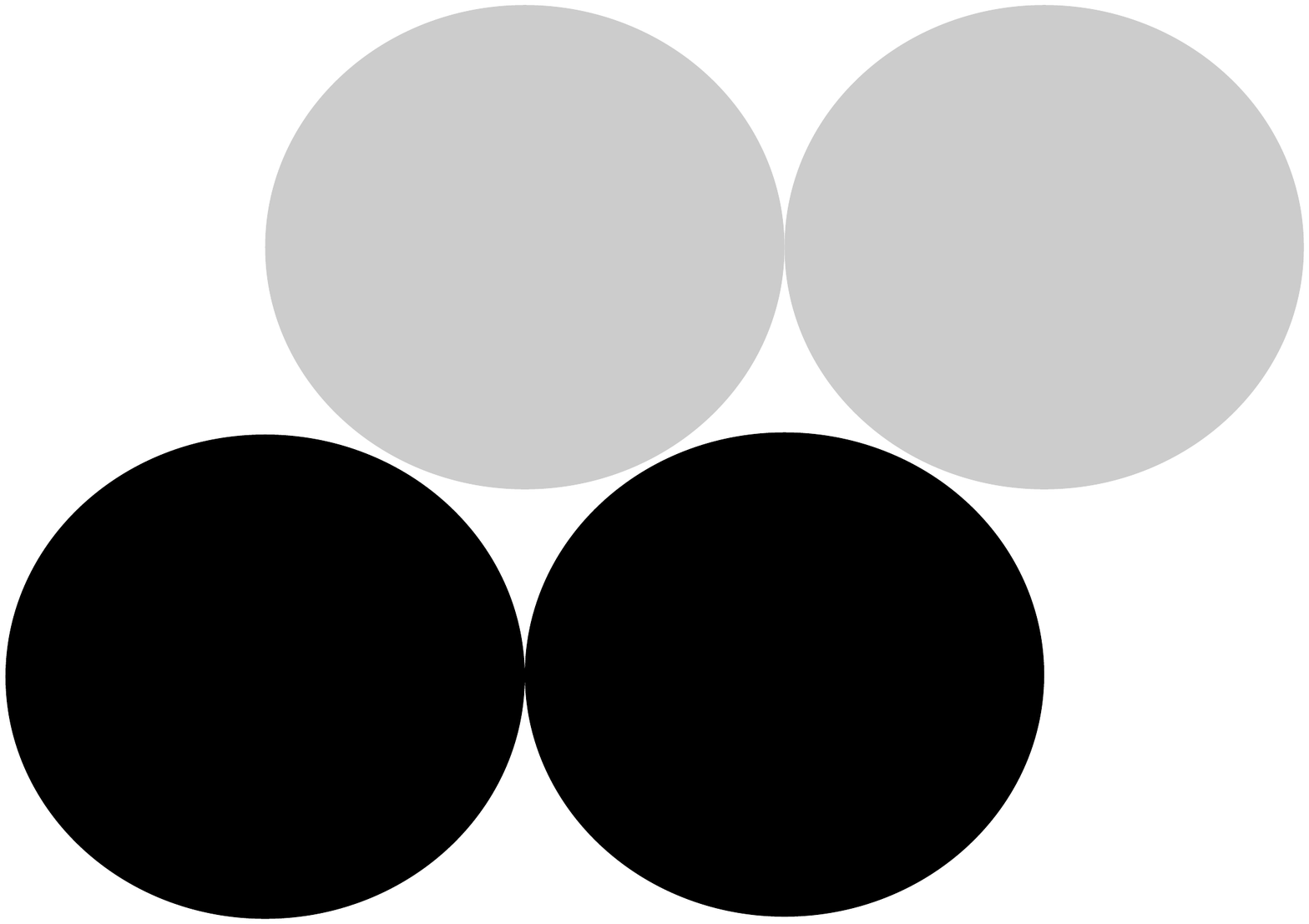}
    \caption{Initial configuration, for $SU(2)$ super Yang-Mills with a Wilson line in the representation $\mathbf{3}$}
    \label{initialSU2-3}
\end{figure}

\begin{figure}[H]
    \centering
    \includegraphics[width=0.5\textwidth]{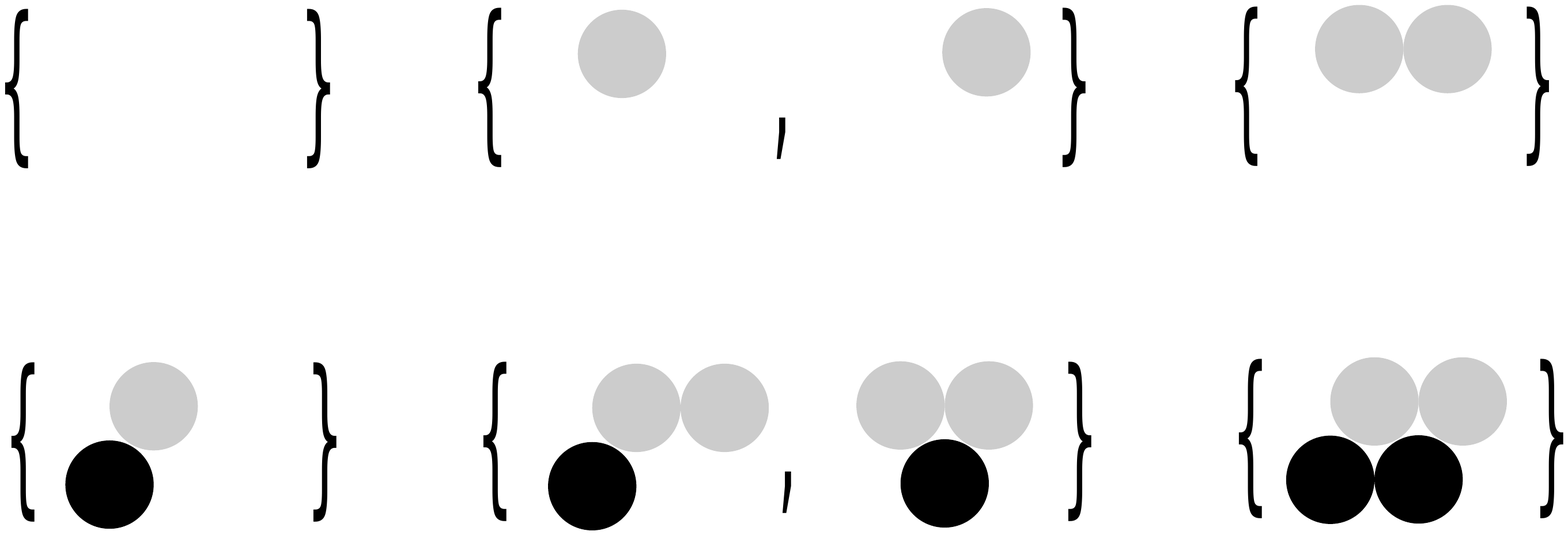}
    \caption{Classification of the fixed points in the case of $SU(2)$ super Yang-Mills with a Wilson line in the representation $\mathbf{3}$}
    \label{fixedSU2-3}
\end{figure}

We proceed now to classify the torus fixed points. If we write $\mathscr{J}_\mathcal{W} = \oplus_n \, \mathscr{J}_{\mathcal{W} , n}$, where the grading is identified with the length of a cyclic modules, we find
\begin{eqnarray}
\mathscr{J}_{\mathcal{W} , 0} &=& \{v \} \, , \cr
\mathscr{J}_{\mathcal{W} , 1} &=& \{ C \,  v ,  \tilde{C} \, v  \} \, , \cr
\mathscr{J}_{\mathcal{W} , 2} &=& \{ A \, \tilde{C} \, v , \tilde{A} \, \tilde{C} \, v = \tilde{A} \, C \, v   \} \, .
\end{eqnarray}
Again the condition that $\dim V_f = 1$ effectively truncates the pyramid arrangement at this point: fixed points containing stones based at the framing node would necessarily imply $\mathrm{dim} V_f > 1$. The classification of fixed points proceeds as in the previous Section. Now there are two arrows starting from the framing node, and therefore two vectors, $C \, v$ and $\tilde{C} \, v$ are selected in $V_\circ$. As a consequence the pyramid arrangement will have two stones in the first layer\footnote{Dealing with situations where the framing involves more arrows is the main reason why we defined cyclic modules starting from a vector in the framing node. Of course ordinary cyclic modules can be defined similarly by dealing with more linearly independent vectors in the unframed quiver specified by more morphisms from the framing node to the unframed quiver \cite{szendroi,Chuang:2008aw}. We find our approach more convenient from a combinatorial viewpoint in the case where there are many framing arrows.}. In Figure \ref{pyramid} we show how this pyramid arrangement is obtained by imposing the F-term relations, which identify $\tilde{A} \, \tilde{C} \, v = \tilde{A} \, C \, v$ and set $A \, C \, v = 0$ in the Jacobian algebra.
To facilitate drawing we will project these structures on the two dimensional plane, as in Figure \ref{initialSU2-3}.
The computation of the framed BPS degeneracies involves the classification of the fixed point configurations. Recall that fixed points correspond to pyramid partitions, configurations of stones to be removed from the pyramid arrangement, defined by the property that if a stone is present in the configuration then all the stones immediately above must be part of the configuration too. Using this definition, it is easy to write down directly all the pyramid partitions which can be obtained from the pyramid arrangement of Figure \ref{initialSU2-3}; they are listed in Figure \ref{fixedSU2-3}.
Since the number of stones in a pyramid partition corresponds to the dimension vectors, we see immediately that the stable framed BPS states are those with $\mathbf{d}$ given by: $(0,0)$, $(1,0)$, $(2,0)$, $(1,1)$, $(2,1)$ and $(2,2)$.

Having classified the fixed points, what remains to be done is to compute their contribution to the framed BPS index using the localization formula. The gauge group $GL (V_\circ) \times GL (V_\bullet)$ has a diagonal torus $\mathbb{C}^*$ which we can use to remove one of the toric parameters. We chose to set to zero the combination $\epsilon_A + \epsilon_B + \epsilon_C = \epsilon_{\tilde{A}} + \epsilon_{\tilde{B}} + \epsilon_{\tilde{C}} = 0$ so that the superpotential $\mathcal{W}$ is invariant. In this case the equivariant deformation complex around a toric fixed point is constructed out of
\begin{eqnarray}
\mathsf{S}^0_\pi &=& \mathrm{Hom}_\mathbb{C} (V_{\circ ,\pi} , V_{\circ,\pi}) \oplus \mathrm{Hom}_\mathbb{C} (V_{\bullet,\pi} , V_{\bullet ,\pi}) \, , \cr
\mathsf{S}^1_\pi &=& \mathrm{Hom}_\mathbb{C} (V_{\circ,\pi} , V_{\bullet,,\pi}) \otimes (t_A + t_{\tilde{A}}) \oplus \mathrm{Hom}_\mathbb{C} (V_{\bullet,\pi} , V_{f,\pi}) \otimes (t_B + t_{\tilde{B}}) 
\cr &&
\oplus \, \mathrm{Hom}_\mathbb{C} (V_{f,\pi} , V_{\circ,\pi}) \otimes (t_{C} + t_{\tilde{C}}) \, , \cr
\mathsf{S}^2_\pi  &=& \mathrm{Hom}_\mathbb{C} (V_{\bullet,\pi} , V_{\circ,\pi}) \otimes (t_A^{-1} + t_{\tilde{A}}^{-1}) \oplus \mathrm{Hom}_\mathbb{C} ( V_{f,\pi}, V_{\bullet,\pi} ) \otimes (t_B^{-1} + t_{\tilde{B}}^{-1}) 
\cr &&
\oplus \, \mathrm{Hom}_\mathbb{C} ( V_{\circ,\pi}, V_{f,\pi} ) \otimes (t_{C}^{-1} + t_{\tilde{C}}^{-1}) \, ,\cr
\mathsf{S}^3_\pi &=& \mathrm{Hom}_\mathbb{C} (V_{\circ,\pi} , V_{\circ,\pi}) \oplus \mathrm{Hom}_\mathbb{C} (V_{\bullet,\pi} , V_{\bullet,\pi}) \, ,
\end{eqnarray}
around a toric fixed point $\pi$. Note that we don't allow gauge transformations (as well as relations between relations) on the vector space $V_{f}$ since it is a framing node. Here $t_X $ stands for the one dimensional $\mathbb{T}_{\mathcal{W}}$-module generated by $\e^{\ii \epsilon_X}$, for any quantum mechanics field $X$. The appearance of the conjugate $t_A^{-1}$ modules is due to the gauge condition $\epsilon_A + \epsilon_B + \epsilon_C = 0$. Since upon imposing this condition the superpotential is $\torus_\cW$ invariant, the arrows transform opposite to the relations.

As explained in the previous Section, we can compute the generating function associated with the quiver quantum mechanics from the character of the deformation complex by using the equivariant localization formula. The equivariant character has the form
\begin{equation}
\mathrm{ch}_{\mathbb{T}_{\mathcal{W}}} (T_\pi^{vir} \mathcal{M}_{\mathbf{d}}^{c} (Q [f_{\mathbf{3}}])) = \mathsf{S}^0_\pi -  \mathsf{S}^1_\pi + \mathsf{S}^2_\pi - \mathsf{S}^3_\pi \, .
\end{equation}
Each module can be decomposed according to the toric action as $\mathsf{S}^a_{\pi} = \sum_i \e^{w^a_i}$. The contribution of each fixed point to the quantum mechanics partition function can be computed via virtual localization with respect to the toric action. The data involved in the localization formula can be obtained from the equivariant character by the standard conversion \cite{Nekrasov:2002qd}
\begin{equation}
\sum_i n_i \e^{w_i} \longrightarrow \prod_i w_i^{n_i} \, ,
\end{equation}
where $n_i = \pm$. Since the complex is self-dual, numerator and denominators cancel up to an overall sign given by $\dim \mathsf{S}^0_\pi + \dim \mathsf{S}^2_\pi$. In particular the dependence on the toric weights drops out. Comparing with (\ref{DTinv}) we see
\begin{equation}
(-1)^{\dim T_\pi \mathcal{M}_{\mathbf{d}}^{c} (Q [f_{\mathbf{3}}])} = (-1)^{ d_\circ^2 + d_\bullet^2 + 2 d_\circ + 2 d_\bullet + 2 d_\circ d_\bullet } \, .
\end{equation}
The framed degeneracies are therefore given by
\begin{equation} \label{DTSU3-3}
{\tt DT}^{c}_{\mathbf{d}} ( W_{\zeta , \mathbf{3}})= \sum_{\pi \in \mathcal{M}_{\mathbf{d}}^{c} (Q [f_{\mathbf{3}}])^{\mathbb{T}_\mathcal{W}}} \  (-1)^{ d_\circ^2 + d_\bullet^2 + 2 d_\circ + 2 d_\bullet + 2 d_\circ d_\bullet } \, .
\end{equation}
We can now compute \eqref{DTSU3-3} explicitly and write down the BPS invariants in the following Table
\begin{equation} \label{SU2-3-table}
\begin{array}{|c|c|c|c|}
\hline
(d_\circ , d_\bullet) & \text{n. fixed points} & (-1)^{\dim T_\pi} & DT \\
\hline (0,0) & 1 & + & +1 \\
(1,0) & 2 & - & -2 \\
(2,0) & 1 & + & +1 \\
(1,1) & 1 & + & +1 \\
(2,1) & 2 & - & -2 \\ 
(2,2) & 1 & + & +1 \\
\hline 
\end{array}
\end{equation}
Finally we can compute the generating function of framed BPS degeneracies, that is the vev of the Wilson line expressed in terms of the Darboux coordinates on the Hitchin moduli space
\begin{eqnarray}
\langle W_{\zeta , \mathbf{3}} \rangle_{q=-1} &=& \sum_{\mathbf{d}} \ {\tt DT}^c_{\mathbf{d}}  ( W_{\zeta , \mathbf{3}}) \ X_{e_{\mathbf{3}} + d_{\bullet} e_{\bullet} + d_{\circ} e_{\circ}}  \cr
&=& X_{-(e_\circ + e_\bullet)} -2 X_{- e_\bullet} + X_{+e_\circ - e_\bullet} + X_{0} -2 X_{e_\circ} + X_{e_\circ + e_\bullet} \ .
\end{eqnarray}
Here $X_0 = 1$ by convention. This spectum consists of four hypermultiplets and two vector multiplets. Note that the Donaldson-Thomas invariant is negative for vector multiplets, as expected. Passing to untwisted coordinates, this expression agrees with the results of \cite{BPSlinesCluster}.

\subsection{$SU(2)$ with a Wilson line in the representation $\mathbf{4}$}

We will now couple $SU(2)$ super Yang-Mills to a Wilson line defect in the representation $\mathbf{4}$ of $SU(2)$. As explained in the previous Section, this is given by the quiver
$Q [f_{\mathbf{4}}]$
\begin{equation}
\xymatrix@C=8mm{
&  \bullet  \ar@{..>}@<-0.5ex>[dl]_{B_1, \cdots, B_3}  \\
f_{\mathbf{4}}  \ar@{..>}@<-0.5ex>[dr]_{C_1, \cdots, C_3}   & \\
& \circ \ar@<-0.5ex>[uu]_{\tilde{A}}  \ar@<0.5ex>[uu]^{A} 
}
\end{equation}
with superpotential
\begin{equation}
\mathcal{W} = A C_1 B_1 + B_2 ( \tilde{A} C_2  - \tilde{A} C_1 ) + B_3 (A C_3 - A C_2)
\end{equation}
The equations of motion can be computed easily and read
\begin{eqnarray}
r_A \ : \ C_1 B_1 + C_3 B_3 - C_2 B_3 = 0, \qquad  r_{\tilde{A}} \ : \ C_2 B_2 - C_1 B_2 = 0, \cr
r_{B_1} \ : \ A C_1 = 0, \ \ r_{B_2} \ : \ \tilde{A} C_2 - \tilde{A} C_1 = 0, \  \ r_{B_3} \ : \ A C_3 - A C_2 = 0, \cr
r_{C_1} \ : \ B_1 A - B_2 \tilde{A}= 0, \ \ r_ {C_2} \ : \ B_2 \tilde{A} - B_3 A = 0, \ \ r_{C_3} \ : \ B_3 A = 0 \, ,
\end{eqnarray}
while the following relations hold
\begin{eqnarray}
rr_{\circ} \ : \ A r_A + \tilde{A} r_{\tilde{A}} - r_{B_1} B_1 - r_{B_2} B_2 - r_{B_3} B_3 = 0 \, , \cr
rr_{\bullet} \ : \ r_A A + r_{\tilde{A}} \tilde{A} - C_1 r_{C_1} - C_2 r_{C_2} -C_3 r_{C_3} = 0 \, .
\end{eqnarray}
As we have explained before, we define a toric action on the matrix quantum mechanics such that each field is rescaled by a phase. The condition that the F-term relations are invariant implies
\begin{equation}
\epsilon_{C_1} = \epsilon_{C_2} = \epsilon_{C_3} \, , \qquad \epsilon_{B_1} = \epsilon_{B_3} \, , \qquad \epsilon_{B_1} + \epsilon_A = \epsilon_{B_2} + \epsilon_{\tilde{A}} \, .
\end{equation}
Furthermore we can impose one extra condition coming from the gauge torus. We choose
\begin{equation} \label{toricSU2-4}
\epsilon_A + \epsilon_{C_1} + \epsilon_{B_1} = 0 \, ,
\end{equation}
so that the superpotential $\mathcal{W}$ is toric invariant. Note that indeed condition (\ref{toricSU2-4}) also implies $\epsilon_{\tilde{A}} + \epsilon_{B_2} + \epsilon_{C_2} = 0$.

\begin{figure}[h]
    \centering
    \includegraphics[width=0.20\textwidth]{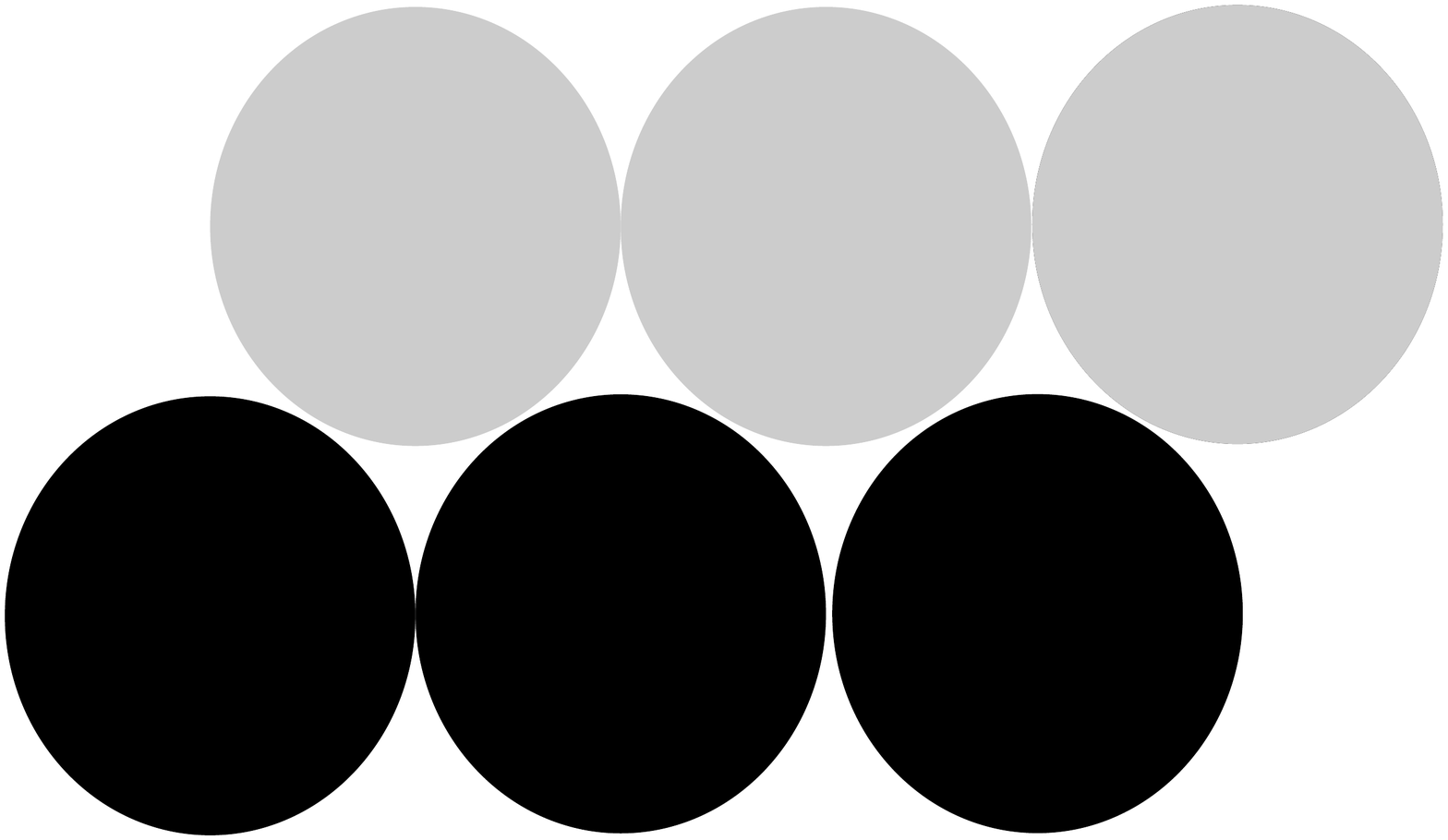}
    \caption{Initial configuration in the case of $SU(2)$ super Yang-Mills coupled to a Wilson line defect in the representation $\mathbf{4}$}
    \label{initialSU2-4}
\end{figure}

\begin{figure}[h]
    \centering
    \includegraphics[width=0.5\textwidth]{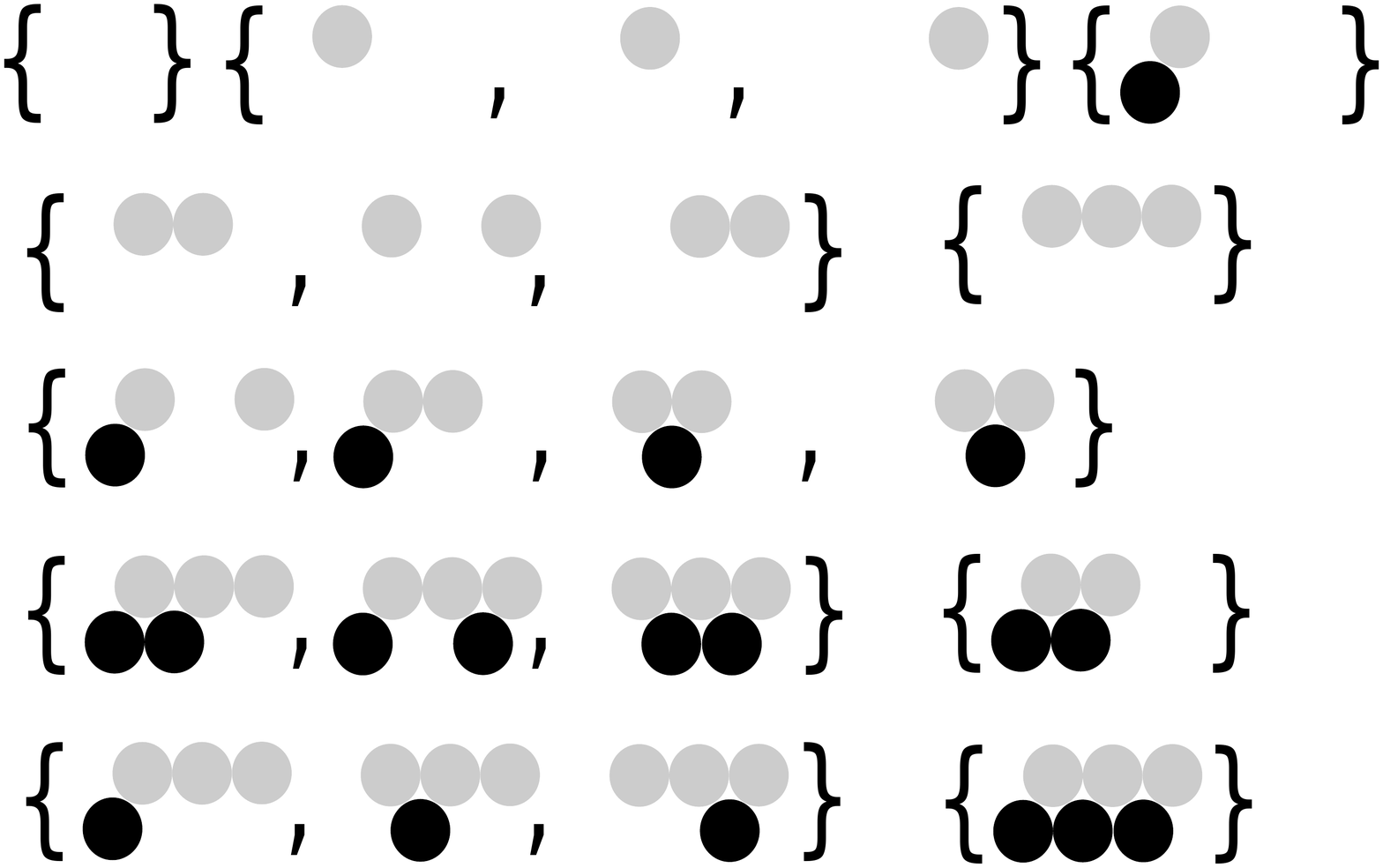}
    \caption{All the pyramid partitions $\pi$ for $SU(2)$ with a Wilson line in the $\mathbf{4}$}
    \label{fixedSU2-4}
\end{figure}

The classification of $\mathbb{T}_{\mathcal{W}}$-fixed points proceeds as in the previous Sections. Now the initial pyramid arrangement is shown in Figure \ref{initialSU2-4}
and we enumerate all the fixed points (that is to say the pyramid partitions obtained from Figure \ref{initialSU2-4}) in Figure  \ref{fixedSU2-4}, by the combinatorial rule that if a stone is part of the pyramid partition $\pi$, then all the stones immediately above are in $\pi$ too.
Now we turn to the contribution of the fixed point configurations. The equivariant quiver deformation complex is again self dual and therefore the parity of the tangent space $(-1)^{\dim T_\pi \mathcal{M}_{\mathbf{d}}^{c} (Q [f_\mathbf{4}])}$ can be computed from 
\begin{equation}
\dim \mathsf{S}^{0}_\pi + \dim \mathsf{S}^2_\pi = d_\circ^2 +d_\bullet^2 + 3 d_\circ + 3 d_\bullet + 2 d_\circ d_\bullet
\end{equation}
The framed BPS degeneracies are now given in compact form by
\begin{equation}
{\tt DT}^{c}_{\mathbf{d}} (W_{\zeta , \mathbf{4}} )= \sum_{\pi \in \mathcal{M}_{\mathbf{d}}^{c} (Q [f_\mathbf{4}])^{\mathbb{T}_\mathcal{W}}} \ (-1)^{d_\circ^2 +d_\bullet^2 + 3 d_\circ + 3 d_\bullet + 2 d_\circ d_\bullet} \ ,
\end{equation}
and are listed in the following Table
\begin{equation}
\begin{array}{|c|c|c|c|}
\hline
(d_\circ , d_\bullet) & \text{n. fixed points} & (-1)^{\dim T_\pi} & DT \\
\hline
(0,0) & 1 & + & +1 \\
(1,0) & 3 & + & +3 \\
(1,1) & 1 & + & +1 \\ 
(2,0) & 3 & + & +3 \\
(2,1) & 4 & + & +4 \\
(2,2) & 1 & + & +1 \\
(3,0) & 1 & + & +1 \\
(3,1) & 3 & + & +3 \\
(3,2) & 3 & + & +3 \\
(3,3) & 1 & + & +1 \\
\hline 
\end{array}
\end{equation}
Finally by putting all the results together, we can compute the generating function for a defect with core charge $\mathbf{RG} [W_{\zeta , \mathbf{4}} ] = e_\mathbf{4} = -\frac32 (e_\circ + e_\bullet)$
\begin{eqnarray}
\langle W_{\zeta , \mathbf{4}} \rangle_{q=-1} &=& \sum_{\mathbf{d}} \ {\tt DT}^{c}_{\mathbf{d}} (W_{\zeta , \mathbf{4}} ) \ X_{e_\mathbf{4} + d_\circ e_\circ + d_\bullet e_\bullet} \cr
&=& X_{-\frac32 (e_\bullet + e_\circ)} + 3 \, X_{- \frac32 e_\bullet - \frac12 e_\circ} + X_{-\frac12 e_\bullet - \frac12 e_\circ } + 3 \, X_{-\frac32 e_\bullet + \frac12 e_\circ} + 4 \, X_{-\frac12 e_\bullet + \frac12 e_\circ}
 \nonumber \\[4pt]  && 
 + X_{\frac12 e_\bullet + \frac12 e_\circ} + X_{- \frac32 e_\bullet + \frac32 e_\circ} +3 \, X_{-\frac12 e_\bullet + \frac32 e_\circ}
+ 3 \, X_{\frac12 e_\bullet + \frac32 e_\circ} + X_{\frac32 e_\bullet + \frac32 e_\circ} \ .
\end{eqnarray}
This result agrees with \cite{BPSlinesCluster,Cordova:2013bza}.

\subsection{$SU(2)$ with a Wilson line in the representation $\mathbf{5}$}

\begin{figure}[h]
    \centering
    \includegraphics[width=0.20\textwidth]{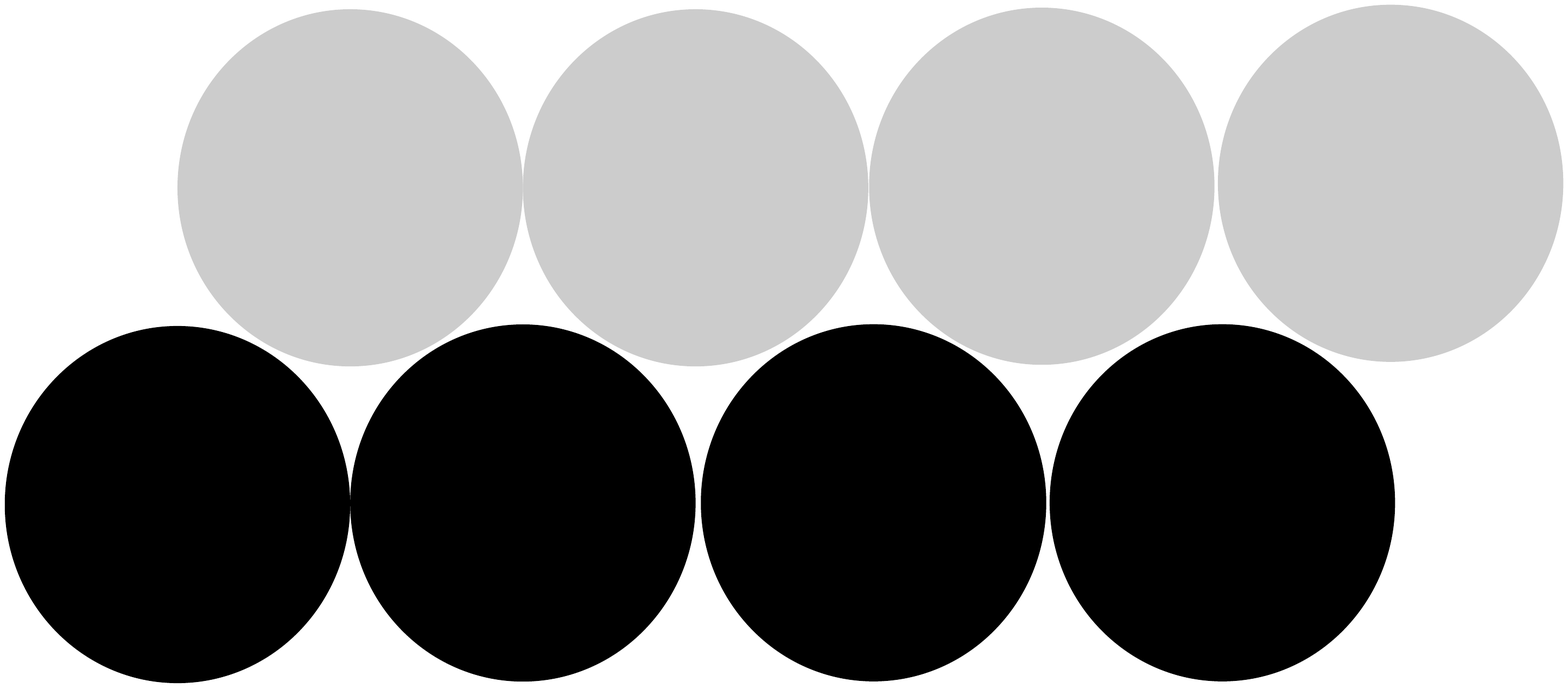}
    \caption{Initial configuration in the case of $SU(2)$ super Yang-Mills coupled to a Wilson line defect in the representation $\mathbf{5}$}
    \label{initialSU2-5}
\end{figure}

\begin{figure}[h]
    \centering
    \includegraphics[width=0.7\textwidth]{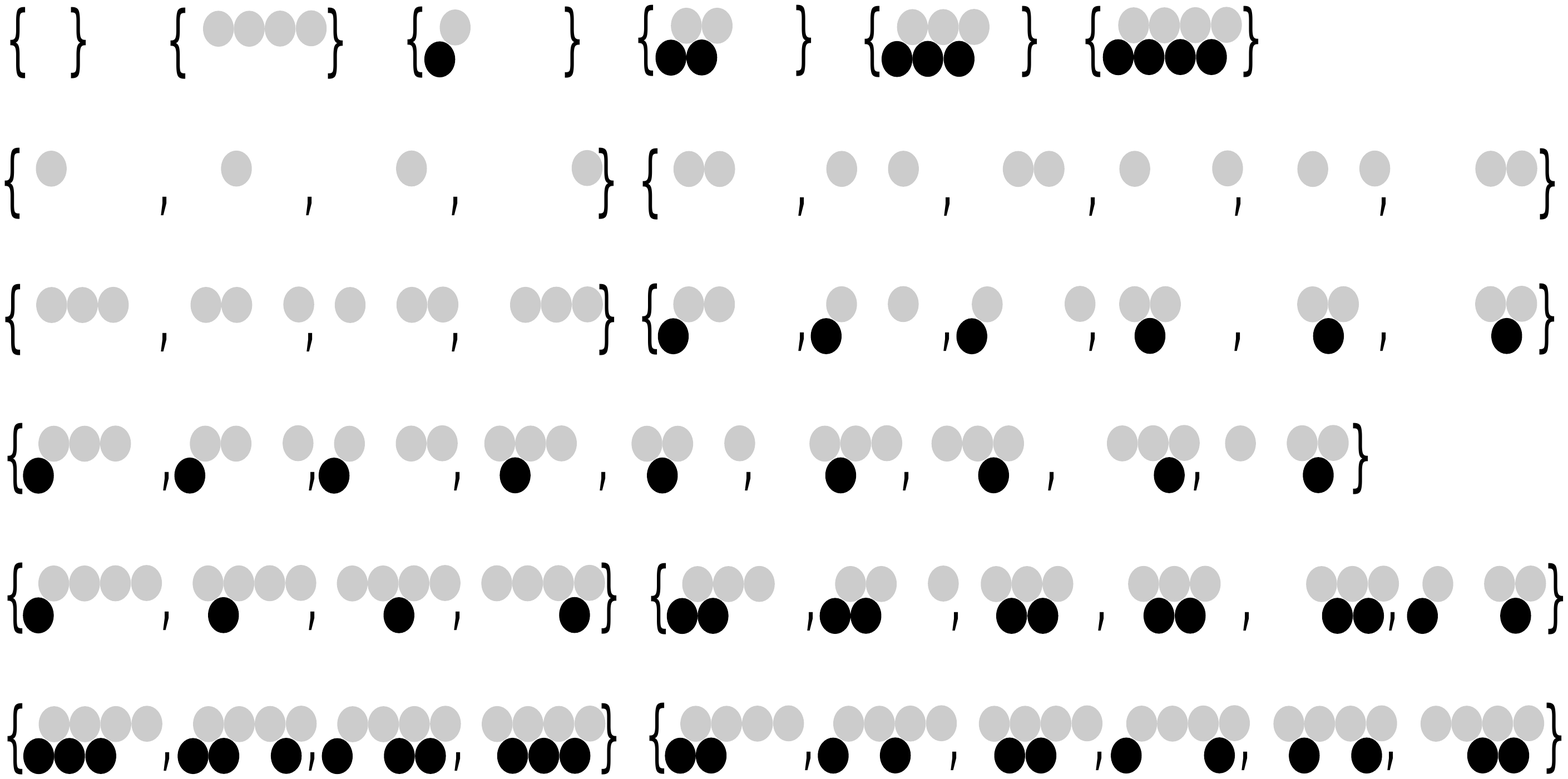}
    \caption{All the fixed points for $SU(2)$ with a Wilson line in the $\mathbf{5}$, classified by pyramid partitions}
    \label{fixedSU2-5}
\end{figure}

In this example we take the coupling of $SU(2)$ super Yang-Mills to a Wilson line in the representation $\mathbf{5}$ of $SU(2)$, given by the framed quiver
\begin{equation}
\xymatrix@C=8mm{
&  \bullet  \ar@{..>}@<-0.5ex>[dl]_{B_1, \cdots, B_4}  \\
f_{\mathbf{5}}  \ar@{..>}@<-0.5ex>[dr]_{C_1, \cdots, C_4}   & \\
& \circ \ar@<-0.5ex>[uu]_{\tilde{A}}  \ar@<0.5ex>[uu]^{A} 
}
\end{equation}
with superpotential
\begin{equation}
\mathcal{W} = A C_1 B_1 + B_2 (\tilde{A} C_2 - \tilde{A} C_1) + B_3 (A C_3 -  A C_2) + B_4 ( \tilde{A} C_4 -\tilde{A} C_3) \ .
\end{equation}
From $\mathcal{W}$ we find the following F-term equations
\begin{eqnarray}
r_A \ : \ C_1 B_1 + C_3 B_3 - C_2 B_3 = 0 \, , \cr  r_{\tilde{A}} \ : \ C_2 B_2 - C_1 B_2 + C_4 B_4 - C_3 B_4= 0 \, , \cr
r_{B_1} \ : \ A C_1 = 0, \ \ r_{B_2} \ : \ \tilde{A} C_2 - \tilde{A} C_1 = 0 \, , \cr  \ r_{B_3} \ : \ A C_3- A C_2 = 0, \ \ r_{B_4} \ : \ \tilde{A} C_4 - \tilde{A} C_3 = 0 \, , \cr
r_{C_1} \ : \ B_1 A -B_2 \tilde{A} = 0, \ \ r_ {C_2} \ : \ B_2 \tilde{A} - B_3 A = 0 \, , \cr \ r_{C_3} \ : \ B_3 A - B_4  \tilde{A} = 0, \ \ r_{C_4} \ : \ B_4 \tilde{A} = 0 \, ,
\end{eqnarray}
and the following relations
\begin{eqnarray}
rr_{\circ} \ : \ A r_A + \tilde{A} r_{\tilde{A}} - r_{B_1} B_1 - r_{B_2} B_2 - r_{B_3} B_3 - r_{B_4} B_4 = 0 \, , \cr
rr_{\bullet} \ : \ r_A A + r_{\tilde{A}} \tilde{A} - C_1 r_{C_1} - C_2 r_{C_2} -C_3 r_{C_3} - C_4 r_{C_4}= 0 \, .
\end{eqnarray}
Consider now the torus $\mathbb{T}_{\mathcal{W}}$. Compatibility with the F-term equations requires $\epsilon_{C_1} = \epsilon_{C_2} = \epsilon_{C_3} = \epsilon_{C_4} $, $\epsilon_{B_1} =\epsilon_{B_3} $, $\epsilon_{B_1} + \epsilon_A = \epsilon_{B_2} + \epsilon_{\tilde{A}}$, and $\epsilon_{B_2} =\epsilon_{B_4} $. Furthermore we can use the gauge torus to impose the extra condition $\epsilon_{A} +\epsilon_{C_1} +\epsilon_{B_1} = 0$. This condition also implies $\epsilon_{\tilde{A}} +\epsilon_{B_2} +\epsilon_{C_2} = 0$ and therefore the superpotential $\mathcal{W}$ is invariant. The initial pyramid arrangement and the pyramid partitions corresponding to the fixed points are shown in Figures \ref{initialSU2-5} and \ref{fixedSU2-5}.
From the localization formula we see that the parity of the tangent space can be read off
\begin{equation}
\dim \mathsf{S}^{0}_\pi + \dim \mathsf{S}^2_\pi = d_\circ^2 +d_\bullet^2 + 4 d_\circ + 4 d_\bullet + 2 d_\circ d_\bullet \, .
\end{equation}
Again the framed BPS degeneracies can be computed according to
\begin{equation}
{\tt DT}^{c}_{\mathbf{d}} (W_{\zeta , \mathbf{5}} ) = \sum_{\pi \in \mathcal{M}_{\mathbf{d}}^{c} (Q [f_{\mathbf{5}}] )^{\mathbb{T}_\mathcal{W}}} \ (-1)^{d_\circ^2 +d_\bullet^2 + 4 d_\circ + 4 d_\bullet + 2 d_\circ d_\bullet} \ .
\end{equation}
The contribution of each fixed point is summarized in the following table
\begin{equation}
\begin{array}{|c|c|c|c|}
\hline
(d_\circ , d_\bullet) & \text{n. fixed points} & (-1)^{\dim T_\pi} & DT \\
\hline
(0,0) & 1 & + & +1 \\
(1,0) & 4 & - & -4 \\
(1,1) & 1 & + & +1 \\ 
(2,0) & 6 & + & +6\\
(2,1) & 6 & - & -6 \\
(2,2) & 1 & + & +1 \\
(3,0) & 4 & - & -4 \\
(3,1) & 9 & + & +9 \\
(3,2) & 6 & - & -6 \\
(3,3) & 1 & + & +1 \\
(4,0) & 1 & + & +1 \\
(4,1) & 4 & - & -4 \\
(4,2) & 6 & + & +6 \\
(4,3) & 4 & - & -4 \\
(4,4) & 1 & + & +1 \\
\hline 
\end{array}
\end{equation}
Finally we can compute the vev of the Wilson line operator, given by the generating function of framed BPS degeneracies
\begin{align}
\langle W_{\zeta , \mathbf{5}} \rangle_{q=-1} &= \sum_{\mathbf{d}} \ {\tt DT}^{c}_{\mathbf{d}} (W_{\zeta , \mathbf{5}} ) \ X_{e_\mathbf{5} + d_\circ e_\bullet + d_\bullet e_\circ} \cr
&= X_{-2 (e_\bullet + e_\circ)} - 4 \, X_{-2 e_\bullet - e_\circ} + X_{- e_\bullet - e_\circ} + 6 \, X_{- 2 e_\bullet}
- 6 \, X_{- e_\bullet} + X_0 - 4 \, X_{e_\circ - 2 e_\bullet} 
 \\[4pt] \nonumber & \ \ 
+ 9 \, X_{e_\circ - e_\bullet} - 6 \, X_{e_\circ} 
+ X_{e_\bullet + e_\circ} + X_{- 2 e_\bullet + 2 e_\circ} - 4 \, X_{2 e_\circ - e_\bullet} + 6 \, X_{2 e_\circ} 
- 4 \, X_{2 e_\circ + e_\bullet} + X_{2 e_\bullet + 2 e_\circ}
\end{align}
for a defect with core charge $\mathbf{RG} (W_{\zeta , \mathbf{5}}) = e_\mathbf{5} = -2 (e_\circ + e_\bullet)$. 

This result agrees with \cite{Cordova:2013bza,BPSlinesCluster}.

\subsection{$SU(2)$ with a Wilson line in the representation $\mathbf{6}$}

To further illustrate the power of our combinatorial solution for the localization computation, we will now consider the case of a Wilson line in the representation $\mathbf{6}$ of $SU(2)$. In this case we limit ourselves to writing down the result directly. The pyramid arrangement is the one in Figure \ref{initialSU2-n} where each line contains $5$ stones. The parity of the tangent space can be read off directly from \eqref{paritySU2-n}. Our prediction for the framed BPS degeneracies is contained in the following table:
\begin{equation}
\begin{array}{|c|c|c|c|}
\hline
(d_\circ , d_\bullet) & \text{n. fixed points} & (-1)^{\dim T_\pi} & DT \\
\hline
(0,0) & 1 & + & +1 \\
(1,0) & 5 & + & +5 \\
(1,1) & 1 & + & +1 \\ 
(2,0) & 10 & + & +10\\
(2,1) & 8 & + & +8 \\
(2,2) & 1 & + & +1 \\
(3,0) & 10 & + & +10 \\
(3,1) & 18 & + & +18 \\
(3,2) & 9 & + & +9 \\
(3,3) & 1 & + & +1 \\
(4,0) & 5 & + & +5 \\
(4,1) & 16 & + & +16 \\
(4,2) & 18 & + & +18 \\
(4,3) & 8 & + & +8 \\
(4,4) & 1 & + & +1 \\
(5,0) &  1&  +& +1  \\
(5,1) &  5&  +& +5  \\
(5,2) &  10& + & +10  \\
(5,3) & 10 & + &  +10 \\
(5,4) &  5 & + &  +5 \\
(5,5) &  1& + &  +1 \\ 
\hline 
\end{array}
\end{equation}
One can check that this is result is correct by computing the OPE
\be
W_{\zeta , \mathbf{2}} * W_{\zeta , \mathbf{5}} = W_{\zeta , \mathbf{6}} + W_{\zeta , \mathbf{4}} 
\ee
where $\langle W_{\zeta , \mathbf{6}} \rangle_{q=-1} $ is the only unknown. Note however that our method does not assume the OPE but rather the OPE is explicitly recovered from the degeneracies computed with localization. Therefore our formalism is in principle a tool to determine the OPE of line defects in full generality. Also note that the combinatorial computation is relatively easy. It would be very interesting to extend our formalism to the full $q$-deformed case. We are currently investigating this issue. 

\subsection{Dyonic defects} \label{SU2-dyons-loca}

We have so far discussed Wilson line defects. We will now briefly consider another class of defects where the core charge is a dyon. In the case of $SU(2)$ super Yang-Mills, dyonic defects of this kind were studied in \cite{Cirafici:2013bha,BPSlinesCluster} using cluster algebra methods. Alternatively their vevs can be computed with the approach of \cite{Gaiotto:2010be}. Now we will reproduce these result by a direct localization computation\footnote{
Technically to reproduce \textit{exactly} these results we should use the untwisted coordinates $Y_\gamma$ in the Hitchin moduli space, related to the coordinates $X_\gamma$ by a quadratic refinement of the pairing between charges. In the examples below this simply amounts in neglecting the minus signs in the BPS invariants
.}. 

Consider the framed BPS quiver
\begin{equation}
\xymatrix@C=8mm{    f_{\bullet}^{[2]} \ar@{..>}[d]_{\beta_1,\beta_2} & \\
 \circ   \ar@<-0.5ex>[r]_{A_1,A_2}   \ar@<0.5ex>[r]  &  \bullet   \ar@{..>}[ul]_{\alpha_1,\dots,\alpha_4}
 }
\end{equation}
with superpotential
\begin{equation}
\mathcal{W}_{f_{\bullet}^{[2]}} = \alpha_1 A_1 \beta_1  + \alpha_2 A_2  \beta_1  + \alpha_3 A_1  \beta_2  + \alpha_4 A_2  \beta_2
\end{equation}
In this case the pyramid arrangement simply consists in two copies of $\circ$, since any other concatenation of arrows is set to zero by the equations of motion. The fixed points have the following dimension vectors $(d_\circ , d_\bullet)$: $(0,0)$, $(1,0)$ (with multiplicity two) and $(2,0)$. It is easy to see that the parity of the tangent space is
\begin{equation}
(-1)^{\mathrm{dim} \, T_\pi \mathcal{M}^c_{\mathbf{d}}} = (-1)^{d_\circ^2 + d_\bullet^2 - 2 d_\circ - 4 d_\bullet - 2 d_\circ d_\bullet } \, .
\end{equation}
Therefore the framed BPS spectrum is
\begin{equation}
\langle L_{\zeta , f_{\bullet}^{[2]}} \rangle_{q=-1} =  X_{-e_\bullet - 2 e_\circ} - 2 X_{- e_\circ - e_\bullet} + X_{- e_\bullet} \, .
\end{equation}
consistent with \cite{Cirafici:2013bha}. 

\begin{figure}[h]
    \centering
    \includegraphics[width=0.2\textwidth]{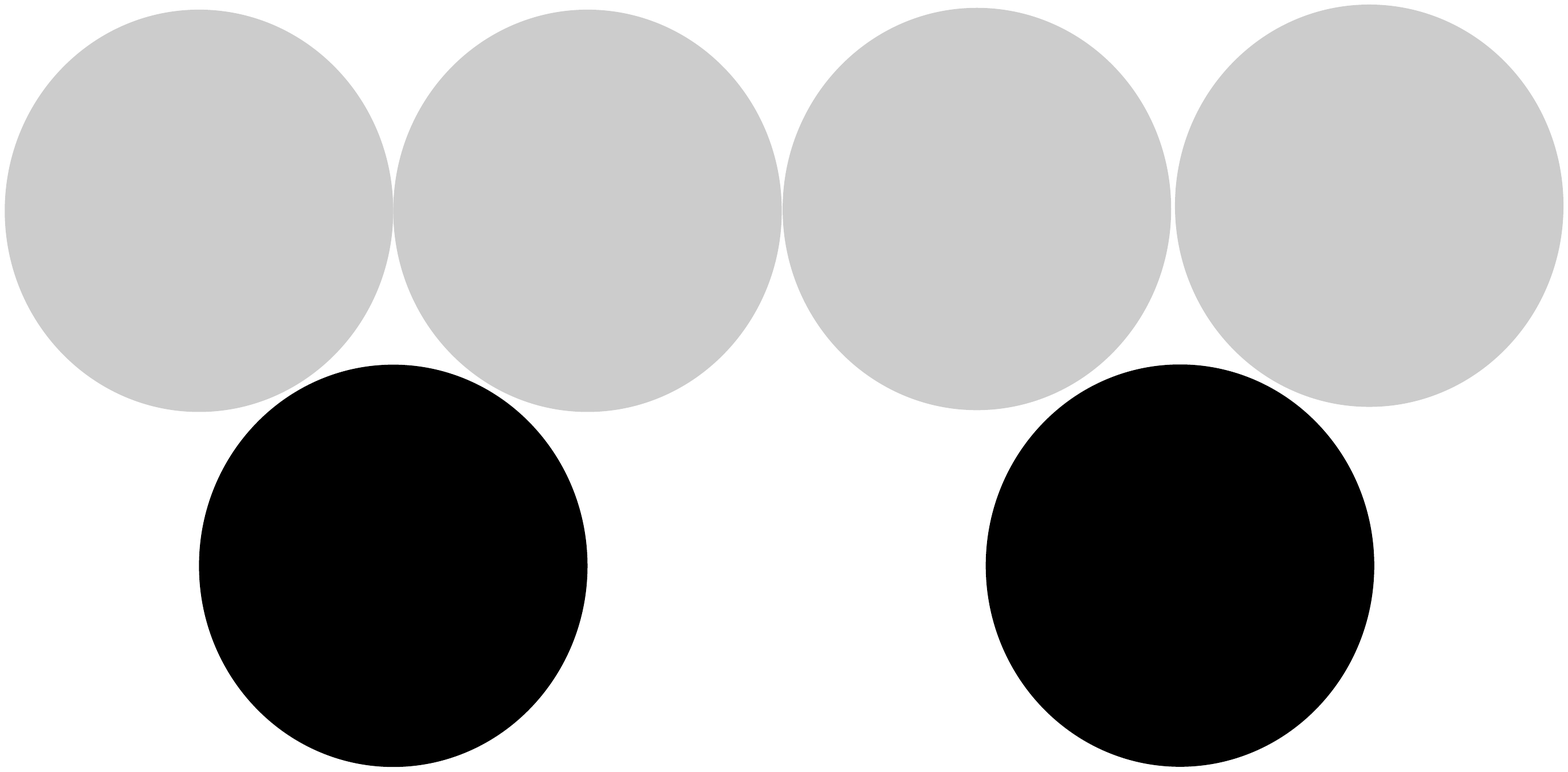}
    \caption{Pyramid arrangement for the defect $ f_{\bullet}^{[3]} $}
    \label{SU2dyon-3}
\end{figure}

Consider now the framed BPS quiver 
\begin{equation}
\xymatrix@C=8mm{    f_{\bullet}^{[3]} \ar@{..>}[d]_{\beta_1,\dots ,\beta_4} & \\
 \circ   \ar@<-0.5ex>[r]_{A_1,A_2}   \ar@<0.5ex>[r]  &  \bullet   \ar@{..>}[ul]_{\alpha_1,\dots,\alpha_6}
 }
\end{equation}
with superpotential
\begin{equation}
\mathcal{W}_{f_{\bullet}^{[3]}}  = \alpha_1 A_1 \beta_1  + \alpha_2 A_1 \beta_2 + \alpha_3 A_1 \beta_3  + \alpha_4 A_1 \beta_4  + \alpha_5 A_2 \beta_1 - \alpha_1 A_2 \beta_2 + \alpha_6 A_2 \beta_3 - \alpha_3 A_2  \beta_4 \, .
\end{equation}
The relevant equations of motions are
\begin{align}
A_1 \beta_1 &= A_2 \beta_2 \, , \cr
A_1 \beta_3 &= A_2 \beta_4 \, , \cr
A_1 \beta_2 &= A_1 \beta_4 = A_2 \beta_1 = A_2 \beta_3 = 0 \, ,
\end{align}
out of which we form the pyramid arrangement in Figure \ref{SU2dyon-3}.
The contribution of each fixed point is simply
\begin{equation}
(-1)^{d_\circ^2 + d_\bullet^2 - 4 d_\circ - 6 d_\bullet  - 2 d_\circ d_\bullet} \, ,
\end{equation}
and the framed BPS indices can be summarized in the following table
\begin{equation}
\begin{array}{|c|c|c|c|c|c|c|c|}
\hline
(d_{\circ} , d_{\bullet} ) & \text{fixed pts} & (-1)^{\dim T_\pi} & {\tt DT} & (d_{\circ} , d_{\bullet}  )& \text{fixed pts} & (-1)^{\dim T_\pi} & {\tt DT} \\
\hline 
(0,0) & 1 & + & +1 & (2,1) & 2 & - & -2 \\
(1,0) & 4 & - & -4  &  (3,1) & 4 & + & +4 \\
(2,0) & 6 & + & +6 & (4,1) & 2 & - & -2\\
(3,0) & 4 & - & -4 & (4,2) & 1 & + & +1 \\
(4,0) & 1 & + & +1  & &  &  &  \\
\hline 
\end{array}
\end{equation}
which again agrees with \cite{Cirafici:2013bha,BPSlinesCluster}.

\section{$SU(3)$ super Yang-Mills} \label{SU3loca}

We will now consider $SU(3)$ super Yang-Mills, coupled to Wilson line or dyonic defects. In this case the combinatorics is slightly more complicated, but we are still able to provide solutions in closed form. These computations agree with the results of \cite{BPSlinesCluster} obtained with cluster algebra methods.

The BPS quiver of pure $SU(3)$ super Yang-Mills is the following \cite{Fiol:2000pd,Alim:2011kw}:
\begin{equation}
\xymatrix@C=8mm{
&  \bullet_1   \ar[rr]^{\tilde{\psi}} & & \ar@<-0.5ex>[dd]_{\tilde{A}_2}  \ar@<0.5ex>[dd]^{A_2}   \circ_2 \\
 & & & \\
& \circ_1 \ar@<-0.5ex>[uu]_{\tilde{A}_1}  \ar@<0.5ex>[uu]^{A_1}& &   \ar[ll]^\psi  \bullet_2
}
\end{equation}
with superpotential
\begin{equation}
\mathcal{W} = \tilde{\psi} A_1 \psi A_2 - \tilde{\psi} \tilde{A}_1 \psi \tilde{A}_2 \, .
\end{equation}

\subsection{$SU(3)$ with a fundamental Wilson line} \label{SU3with3}

We will begin with the case of a Wilson line in the fundamental representation $\mathbf{3}$ of $SU(3)$. Such a defect has core charge $\mathbf{RG} (W_{\zeta , \mathbf{3}}) = -\frac23 (e_{\bullet_1} + e_{\circ_1}) - \frac13 (e_{\bullet_2} + e_{\circ_2})$, which indeed corresponds to the highest weight of the fundamental representation of SU(3) (see \cite{BPSlinesCluster}). The BPS quiver describing the coupled system is
\begin{equation}
\xymatrix@C=8mm{
&  \bullet_1 \ar@{..>}[dl]^{B}  \ar[rr]^{\tilde{\psi}} & & \ar@<-0.5ex>[dd]_{\tilde{A}_2}  \ar@<0.5ex>[dd]^{A_2}   \circ_2 \\
f_{\mathbf{3}}   \ar@{..>}[dr]_{C}  & & & \\
& \circ_1 \ar@<-0.5ex>[uu]_{\tilde{A}_1}  \ar@<0.5ex>[uu]^{A_1}& &   \ar[ll]^\psi  \bullet_2
}
\end{equation}
with superpotential
\begin{equation} \label{SU3-3-W}
\mathcal{W} = CBA_1 + \tilde{\psi} A_1 \psi A_2 - \tilde{\psi} \tilde{A}_1 \psi \tilde{A}_2 \, .
\end{equation}
The equations of motion are
\begin{align}
r_B \ &: \ A_1 \, C = 0 \, , \cr
r_C \ &: \ B \, A = 0 \, , \cr
r_{A_1} \ &: \ C \, B + \psi A_2 \tilde{\psi} = 0 \, , \cr
r_{\tilde{A}_1} \ &: \ \psi \, \tilde{A}_2 \tilde{\psi} = 0 \, , \cr
r_{A_2} \ &: \ \tilde{\psi} \, A_1 \psi = 0 \, , \cr
r_{\tilde{A}_2} \ &: \ \tilde{\psi} \tilde{A}_1 \psi = 0 \, , \cr
r_\psi \ &: \ A_2 \tilde{\psi} A_1 - \tilde{A}_2 \tilde{\psi} \tilde{A}_1 = 0 \, , \cr
r_{\tilde{\psi}} \ &: \ A_1 \psi A_2 - \tilde{A}_1 \psi \tilde{A}_2 = 0 \ .
\end{align}
The natural toric action is compatible with the F-term equations if
\begin{align}
\epsilon_C + \epsilon_B &= \epsilon_\psi + \epsilon_{A_2} + \epsilon_{\tilde{\psi}} \, , \cr
\epsilon_{A_1} + \epsilon_{A_2} &= \epsilon_{\tilde{A_1}} + \epsilon_{\tilde{A}_2} \, .
\end{align}
Once again to construct the pyramid arrangement we have to look at the relevant terms in the Jacobian algebra $\mathscr{J}_\mathcal{W} = \oplus_{n\ge 0} \mathscr{J}_{\mathcal{W}, n}$:
\begin{align}
\mathscr{J}_{\mathcal{W} , 0} &= \{ v \} \, , \cr
\mathscr{J}_{\mathcal{W} , 1} &= \{C \, v \} \, , \cr
\mathscr{J}_{\mathcal{W} , 2} &= \{\tilde{A}_1 \, C \, v \} \, , \cr
\mathscr{J}_{\mathcal{W} , 3} &= \{\tilde{\psi} \, \tilde{A}_1 \, C \, v \} \, , \cr
\mathscr{J}_{\mathcal{W} , 4} &= \{A_2 \, \tilde{\psi} \, \tilde{A}_1 \, C \, v \} \, .
\end{align}
The pyramid arrangement truncates due to the equation $C \, B + \psi \, A_2 \, \tilde{\psi} = 0$ restricted to the set of field configurations such that $B=0$, the only regions of the moduli space which can contribute to the fixed points due to the condition $\dim \, V_f = 1$.  As we have explained previously configurations with $B \neq 0$ do not contribute to the set of relevant fixed points (for which $\mathrm{dim} \, V_{f_{\mathbf{3}}} = 1$) and therefore, while they are generically part of the moduli space, can be excluded from the analysis. In this case the pyramid arrangement is trivial, and only consists in the sequence of stones: $\circ_1 , \bullet_1 , \circ_2 , \bullet_2$. It is therefore a routine application of our formalism to write down the fixed points.

To simplify the computations we can use the gauge torus $\mathbb{T}_G = (\mathbb{C}^*)^3$ to impose three conditions on the toric weights. In this case 
\begin{align} \label{TGSU3-3}
\epsilon_{A_1} + \epsilon_C + \epsilon_B &= 0 \, , \cr
\epsilon_{\tilde{\psi}} + \epsilon_{A_1} + \epsilon_{\psi} + \epsilon_{A_2} &= 0 \, .
\end{align}
are enough to make the superpotential invariant, and we can use the remaining condition to set one of the toric weights to zero. Around each fixed point $\pi$, the local structure of the moduli space is captured by the deformation complex
\begin{equation}
\xymatrix@C=8mm{  0 \ar[r] & \mathsf{S}^0_{\pi} \ar[r]^{\delta_0} & \mathsf{S}^1_\pi \ar[r]^{\delta_1} & \mathsf{S}^2_\pi \ar[r]^{\delta_2} & \mathsf{S}^3_\pi \ar[r] & 0
} \ .
\end{equation}
Due to the equations \eqref{TGSU3-3} the complex is self-dual, and the relevant terms are
\begin{align}
\mathsf{S}^0_\pi &=\mathrm{Hom}_{\mathbb{C}} (V_{\circ_1 , \pi} , V_{\circ_1 , \pi}) \oplus \mathrm{Hom}_{\mathbb{C}} (V_{\bullet_1 , \pi} , V_{\bullet_1 , \pi}) \oplus
\mathrm{Hom}_{\mathbb{C}} (V_{\circ_2 , \pi} , V_{\circ_2 , \pi}) \oplus \mathrm{Hom}_{\mathbb{C}} (V_{\bullet_2 , \pi} , V_{\bullet_2 , \pi}) \, ,
\cr
\mathsf{S}^1_\pi &= \mathrm{Hom}_{\mathbb{C}} (V_{\circ_1 , \pi} , V_{\bullet_1 , \pi}) \otimes (t_{A_1} + t_{\tilde{A}_1}) \oplus \mathrm{Hom}_{\mathbb{C}} (V_{\bullet_1 , \pi} , V_{f , \pi}) \otimes t_B  
\cr &
\oplus \mathrm{Hom}_{\mathbb{C}} (V_{f , \pi} , V_{\circ_1 , \pi}) \otimes t_{C} \oplus \mathrm{Hom}_\mathbb{C} (V_{\bullet_2 , \pi} , V_{\circ_1 , \pi}) \otimes t_{\psi} \oplus \mathrm{Hom}_\mathbb{C} (V_{\bullet_1, \pi} , V_{\circ_2 , \pi}) \otimes t_{\tilde{\psi}}
\cr &
\oplus \mathrm{Hom}_\mathbb{C} (V_{\circ_2 , \pi} , V_{\bullet_2 , \pi}) \otimes (t_{A_2} + t_{\tilde{A}_2}) \, .
\end{align}
The contribution to each fixed point can be written down immediately
\begin{equation}
(-1)^{\mathrm{dim} \, T_\pi \mathcal{M}^c_{\mathbf{d}}} = (-1)^{d_{\bullet_1}^2+d_{\circ_1}^2+d_{\bullet_2}^2+d_{\circ_2}^2-d_{\bullet_1}-d_{\circ_1}-2 d_{\bullet_1} d_{\circ_1}-d_{\bullet_1} d_{\circ_2}-d_{\circ_1} d_{\bullet_2}-2 d_{\bullet_2} d_{\circ_2}} \, .
\end{equation}
It is easy to see that this is always positive for the dimension vectors corresponding to toric fixed modules. Therefore the framed BPS spectrum consists of only hypermultiplets, and its generating function is
\begin{align}
\langle W_{ \zeta , \mathbf{3}} \rangle_{q=-1} &= \sum_{\mathbf{d}} \ \mathtt{DT}^c_{\mathbf{d}} (W_{ \zeta , \mathbf{3}}) \ X_{e_c + d_{\circ_1} e_{\circ_1} + d_{\bullet_1} e_{\bullet_1}+d_{\circ_2} e_{\circ_2}+d_{\bullet_2} e_{\bullet_2} } \cr
&= X_{-\frac23 e_{\bullet_1} - \frac23 e_{\circ_1} - \frac13 e_{\bullet_2} - \frac13 e_{\circ_2}} + 
X_{-\frac23 e_{\bullet_1} + \frac13 e_{\circ_1} - \frac13 e_{\bullet_2} - \frac13 e_{\circ_2}} + 
X_{+\frac13 e_{\bullet_1} + \frac13 e_{\circ_1} - \frac13 e_{\bullet_2} - \frac13 e_{\circ_2}} \cr & \ + 
X_{+\frac13 e_{\bullet_1} + \frac13 e_{\circ_1} - \frac13 e_{\bullet_2} + \frac23 e_{\circ_2}} + 
X_{+\frac13 e_{\bullet_1} + \frac13 e_{\circ_1} + \frac23 e_{\bullet_2} + \frac23 e_{\circ_2}} \, .
\end{align}
This results indeed agrees with \cite{BPSlinesCluster}, based on cluster algebra methods, as well as with \cite{Williams:2014efa}.

\subsection{$SU(3)$ with a Wilson line in the representation $\mathbf{6}$}

Now we consider the case of $SU(3)$ with a Wilson line in the representation $\mathbf{6}$. This case is modelled on the framed quiver
\begin{equation}
\xymatrix@C=8mm{
&  \bullet_1 \ar@{..>}@<-0.5ex>[dl]_{\tilde{B}}  \ar@{..>}@<0.5ex>[dl]^{B}  \ar[rr]^{\tilde{\psi}} & & \ar@<-0.5ex>[dd]_{\tilde{A}_2}  \ar@<0.5ex>[dd]^{A_2}   \circ_2 \\
f_{\mathbf{6}}   \ar@{..>}@<-0.5ex>[dr]_{\tilde{C}}  \ar@{..>}@<0.5ex>[dr]^{C}   & & & \\
& \circ_1 \ar@<-0.5ex>[uu]_{\tilde{A}_1}  \ar@<0.5ex>[uu]^{A_1}& &   \ar[ll]^\psi  \bullet_2
} \, .
\end{equation}
We take the superpotential
\begin{equation} \label{W-SU3-6}
\mathcal{W} = CBA_1-\tilde{C} B A_1+ \tilde{C} \tilde{B} A_1 - C \tilde{B} \tilde{A}_1 + \tilde{\psi} A_1 \psi A_2 - \tilde{\psi} \tilde{A}_1 \psi \tilde{A}_2 \, .
\end{equation}
It is easy to derive the equations of motion
\begin{eqnarray}
r_{A_1} \ &:& \ C B-\tilde{C} B+\tilde{C} \tilde{B} + \psi A_2 \tilde{\psi} = 0 \, ,  \label{rA1-SU3-6}  \\
r_B \ &:& \ A_1 C - A_1 \tilde{C} = 0 \, ,   \label{rB-SU3-6} \\
r_C \ &:& \ B A_1 - \tilde{B} \tilde{A}_1 = 0 \, ,  \\
r_{\tilde{A}_1} \ &:& \ - C \tilde{B} - \psi \tilde{A}_2 \tilde{\psi} = 0 \, ,  \label{rA1t-SU3-6}  \\
r_{\tilde{B}} \ &:& \  A_1 \tilde{C} - \tilde{A}_1 C= 0 \, , \label{rBt-SU3-6}   \\
r_{\tilde{C}} \ &:& \ - B A_1 + \tilde{B} A_1 = 0 \, , \\
r_{A_2} \ &:& \ \tilde{\psi} A_1 \psi = 0\, ,  \\
r_{\tilde{A_2}} \ &:& \ \tilde{\psi} \tilde{A}_1 \psi = 0 \, ,  \\
r_{\psi} \ &:& \ A_2  \tilde{\psi} A_1 - \tilde{A}_2 \tilde{\psi} \tilde{A}_1 = 0 \, ,  \\
r_{\tilde{\psi}} \ &:& \ A_1 \psi A_2 - \tilde{A}_1 \psi  \tilde{A}_2 = 0 \, .  \label{rpsit-SU3-6}
\end{eqnarray}
As explained in the previous Sections, for the classification of the fixed points we are only interested in solutions of the equations of motions for which $B = \tilde{B} = 0$.

We let $\mathbb{T}$ act by rescaling each field. Compatibility with the $\partial \, \mathcal{W} = 0$ relations implies
\begin{eqnarray}
\epsilon_C = \epsilon_{\tilde{C}} \, , \qquad \epsilon_{A_1} = \epsilon_{\tilde{A}_1} \, , \qquad \epsilon_B = \epsilon_{\tilde{B}} \, , \qquad \epsilon_{A_2} = \epsilon_{\tilde{A}_2}  \, .
\end{eqnarray}
If we furthermore choose to use the gauge torus $\mathbb{T}_G = (\mathbb{C}^*)^3$ to impose the following conditions
\begin{eqnarray}
\epsilon_{A_1} + \epsilon_C + \epsilon_B = 0 \, , \\
\epsilon_{\tilde{\psi}} + \epsilon_{A_1} + \epsilon_{\psi} + \epsilon_{A_2} = 0 \, ,
\end{eqnarray}
then the superpotential $\mathcal{W}$ is invariant under the toric action. We have one condition left, which we can use to set, say $\epsilon_{C} = 0$.

\begin{figure}[H]
    \centering
    \includegraphics[width=0.20\textwidth]{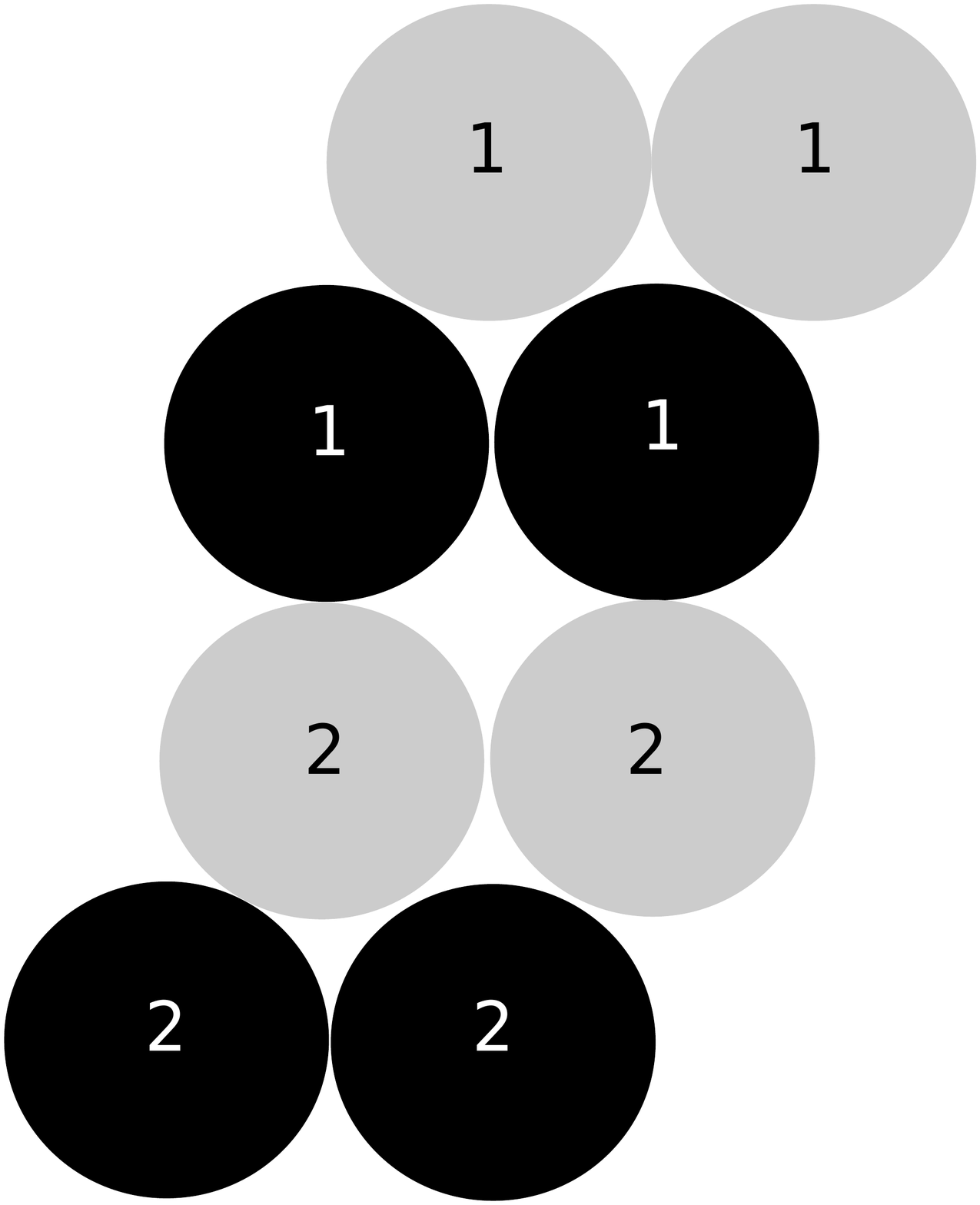}
    \caption{The pyramid arrangement for the case of $SU(3)$ Yang-Mills coupled to a Wilson line in the representation $\mathbf{6}$ of $SU(3)$}
    \label{pyramid-SU3-6}
\end{figure}

\begin{figure}[H]
    \centering
    \includegraphics[width=0.46\textwidth]{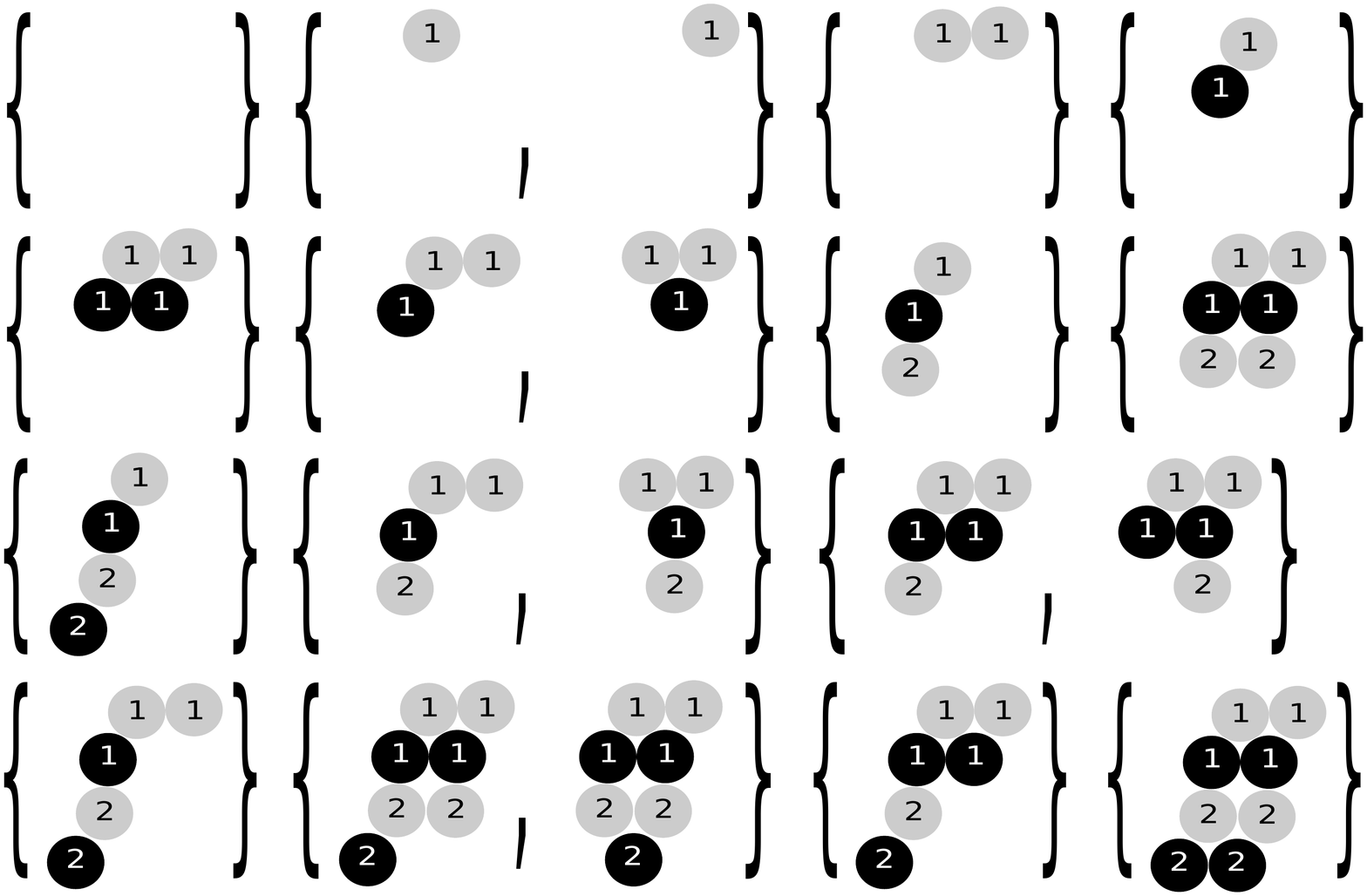}
    \caption{All the pyramid partitions corresponding to fixed points for $SU(3)$ with a Wilson line in the $\mathbf{6}$}
    \label{fixed-SU3-6}
\end{figure}

As in the previous Sections, the fact that the superpotential is invariant under the toric action implies that the deformation complex 
\begin{equation}
\xymatrix@C=8mm{  0 \ar[r] & \mathsf{S}^0_{\pi} \ar[r]^{\delta_0} & \mathsf{S}^1_\pi \ar[r]^{\delta_1} & \mathsf{S}^2_\pi \ar[r]^{\delta_2} & \mathsf{S}^3_\pi \ar[r] & 0 
} \, ,
\end{equation}
is self-dual and therefore to compute the contribution of each fixed point we only need the first two terms
\begin{eqnarray}
\mathsf{S}^0_\pi &=&\mathrm{Hom}_{\mathbb{C}} (V_{\circ_1 , \pi} , V_{\circ_1 , \pi}) \oplus \mathrm{Hom}_{\mathbb{C}} (V_{\bullet_1 , \pi} , V_{\bullet_1 , \pi}) \oplus
\mathrm{Hom}_{\mathbb{C}} (V_{\circ_2 , \pi} , V_{\circ_2 , \pi}) \oplus \mathrm{Hom}_{\mathbb{C}} (V_{\bullet_2 , \pi} , V_{\bullet_2 , \pi}) \, ,
\cr
\mathsf{S}^1_\pi &=& \mathrm{Hom}_{\mathbb{C}} (V_{\circ_1 , \pi} , V_{\bullet_1 , \pi}) \otimes (t_{A_1} + t_{\tilde{A}_1}) \oplus \mathrm{Hom}_{\mathbb{C}} (V_{\bullet_1 , \pi} , V_{f , \pi}) \otimes (t_B + t_{\tilde{B}}) 
\cr &&
\oplus \mathrm{Hom}_{\mathbb{C}} (V_{f , \pi} , V_{\circ_1 , \pi}) \otimes (t_{C} + t_{\tilde{C}} ) \oplus \mathrm{Hom}_\mathbb{C} (V_{\bullet_2 , \pi} , V_{\circ_1 , \pi}) \otimes t_{\psi} \oplus \mathrm{Hom}_\mathbb{C} (V_{\bullet_1, \pi} , V_{\circ_2 , \pi}) \otimes t_{\tilde{\psi}}
\cr &&
\oplus \mathrm{Hom}_\mathbb{C} (V_{\circ_2 , \pi} , V_{\bullet_2 , \pi}) \otimes (t_{A_2} + t_{\tilde{A}_2}) \, .
\end{eqnarray}
In the complex above $\delta_2$ is the linearization of the following relations between the F-term conditions
\begin{eqnarray}
rr_{\bullet_1} \ & :&  \ A_1 r_{A_1} + \tilde{A}_1 r_{\tilde{A}_1} - r_B B - r_{\tilde{B}} \tilde{B} - r_{\tilde{\psi}} \tilde{\psi} = 0 \, , \cr
rr_{\circ_1} \ & : & \  C r_C + \tilde{C} r_{\tilde{C}} + \psi r_\psi - r_{A_1} A_1 - r_{\tilde{A}_1} \tilde{A}_1 = 0 \, , \cr
rr_{\circ_2} \ & : & \  \tilde{\psi} r_{\tilde{\psi}} -r_{A_2} A_2 + r_{\tilde{A}_2} \tilde{A}_2 = 0 \, , \cr
rr_{\bullet_2} \ & : & \ A_2 r_{A_2} - \tilde{A}_2 r_{\tilde{A}_2} - r_\psi \psi = 0 \, .
\end{eqnarray}
Therefore at each fixed point
\begin{equation}
(-1)^{\dim T_\pi} = (-1)^{d_{\bullet_1}^2+d_{\circ_1}^2+d_{\bullet_2}^2+d_{\circ_2}^2-2 d_{\bullet_1}-2 d_{\circ_1}-2 d_{\bullet_1} d_{\circ_1}-d_{\bullet_1} d_{\circ_2}-d_{\circ_1} d_{\bullet_2}-2 d_{\bullet_2} d_{\circ_2}} \, .
\end{equation}
Now we write down the Jacobian algebra elements corresponding to cyclic modules $\mathscr{J}_{\mathcal{W}} = \oplus_{n=0}^\infty \mathscr{J}_{\mathcal{W} , n}$
\begin{eqnarray}
\mathscr{J}_{\mathcal{W} , 0} &=& \{ v  \} \, , \cr
\mathscr{J}_{\mathcal{W} , 1} &=& \{ C v \, ,\tilde{C} v  \} \, , \cr
\mathscr{J}_{\mathcal{W} , 2} &=& \{A_1 C v = A_1 \tilde{C} v = \tilde{A}_1 C v \ , \  \tilde{A}_1 \tilde{C} v \} \, , \cr
\mathscr{J}_{\mathcal{W} , 3} &=& \{ \tilde{\psi} A_1 C v \ , \ \tilde{\psi} \tilde{A}_1 \tilde{C} v \} \, , \cr
\mathscr{J}_{\mathcal{W} , 4} &=& \{ A_2 \tilde{\psi} A_1 C v = \tilde{A}_2 \tilde{\psi} \tilde{A}_1 \tilde{C} v = \tilde{A}_2 \tilde{\psi} A_1 C v \ , \ A_2 \tilde{\psi} \tilde{A}_1 \tilde{C} v \} \, ,
\end{eqnarray}
and summands with higher grading do not contribute to the classification of the fixed points due to the condition $\dim \, V_f = 1$ and equations (\ref{rA1-SU3-6}) and (\ref{rA1t-SU3-6}) evaluated at $B=\tilde{B}=0$. To write down $\mathscr{J}_{\mathcal{W} , 4}$ we have used
\begin{equation}
A_2 \tilde{\psi} A_1 C v = A_2 \tilde{\psi} A_1 \tilde{C} v = \tilde{A}_2 \tilde{\psi} \tilde{A}_1 \tilde{C} v \, ,
\end{equation}
thanks to (\ref{rB-SU3-6}) and (\ref{rpsit-SU3-6}) respectively, and 
\begin{equation}
A_2 \tilde{\psi} A_1 C v = \tilde{A}_2 \psi \tilde{A}_1 C v = \tilde{A}_2 \psi A_1 \tilde{C} v = \tilde{A}_2 \psi A_1 C v \, ,
\end{equation}
thanks to  (\ref{rpsit-SU3-6}), (\ref{rBt-SU3-6}) and (\ref{rB-SU3-6}) respectively. Now we would like to have a classification of fixed points, in terms of pyramid partitions extracted out of a pyramid arrangement. These are simply given by configurations of stones $\pi$ such that, if a stone is in $\pi$, then so are all the stones immediately above. Given the form of the path algebra, the pyramid arrangement for this quiver is the one shown in Figure \ref{pyramid-SU3-6}
The pyramid partitions of  Figure \ref{pyramid-SU3-6} are listed in Figure \ref{fixed-SU3-6}
At this point the associated enumerative invariants can be easily computed and are listed in the following Table
\begin{equation}
\begin{array}{|c|c|c|c|c|c|c|c|}
\hline
(d_{\circ_1} , d_{\bullet_1} , d_{\circ_2} , d_{\bullet_2} ) & \text{fixed pts} & (-1)^{\dim T_\pi} & {\tt DT} & (d_{\circ_1} , d_{\bullet_1} , d_{\circ_2} , d_{\bullet_2} )& \text{fixed pts} & (-1)^{\dim T_\pi} & {\tt DT} \\
\hline 
(0,0,0,0) & 1 & + & +1 & (2,2,1,0) & 2 & - & -2 \\
(1,0,0,0) & 2 & - & -2  &  (2,2,2,0) & 1 & + & +1 \\
(2,0,0,0) & 1 & + & +1 & (1,1,1,1) & 1 & + & +1 \\
(1,1,0,0) & 1 & + & +1 & (2,1,1,1) & 1 & + & +1 \\
(2,1,0,0) & 2 & - & -2  & (2,2,1,1) & 1 & + & +1 \\
(2,2,0,0) & 1 & + & +1 & (2,2,2,1) & 2 & - & -2 \\
(1,1,1,0) & 1 & + & +1 & (2,2,2,2) & 1 & + & +1 \\
(2,1,1,0) & 2 & - & -2  &  &  &  &  \\
\hline 
\end{array}
\end{equation}
Putting everything together we have the following prediction for the vev of the Wilson line operator
\begin{eqnarray}
\langle W_{\zeta , \mathbf{6}} \rangle_{q=-1} &=& \sum_{\mathbf{d}} {\tt DT}_{\mathbf{d}} (W_{\zeta , \mathbf{6}}) \ X_{e_{\mathbf{6}} + \mathbf{ e \cdot d}}
\cr &=&
X_{-\frac{4 e_{\circ_1}}{3}-\frac{2 e_{\circ_2}}{3}-\frac{4 e_{\bullet_1}}{3}-\frac{2
   e_{\bullet_2}}{3}}-2 X_{-\frac{e_{\circ_1}}{3}-\frac{2 e_{\circ_2}}{3}-\frac{4
   e_{\bullet_1}}{3}-\frac{2 e_{\bullet_2}}{3}}+X_{\frac{2 e_{\circ_1}}{3}-\frac{2
   e_{\circ_2}}{3}-\frac{4 e_{\bullet_1}}{3}-\frac{2
   e_{\bullet_2}}{3}}
    \cr &&
   +X_{-\frac{e_{\circ_1}}{3}-\frac{2
   e_{\circ_2}}{3}-\frac{e_{\bullet_1}}{3}-\frac{2 e_{\bullet_2}}{3}}
   -2 X_{\frac{2
   e_{\circ_1}}{3}-\frac{2 e_{\circ_2}}{3}-\frac{e_{\bullet_1}}{3}-\frac{2
   e_{\bullet_2}}{3}}+X_{-\frac{e_{\circ_1}}{3}+\frac{e_{\circ_2}}{3}-\frac{e_{\bullet_1}
   }{3}-\frac{2 e_{\bullet_2}}{3}}
    \cr &&
   -2 X_{\frac{2
   e_{\circ_1}}{3}+\frac{e_{\circ_2}}{3}-\frac{e_{\bullet_1}}{3}-\frac{2
   e_{\bullet_2}}{3}}+X_{\frac{2 e_{\circ_1}}{3}-\frac{2 e_{\circ_2}}{3}+\frac{2
   e_{\bullet_1}}{3}-\frac{2 e_{\bullet_2}}{3}}
   -2 X_{\frac{2
   e_{\circ_1}}{3}+\frac{e_{\circ_2}}{3}+\frac{2 e_{\bullet_1}}{3}-\frac{2
   e_{\bullet_2}}{3}}
    \cr &&
   +X_{\frac{2 e_{\circ_1}}{3}+\frac{4 e_{\circ_2}}{3}+\frac{2
   e_{\bullet_1}}{3}-\frac{2
   e_{\bullet_2}}{3}}+X_{-\frac{e_{\circ_1}}{3}+\frac{e_{\circ_2}}{3}-\frac{e_{\bullet_1}
   }{3}+\frac{e_{\bullet_2}}{3}}+X_{\frac{2
   e_{\circ_1}}{3}+\frac{e_{\circ_2}}{3}-\frac{e_{\bullet_1}}{3}+\frac{e_{\bullet_2}}{3}}
   \cr &&
   +X_{\frac{2 e_{\circ_1}}{3}+\frac{e_{\circ_2}}{3}+\frac{2
   e_{\bullet_1}}{3}+\frac{e_{\bullet_2}}{3}}-2 X_{\frac{2 e_{\circ_1}}{3}+\frac{4
   e_{\circ_2}}{3}+\frac{2 e_{\bullet_1}}{3}+\frac{e_{\bullet_2}}{3}}+X_{\frac{2
   e_{\circ_1}}{3}+\frac{4 e_{\circ_2}}{3}+\frac{2 e_{\bullet_1}}{3}+\frac{4
   e_{\bullet_2}}{3}}
\end{eqnarray}
This prediction was confirmed in the $q \longrightarrow +1$ limit in \cite{BPSlinesCluster} imposing that vevs of Wilson lines obey an OPE derived from the tensor product decomposition of SU(3) representations. Here we find the same result, however \textit{without} imposing the OPE, and keeping track of certain spin information.

\subsection{$SU(3)$ with a Wilson line in the representation $\mathbf{10}$}

\begin{figure}[H]
    \centering
    \includegraphics[width=0.30\textwidth]{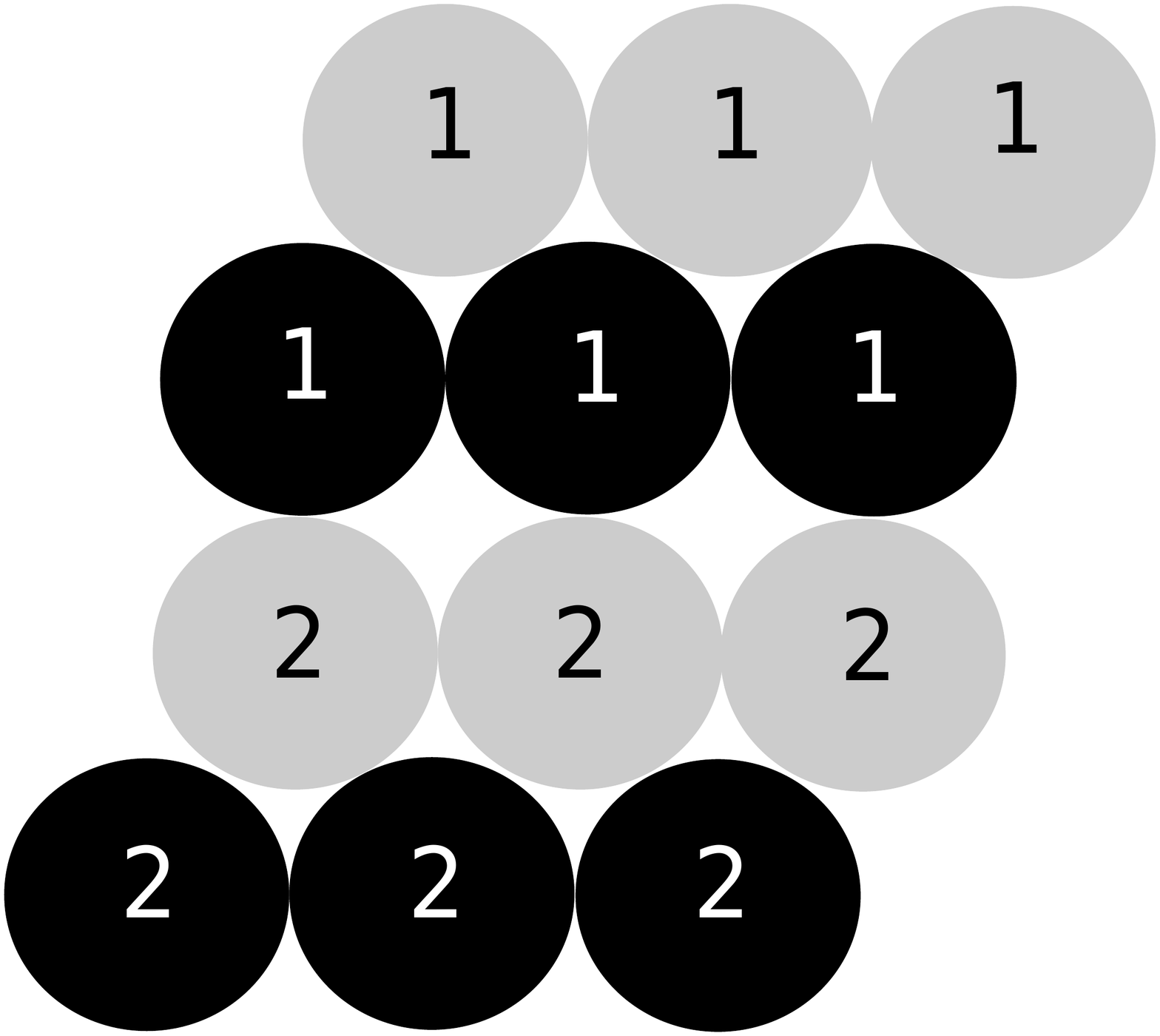}
    \caption{Pyramid arrangement for $SU(3)$ Yang-Mills coupled to a Wilson line in the representation $\mathbf{10}$}
    \label{pyramid-SU3-10}
\end{figure}

Finally we couple the SU(3) BPS quiver to a line operator with core charge $\mathbf{RG} (W_{\zeta , \mathbf{10}}) = - 2 (e_{\bullet_1} + e_{\circ_1}) - (e_{\bullet_2} + e_{\circ_2})$. 
In this case the relevant quiver is given by
\begin{equation}
\xymatrix@C=8mm{
&  \bullet_1 \ar@{..>}@<-0.5ex>[dl]_{B_1, \cdots, B_3}   \ar[rr]^{\tilde{\psi}} & & \ar@<-0.5ex>[dd]_{\tilde{A}_2}  \ar@<0.5ex>[dd]^{A_2}   \circ_2 \\
f_{\mathbf{10}}   \ar@{..>}@<-0.5ex>[dr]_{C_1, \cdots, C_3}    & & & \\
& \circ_1 \ar@<-0.5ex>[uu]_{\tilde{A}_1}  \ar@<0.5ex>[uu]^{A_1}& &   \ar[ll]^\psi  \bullet_2
}
\end{equation}
and we take the superpotential
\begin{equation}
\mathcal{W} = B_1 ( A_1 C_1 - A_1 C_2) + B_2 (\tilde{A}_1 C_2 - \tilde{A}_1 C_3) + B_3 (A_1 C_3- \tilde{A}_1 C_2) + \tilde{\psi} A_1 \psi A_2 - \tilde{\psi} \tilde{A}_1 \psi \tilde{A}_2 \, .
\end{equation}
We write only the equations of motions that we will need in the following:
\begin{eqnarray}
r_{B_1} \ & : & \   A_1 C_1  - A_1 C_2 = 0 \, , \cr
r_{B_2} \ & : & \ \tilde{A}_1 C_2 - \tilde{A}_1 C_3 = 0 \, , \cr
r_{B_3} \ & : & \ A_1 C_3 - \tilde{A}_1 C_2 = 0 \, , \cr
r_{\psi} \ & : & \  A_2 \tilde{\psi} A_1 -  \tilde{A}_2 \tilde{\psi} \tilde{A}_1 = 0 \, , \cr
r_{A_2} \ & : & \ \tilde{\psi} A_1 \psi = 0 \, , \cr
r_{\tilde{A}_2} \ & : & \  \tilde{\psi} \tilde{A}_1 \psi = 0 \, .
\end{eqnarray}
Using these equations we find the following identifications
\begin{align}
A_1 C_1 v & = A_1 C_2 v \, , \\
\tilde{A}_1 C_2 v & = \tilde{A}_1 C_3 v = A_1 C_3 v \, , 
\end{align}
as well as
\begin{align}
\tilde{A}_2 \tilde{\psi} A_1 C_1 v &= \tilde{A}_2 \tilde{\psi} A_1 C_2 v \, , \\
A_2 \tilde{\psi} A_1 C_1 v &= \tilde{A}_2 \tilde{\psi} \tilde{A}_1 C_1 v = A_2 \tilde{\psi} A_1 C_2 v = \tilde{A}_2 \tilde{\psi} \tilde{A}_1 C_2 v \cr
&= \tilde{A}_2 \tilde{\psi} A_1 C_3 v = \tilde{A}_2 \tilde{\psi} \tilde{A}_1 C_3 v = A_2 \tilde{\psi} A_1 C_3 v \cr 
&= A_2 \tilde{\psi} \tilde{A}_1 C_2 v = A_2 \tilde{\psi} \tilde{A}_1 C_3 v \, .
\end{align}
We can write the cyclic elements of the Jacobian algebra as
\begin{align}
\mathscr{J}_{\mathcal{W} , 0} & = \{ v \} \, , \cr
\mathscr{J}_{\mathcal{W} , 1} &= \{ C_1 v , \, C_2 v , \, C_3 v \} \, , \cr
\mathscr{J}_{\mathcal{W} , 2} &=  \{\tilde{A}_1 C_1 v , \, A_1 C_2 v , \, A_1 C_3 v \} \, , \cr
\mathscr{J}_{\mathcal{W} , 3} &= \{ \tilde{\psi} \tilde{A}_1 C_1 v , \, \tilde{\psi} A_1 C_2 v , \, \tilde{\psi} A_1 C_3 v\} \, , \cr
\mathscr{J}_{\mathcal{W} , 4} &=  \{A_2 \tilde{\psi} \tilde{A}_1 C_1 v , \, \tilde{A}_2 \tilde{\psi} A_1 C_2 v  , \, A_2 \tilde{\psi} A_1 C_3 v \} \, .
\end{align}
The fixed point configuration can be described in terms of the pyramid arrangement in Figure \ref{pyramid-SU3-10}.
The contribution of each fixed point is given by
\begin{equation}
(-1)^{\dim T_\pi} = (-1)^{d_{\bullet_1}^2+d_{\circ_1}^2+d_{\bullet_2}^2+d_{\circ_2}^2-3 d_{\bullet_1}-3 d_{\circ_1}-2 d_{\bullet_1} d_{\circ_1}-d_{\bullet_1} d_{\circ_2}-d_{\circ_1} d_{\bullet_2}-2 d_{\bullet_2} d_{\circ_2}} \, .
\end{equation}
We assemble all the invariants in the following table
\begin{equation}
\begin{array}{|c|c|c|c|c|c|c|c|}
\hline
(d_{\circ_1} , d_{\bullet_1} , d_{\circ_2} , d_{\bullet_2} ) & \text{fixed pts} & (-1)^{\dim T_\pi} & {\tt DT} &(d_{\circ_1} , d_{\bullet_1} , d_{\circ_2} , d_{\bullet_2} ) & \text{fixed pts} & (-1)^{\dim T_\pi} & {\tt DT} \\
\hline 
(0,0,0,0) & 1 & + & +1 & (3,3,2,0) & 3 & + & +3 \\ 
(1,0,0,0) & 3 & + & +3 & (3,3,3,0) & 1 & + & +1 \\
(2,0,0,0) & 3 & + & +3 & (1,1,1,1) & 1 & + & +1 \\
(3,0,0,0) & 1 & + & +1 & (2,1,1,1) & 2 & - & -2 \\
(1,1,0,0) & 1 & + & +1 & (2,2,1,1) & 1 & + & +1 \\
(2,1,0,0) & 4 & + & +4 & (2,2,2,1) & 2 & - & -2 \\
(2,2,0,0) & 1 & + & +1 & (2,2,2,2) & 1 & + & +1 \\
(3,1,0,0) & 3 & + & +3 & (3,2,1,1) & 2 & - & -2 \\
(3,2,0,0) & 3 & + & +3 & (3,2,2,1) & 4 & + & +4 \\
(3,3,0,0) & 1 & + & +1 & (3,2,2,2) & 1 & + & +1 \\
(1,1,1,0) & 1 & + & +1 & (3,1,1,1) & 1 & + & +1 \\
(2,1,1,0) & 4 & + & +4 & (3,3,1,1) & 1 & + & +1 \\
(2,2,1,0) & 2 & - & -2 & (3,3,2,1) & 4 & + & +4 \\
(2,2,2,0) & 1 & + & +1 & (3,3,2,2) & 1 & + & +1 \\
(3,2,1,0) & 6 & - & -6 & (3,3,3,1) & 3 & + & +3 \\
(3,2,2,0) & 3 & + & +3 & (3,3,3,2) & 3 & + & +3 \\
(3,1,1,0) & 3 & + & +3 & (3,3,3,3) & 1 & + & +1 \\
(3,3,1,0) & 3 & + & +3 &  &  &  &  \\
\hline 
\end{array}
\end{equation}
Finally we have the following prediction for the Wilson line vev
\begin{eqnarray}
\langle W_{\zeta , \mathbf{10}} \rangle_{q=-1} &=&
X_{-2 e_{\bullet_1}-e_{\circ_2}-2 e_{\circ_1}-e_{\bullet_2}}+3 X_{-e_{\bullet_1}-e_{\circ_2}-2
   e_{\circ_1}-e_{\bullet_2}}+X_{e_{\bullet_1}-e_{\circ_2}-2
   e_{\circ_1}-e_{\bullet_2}}
    \cr &&
   +X_{-e_{\bullet_1}-e_{\circ_2}-e_{\circ_1}-e_{\bullet_2}}
   +3
   X_{e_{\bullet_1}-e_{\circ_2}-e_{\circ_1}-e_{\bullet_2}}
   +X_{e_{\bullet_1}-e_{\circ_2}+e_{\circ_1}-e_{\bullet_2}
   }+3 X_{e_{\bullet_1}+e_{\circ_2}+e_{\circ_1}-e_{\bullet_2}}
    \cr &&
   +X_{e_{\bullet_1}+2
   e_{\circ_2}+e_{\circ_1}-e_{\bullet_2}}+X_{e_{\bullet_1}+e_{\circ_2}+e_{\circ_1}+e_{\bullet_2}}+3
   X_{e_{\bullet_1}+2 e_{\circ_2}+e_{\circ_1}+e_{\bullet_2}}
   \cr &&
   +X_{e_{\bullet_1}+2 e_{\circ_2}+e_{\circ_1}+2
   e_{\bullet_2}}+4 X_{e_{\bullet_1}+e_{\circ_2}+e_{\circ_1}}+3 X_{e_{\bullet_1}+2 e_{\circ_2}+e_{\circ_1}}+3
   X_{e_{\bullet_1}-e_{\circ_2}-e_{\bullet_2}}
    \cr &&
   +3
   X_{e_{\bullet_1}+e_{\circ_2}-e_{\bullet_2}}
   +X_{e_{\bullet_1}+e_{\circ_2}+e_{\bullet_2}}+4
   X_{e_{\bullet_1}+e_{\circ_2}}+X_{-e_{\bullet_1}-e_{\circ_1}-e_{\bullet_2}}
    \cr &&
   +3
   X_{e_{\bullet_1}-e_{\circ_1}-e_{\bullet_2}}+3
   X_{e_{\bullet_1}+e_{\circ_1}-e_{\bullet_2}}
   +X_{-e_{\bullet_1}-e_{\circ_1}}+X_{e_{\bullet_1}-e_{\circ_1}}+X_{e_{\bullet_1}+e_{\circ_1}}
    \cr &&
   -6 X_{e_{\bullet_1}-e_{\bullet_2}}-2 X_{e_{\bullet_1}}+3 X_{-e_{\circ_2}-2
   e_{\circ_1}-e_{\bullet_2}}+4
   X_{-e_{\circ_2}-e_{\circ_1}-e_{\bullet_2}}
   +X_{-e_{\circ_2}-e_{\bullet_2}}+X_{e_{\circ_2}-e_{\bullet_2}}
    \cr &&
   +X_{e_{\circ_2}+e_{\bullet_2}}-2 X_{e_{\circ_2}}+4 X_{-e_{\circ_1}-e_{\bullet_2}}-2 X_{-e_{\circ_1}}-2
   X_{-e_{\bullet_2}}+X_{0} \ .
\end{eqnarray}
Again it can be easily seen that this result is compatible with the OPE induced by the tensor product decomposition of SU(3) representations $\mathbf{3} * \mathbf{6} = \mathbf{8} + \mathbf{10}$. Equivalently we can claim that we have \textit{derived} this OPE by a direct computation.

\subsection{Dyonic defects} \label{SU3dyon-loca}

Now we will consider another class of defects, where the core charge is dyonic.
Consider first the defect
\begin{equation}
\xymatrix@C=8mm{
f_{\star_1}^{[1]} \ar@{..>}[dr]_{\beta^i}  & \ar@<0.5ex>@{..>}[l]^{\alpha_1^i} \ar@<-0.5ex>@{..>}[l]_{\alpha_2^i} \bullet_1 \ar[rr]^{\psi} & & \circ_2 \ar@<-0.5ex>[d]_{\tilde{A}_1}  \ar@<0.5ex>[d]^{\tilde{A_2}} \\
& \circ_1 \ar@<-0.5ex>[u]_{A_1}  \ar@<0.5ex>[u]^{A_2}  & & \bullet_2  \ar[ll]_{\phi}
 }
\end{equation}
with superpotential
\begin{equation}
\mathcal{W} =\sum_{i=1}^3 ( \beta^i \alpha^i_2 A_2 + \beta^i \alpha^i_1 A_1) +\tilde{A_1} \psi A_1 \phi + \tilde{A_2} \psi A_2 \phi \, .
\end{equation}
Due to the equations of motion, it is immediate to see that the pyramid arrangement consists of three copies of $\circ$ and that the fixed points have dimension vectors $(d_{\circ_1} , d_{\bullet_1} , d_{\circ_2} , d_{\bullet_2})$ as follows: $(0,0,0,0)$, $(1,0,0,0)$ with multiplicity $3$, $(2,0,0,0)$, with multiplicity $3$, and $(3,0,0,0)$ with multiplicity $1$. The parity of the tangent space is always positive. Therefore we have reproduced the result of  \cite{BPSlinesCluster}.

\begin{figure}[h]
    \centering
    \includegraphics[width=0.30\textwidth]{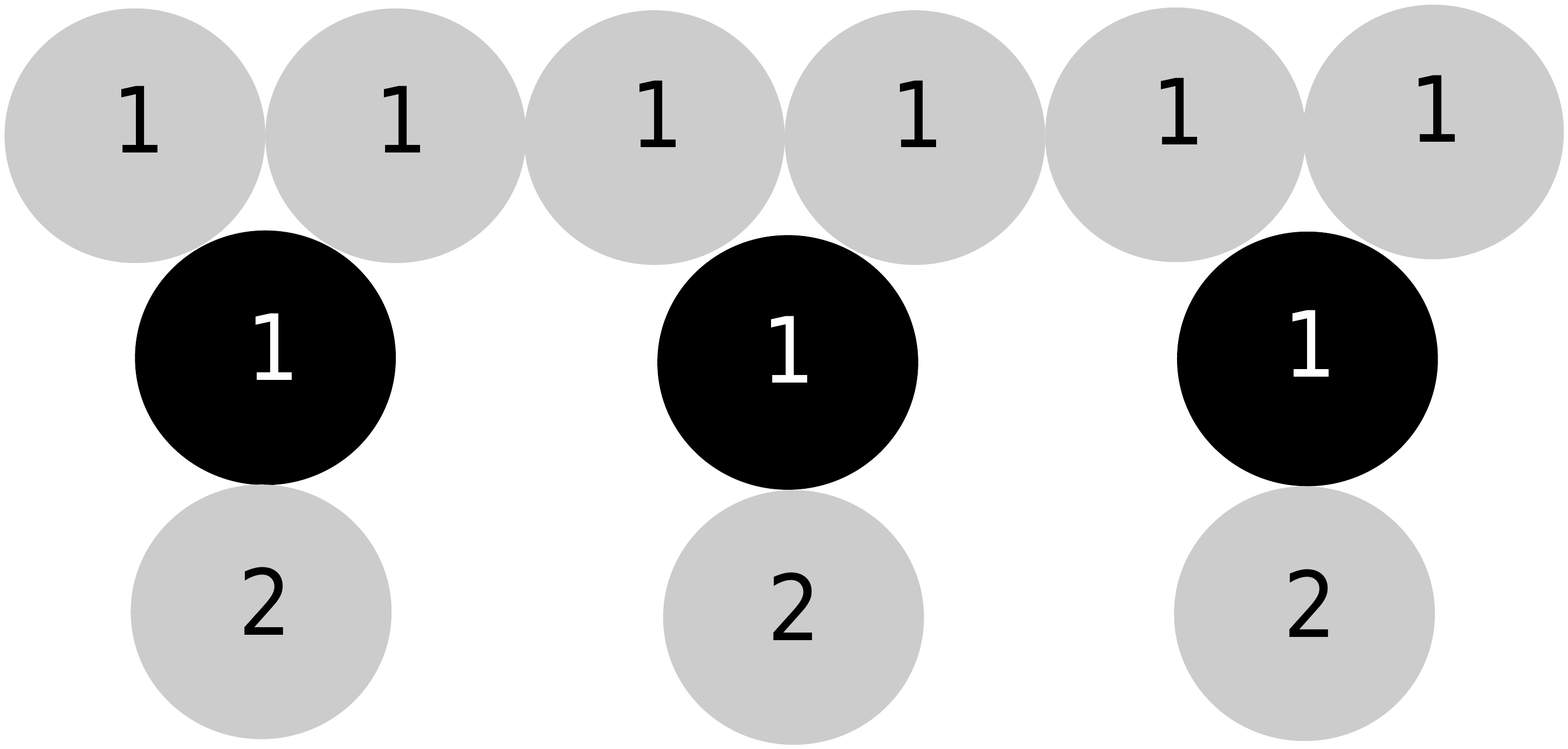}
    \caption{Pyramid arrangement for the defect  $f_{\star_1}^{[2]} $ in $SU(3)$}
    \label{dyonSU3-2}
\end{figure}

Consider now the less trivial defect
\begin{equation}
\xymatrix@C=20mm{
f_{\star_1}^{[2]} \ar@<0.5ex>@{..>}[dr] \ar@<-0.5ex>@{..>}[dr]_{\beta^i_1,\beta^i_2}  &\ar@{..>}[l]\ar@<-0.3pc>@{..>}[l]\ar@<0.3pc>@{..>}[l]^{\a^i_1 , \a^i_2 ,\a^i_3}\bullet_1 \ar[rr]^{\psi} & & \circ_2 \ar@<-0.5ex>[d]_{\tilde{A_1}}  \ar@<0.5ex>[d]^{\tilde{A_2}}  \\
& \circ_1 \ar@<-0.5ex>[u]_{A_1}  \ar@<0.5ex>[u]^{A_2}  & & \bullet_2  \ar[ll]_{\phi}
 } 
\end{equation}
with superpotential
\begin{equation}
\mathcal{W}=  \sum_{i=1}^3 ( A_1 \beta_1^i \alpha_1^i + A_2 \beta_2^i \alpha_2^i + \alpha_3^i (A_1 \beta_2^i - A_2 \beta_1^i))  +\tilde{A_1} \psi A_1 \phi + \tilde{A_2} \psi A_2 \phi \, .
\end{equation}
The relevant equations of motion are
\begin{align}
A_1 \beta_1^i &= A_2 \beta_2^i = 0 \, , \cr
A_1 \beta_2^i &= A_2 \beta_1^i \, , \cr
\tilde{A}_1 \psi A_1 &= - \tilde{A}_2 \psi A_2 \, ,
\end{align}
for $i=1,2,3$. The associated pyramid arrangement takes the form of Figure \ref{dyonSU3-2}.
This shape follows directly from the equations of motion. For example one has that
\begin{equation}
\tilde{A}_2 \psi A_1 \beta_1^i = \tilde{A}_1 \psi A_1 \beta_1 = 0 \, ,
\end{equation}
and similarly
\begin{equation}
\tilde{A}_1 \psi A_2 \beta_1^i = \tilde{A}_1 \psi A_1 \beta_2^i = \tilde{A}_2 \psi A_2 \beta_2 = 0 \, .
\end{equation}
The parity of the tangent space is given by the combination
\begin{equation}
(-1)^{d_{\circ_1}^2 + d_{\bullet_1}^2 + d_{\circ_2}^2+d_{\bullet_2}^2 - 6 d_{\circ_1} - 9 d_{\bullet_1} - 2 d_{\circ_1} d_{\bullet_1} - d_{\bullet_1} d_{\circ_2} - d_{\circ_1} d_{\bullet_2} - 2 d_{\circ_2} d_{\bullet_2}} \, .
\end{equation}
It is now straightforward to compute the framed BPS degeneracies, and the results are contained in the following table
\begin{equation}
\begin{array}{|c|c|c|c|c|c|c|c|}
\hline
(d_{\circ_1} , d_{\bullet_1} , d_{\circ_2} , d_{\bullet_2} ) & \text{fixed pts} & (-1)^{\dim T_\pi} & {\tt DT} &(d_{\circ_1} , d_{\bullet_1} , d_{\circ_2} , d_{\bullet_2} ) & \text{fixed pts} & (-1)^{\dim T_\pi} & {\tt DT} \\
\hline 
(0,0,0,0) & 1 & + & +1 & (6,3,0,0) & 1 & + & +1 \\ 
(1,0,0,0) & 6 & - & -6 & (2,1,1,0) & 3 & + & +3 \\
(2,0,0,0) & 15 & + & +15 & (3,1,1,0) & 12 & - & -12 \\
(3,0,0,0) & 20 & - & -20 & (4,1,1,0) & 18 & + & +18 \\
(4,0,0,0) & 15 & + & +15 & (5,1,1,0) & 12 & - & -12 \\
(5,0,0,0) & 6 & - & -6 & (6,1,1,0) & 3 & + & +3 \\
(6,0,0,0) & 1 & + & +1 & (4,2,1,0) & 6 & - & -6 \\
(2,1,0,0) & 3 & + & +3 & (5,2,1,0) & 12 & + & +12 \\
(3,1,0,0) & 12 & - & -12 & (6,2,1,0) & 6 & - & -6 \\
(4,1,0,0) & 18 & + & +18 & (6,3,1,0) & 3 & + & +3 \\
(5,1,0,0) & 12 & - & -12 & (4,2,2,0) & 3 & + & +3 \\
(6,1,0,0) & 3 & + & +3 & (5,2,2,0) & 6 & - & -6 \\
(4,2,0,0) & 3 & + & +3 & (6,2,2,0) & 3 & + & +3 \\
(5,2,0,0) & 6 & - & -6 & (6,3,2,0) & 3 & + & +3 \\
(6,2,0,0) & 3 & + & +3 & (6,3,3,0) & 1 & + & +1 \\
\hline 
\end{array}
\end{equation}
This framed BPS spectrum agrees with the one computed \cite{BPSlinesCluster} using untwisted variables $Y_\gamma$, if we neglect the signs. The above result goes beyond and also contain spin information.

\section{$SO(8)$ super Yang-Mills} \label{SO8loca}

As the last example, we will now apply our formalism to the case of $SO(8)$ super Yang-Mills. To our knowledge no framed BPS spectra are known for this theory. This case is somewhat more technical than the previous Sections. However we will see that our formalism is very powerful and allows for a complete solution of the framed BPS spectra in many cases. The unframed quiver which describes the BPS spectrum of $SO(8)$ super Yang-Mills can be chosen as \cite{Alim:2011kw}
\begin{equation} \label{BPSquiverSO8}
\xymatrix@C=8mm{
  & & & \circ_2  \ar@<-0.5ex>[dd]_{A_2}  \ar@<0.5ex>[dd]^{\tilde{A}_2} &  \\ 
   \circ_1  \ar@<-0.5ex>[dd]_{A_1}  \ar@<0.5ex>[dd]^{\tilde{A}_1} & & \bullet_3 \ar[ll]^{\tilde{\phi}} \ar[rr]_{\tilde{\lambda}} \ar[ur]^{\psi} &  & \circ_4 \ar@<-0.5ex>[dd]_{A_4}  \ar@<0.5ex>[dd]^{\tilde{A}_4} \\
  & & & \bullet_2 \ar[dl]_{\tilde{\psi}} &  \\
 \bullet_1 \ar[rr]^{\phi} & & \circ_3  \ar@<-0.5ex>[uu]_{A_3}  \ar@<0.5ex>[uu]^{\tilde{A}_3} & & \bullet_4 \ar[ll]_{\lambda}
 } 
\end{equation}
which has the structure of the product of a $D_4$ Dynkin diagram with a Kronecker quiver. The spectrum has a $\mathbb{Z}_{12}$ symmetry, for certain values of the physical parameters, corresponding to the $1/12$-monodromy $\mathbf{r}^+ = \bullet_1 , \bullet_2 , \bullet_3 , \bullet_4$ with permutation $\sigma = \{ (\circ_1 , \bullet_1) , (\circ_2 , \bullet_2) , (\circ_3 , \bullet_3) , (\circ_4 , \bullet_4) \} $.

We will now couple it to Wilson lines in the representations $\mathbf{8}_s$, $\mathbf{8}_c$ and $\mathbf{8}_v$. These framings are related to each other by the triality of $D_4$. Indeed it is enough to pick one representation, say $\mathbf{8}_s$, and the other results can be obtained by permutations. The Wilson line defect in the $\mathbf{8}_s$ has core charge $\mathbf{RG} (W_{\zeta , \mathbf{8}_s}) = - (e_{\bullet_1} + e_{\bullet_3} + e_{\circ_1} + e_{\circ_3}) - \frac12 (e_{\bullet_2} + e_{\bullet_4} + e_{\circ_2} + e_{\circ_4})$. We therefore consider the quiver
\begin{equation}
\xymatrix@C=8mm{
 & & & & \circ_2  \ar@<-0.5ex>[dd]_{A_2}  \ar@<0.5ex>[dd]^{\tilde{A}_2} &  \\ 
& \circ_1  \ar@<-0.5ex>[dd]_{A_1}  \ar@<0.5ex>[dd]^{\tilde{A}_1} & & \bullet_3 \ar[ll]^{\tilde{\phi}} \ar[rr]_{\tilde{\lambda}} \ar[ur]^{\psi} &  & \circ_4 \ar@<-0.5ex>[dd]_{A_4}  \ar@<0.5ex>[dd]^{\tilde{A}_4} \\
f_{\mathbf{8}_s} \ar@{..>}[ru]^C  & & & & \bullet_2 \ar[dl]_{\tilde{\psi}} &  \\
& \bullet_1 \ar[rr]^{\phi} \ar@{..>}[ul]_B  & & \circ_3  \ar@<-0.5ex>[uu]_{A_3}  \ar@<0.5ex>[uu]^{\tilde{A}_3} & & \bullet_4 \ar[ll]_{\lambda}
 }
\end{equation}
with superpotential
\begin{align}
\mathcal{W} =& C \, B \, A_1 + \tilde{\phi} \, A_3 \, \phi \, A_1 - \tilde{\phi} \, \tilde{A}_3 \, \phi \, \tilde{A}_1 + \tilde{\psi} \, A_2 \, \psi \, A_1 - \tilde{\psi} \, \tilde{A}_2 \, \psi \, \tilde{A}_3 \cr
& + \lambda \, A_4 \, \tilde{\lambda} \, A_3 - \lambda \,  \tilde{A}_4 \, \tilde{\lambda} \, \tilde{A}_3 \, .
\end{align}
The framed BPS quivers for line defects in the $\mathbf{8}_c$ and in the $\mathbf{8}_v$ are obtained from this one by coupling the defect in the same fashion, on the two other external Kronecker subquivers (with nodes labelled $\circ_2$ and $\circ_4$ respectively). Note that in the above superpotential the coupling to the Wilson line is the same as in the case of Section \ref{SU3with3}, while the unframed part is given in \cite{Alim:2011kw}.

\begin{figure}[H]
    \centering
    \includegraphics[width=0.40\textwidth]{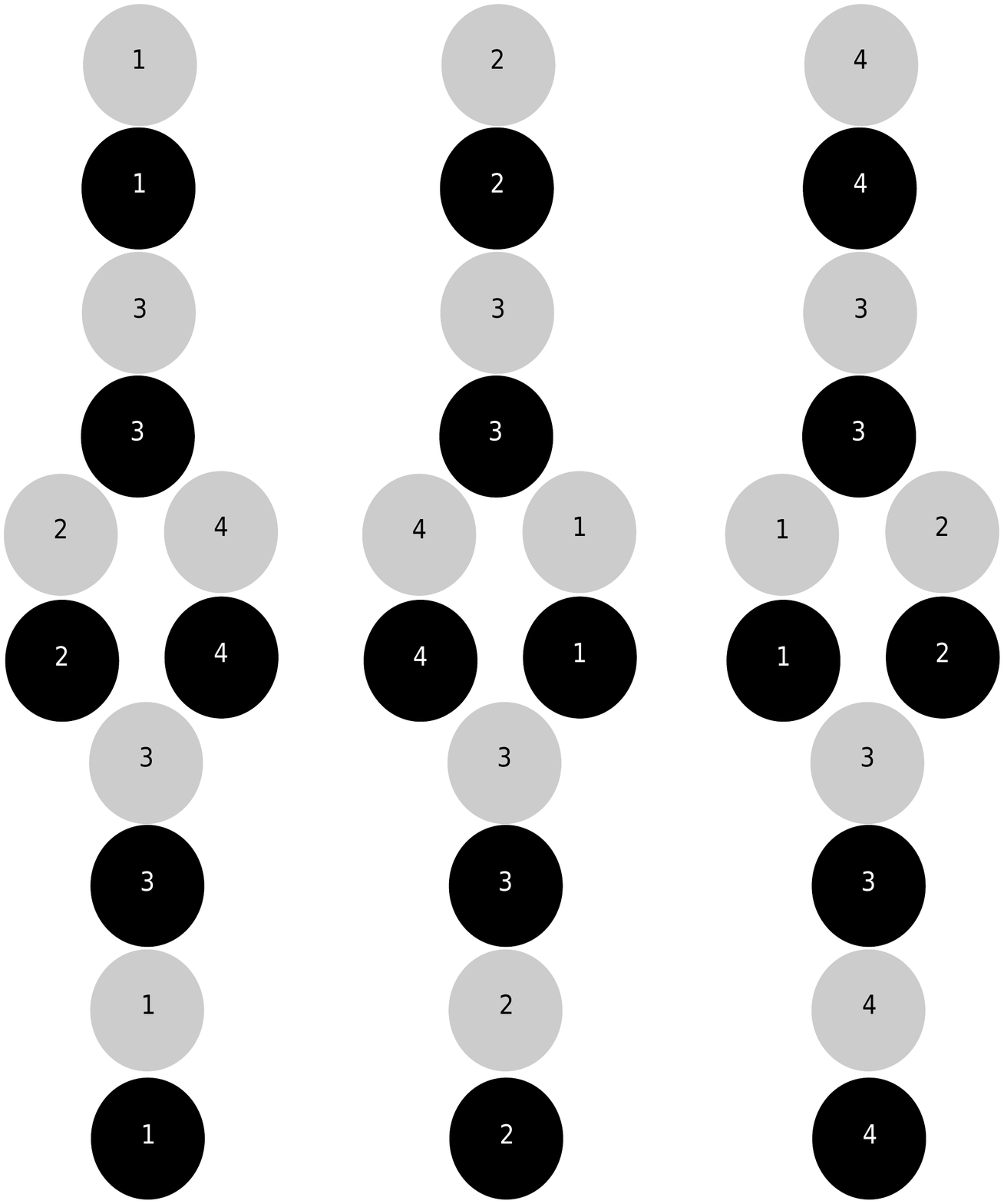}
    \caption{Pyramid arrangements for $SO(8)$ super Yang-Mills coupled to Wilson lines in the representations $\mathbf{8}_s$, $\mathbf{8}_c$ and $\mathbf{8}_v$. The label of the stones corresponds to the label of the nodes of the quiver.}
    \label{pyramid-SO8}
\end{figure}

Since we have discussed several examples so far, we will be brief. The first step is to construct the relevant truncated pyramid arrangement associated with this quiver, and then study the fixed points of the natural toric action, which rescales all the fields while preserving the F-term relations. To construct the shape of the pyramid arrangement we must look at the Jacobian algebra generated by the cyclic vector $v$. As usual this is graded $\mathscr{J}_{\mathcal{W} }= \bigoplus_{n \ge 0} \,\mathscr{J}_{\mathcal{W} , n}$ and the first few terms are
\begin{align}
\mathscr{J}_{\mathcal{W} , 0} =& \{ v \} \, , \cr
\mathscr{J}_{\mathcal{W} , 1} =& \{ C \, v \} \, , \cr
\mathscr{J}_{\mathcal{W} , 2} =& \{ \tilde{A}_1 \, C \, v \} \, , \cr
\mathscr{J}_{\mathcal{W} , 3} =& \{ \phi \, \tilde{A}_1 \, C \, v \} \, , \cr
\mathscr{J}_{\mathcal{W} , 4} =& \{ A_3 \, \phi \, \tilde{A}_1 \, C \, v \} \, , \cr
\mathscr{J}_{\mathcal{W} , 5} =& \{ \psi \, A_3 \, \phi \, \tilde{A}_1 \, C \, v \, , \tilde{\lambda} \, A_3 \, \phi \, \tilde{A}_1 \, C \, v \} \, . 
\end{align}
To go one step further observe that
\begin{align}
A_2 \, \psi \, A_3 \, \phi \, \tilde{A}_1 \, C \, v = \tilde{A}_2 \, \psi \, \tilde{A}_3 \, \phi \, \tilde{A}_1 \, C \, v = \tilde{A}_2 \, \psi \, A_3 \, \phi \, A_1 \, C \, v = 0 \, ,
\end{align}
where we have used the F-term equations $\partial_{\tilde{\psi}} \mathcal{W} = 0$, $\partial_{\tilde{\phi}} \mathcal{W} = 0$ and $\partial_{B} \mathcal{W} = 0$. In the same fashion one concludes that
\begin{equation}
A_4 \, \tilde{\lambda} \,  A_3 \, \phi \, \tilde{A}_1 \, C \, v = 0 \, .
\end{equation}
Therefore we have shown that
\begin{equation}
\mathscr{J}_{\mathcal{W} , 6} =  \{ \tilde{A}_2 \, \psi \, A_3 \, \phi \, \tilde{A}_1 \, C \, v \, , \tilde{A}_4 \, \tilde{\lambda} \, A_3 \, \phi \, \tilde{A}_1 \, C \, v \} \, . 
\end{equation}
At the next level we have
\begin{equation} \label{SO8A7}
\mathscr{J}_{\mathcal{W} , 7} =  \{ \tilde{\psi} \, \tilde{A}_2 \, \psi \, A_3 \, \phi \, \tilde{A}_1 \, C \, v = - \lambda \,  \tilde{A}_4 \, \tilde{\lambda} \, A_3 \, \phi \, \tilde{A}_1 \, C \, v \} \, . 
\end{equation}
This follows from the equation $\partial_{\tilde{A}_3} \mathcal{W} = 0$:
\begin{equation} \label{SO3-eom-At3}
\phi \, \tilde{A}_1 \, \tilde{\phi} + \tilde{\psi} \, \tilde{A}_2 \, \psi + \lambda \, \tilde{A}_4 \, \tilde{\lambda} = 0 \ ,
\end{equation}
which implies
\begin{equation}
\phi \, \tilde{A}_1 \, \tilde{\phi} \,  A_3 \, \phi \, \tilde{A}_1 \, C \, v + \tilde{\psi} \, \tilde{A}_2 \, \psi \,  A_3 \, \phi \, \tilde{A}_1 \, C \, v + \lambda \, \tilde{A}_4 \, \tilde{\lambda} \,  A_3 \, \phi \, \tilde{A}_1 \, C \, v = 0 \ .
\end{equation}
Now \eqref{SO8A7} follows from the condition $\partial_{A_1} \mathcal{W} = 0$ evaluated at $B=0$, a condition which can always be imposed when classifying fixed points, due to the constraint $\dim \, V_f = 1$. Similarly one can see that
\begin{align}
\mathscr{J}_{\mathcal{W} , 8} =& \{ A_3 \,  \tilde{\psi} \, \tilde{A}_2 \, \psi \, A_3 \, \phi \, \tilde{A}_1 \, C \, v = -A_3 \,  \lambda \,  \tilde{A}_4 \, \tilde{\lambda} \, A_3 \, \phi \, \tilde{A}_1 \, C \, v \} \, , \cr
\mathscr{J}_{\mathcal{W} , 9} =& \{ \tilde{\phi} \, A_3 \,  \tilde{\psi} \, \tilde{A}_2 \, \psi \, A_3 \, \phi \, \tilde{A}_1 \, C \, v \} \ ,
\end{align}
where we have used $\partial_{A_2} \mathcal{W} = 0$ and $\partial_{A_4} \mathcal{W} = 0$, and finally
\begin{equation}
\mathscr{J}_{\mathcal{W} , 10} = \{ \tilde{A}_1 \, \tilde{\phi} \, A_3 \,  \tilde{\psi} \, \tilde{A}_2 \, \psi \, A_3 \, \phi \, \tilde{A}_1 \, C \, v \} \ .
\end{equation}
Indeed
\begin{equation}
\phi \,  \tilde{A}_1 \, \tilde{\phi} \, A_3 \,  \tilde{\psi} \, \tilde{A}_2 \, \psi \, A_3 \, \phi \, \tilde{A}_1 \, C \, v = 0 \ ,
\end{equation}
which follows from \eqref{SO3-eom-At3}, upon imposing $\partial_{A_2} \mathcal{W} = 0$ and $\partial_{A_4} \mathcal{W} = 0$. Therefore the pyramid arrangement assumes the simple form of Figure \ref{pyramid-SO8}.
We have also taken a step further, and used the triality of the $D_4$ Dynkin graph to write down the pyramid arrangements also for the $\mathbf{8}_c$ and $\mathbf{8}_v$ Wilson lines. Fixed points are now in correspondence with pyramid partitions which can be removed from the pyramid arrangement. It can be easily checked that the parity of the tangent space is positive at each fixed point. Therefore we can write down directly the Wilson line vevs simply by enumerating configurations:
\begin{align}
\langle W_{\zeta , \mathbf{8}_s} \rangle =& \Big[ \frac{1}{X_{\bullet_1} X_{\bullet_3} X_{\circ_1} X_{\circ_3} (X_{\bullet_2} X_{\bullet_4} X_{\circ_2} X_{\circ_4})^{1/2}}+\frac{1}{X_{\bullet_3} X_{\circ_3}  (X_{\bullet_2} X_{\bullet_4} X_{\circ_2} X_{\circ_4})^{1/2}}+\frac{1}{(X_{\bullet_2} X_{\bullet_4} X_{\circ_2} X_{\circ_4})^{1/2}}
\nonumber \\[4pt] & \ 
+\left( \frac{X_{\bullet_4} X_{\circ_4}}{X_{\bullet_2} X_{\circ_2}} \right)^{1/2}+\left( \frac{X_{\bullet_2} X_{\circ_2}}{X_{\bullet_4} X_{\circ_4}} \right)^{1/2}+\left( X_{\bullet_2} X_{\bullet_4} X_{\circ_2} X_{\circ_4} \right)^{1/2}+X_{\bullet_3} X_{\circ_3} \left(  X_{\bullet_2} X_{\bullet_4} X_{\circ_2} X_{\circ_4} \right)^{1/2}
\nonumber \\[4pt] & \ 
+X_{\bullet_1} X_{\bullet_3}  X_{\circ_1}  X_{\circ_3} \left( X_{\bullet_2} X_{\bullet_4} X_{\circ_2} X_{\circ_4} \right)^{1/2} \Big]
+ 
\frac{1}{X_{\bullet_3}  \left( X_{\bullet_2} X_{\bullet_4} X_{\circ_2} X_{\circ_4} \right)^{1/2}}+\left( \frac{X_{\circ_2}}{ X_{\bullet_2} X_{\bullet_4} X_{\circ_4}} \right)^{1/2}
\nonumber \\[4pt] & \ 
+\frac{1}{X_{\bullet_1}  X_{\bullet_3}  X_{\circ_3} \left( X_{\bullet_2} X_{\bullet_4} X_{\circ_2} X_{\circ_4} \right)^{1/2}}+\left( \frac{X_{\circ_4}}{X_{\bullet_2} X_{\bullet_4} X_{\circ_2}} \right)^{1/2}+\left( \frac{X_{\circ_2} X_{\circ_4}}{X_{\bullet_2} X_{\bullet_4}} \right)^{1/2}+\left( \frac{ X_{\bullet_2} X_{\circ_2} X_{\circ_4}}{ X_{\bullet_4}} \right)^{1/2}
\nonumber \\[4pt] & \ 
+\left( \frac{ X_{\bullet_4} X_{\circ_2}  X_{\circ_4}}{ X_{\bullet_2}} \right)^{1/2}+ X_{\circ_3}  \left( X_{\bullet_2} X_{\bullet_4} X_{\circ_2} X_{\circ_4} \right)^{1/2}+ X_{\bullet_3} X_{\circ_1}X_{\circ_3}  \left(  X_{\bullet_2} X_{\bullet_4} X_{\circ_2} X_{\circ_4} \right)^{1/2}
\nonumber \\[4pt] 
\langle W_{\zeta , \mathbf{8}_c} \rangle =& \langle W_{\zeta , \mathbf{8}_s} \rangle \vert_{\text{cyclic perm.} \{ \circ_1 , \circ_2 , \circ_4\} \,  , \{ \bullet_1 , \bullet_2 , \bullet_4 \} }
\nonumber \\[4pt]
\langle W_{\zeta , \mathbf{8}_v} \rangle =& \langle W_{\zeta , \mathbf{8}_c} \rangle \vert_{\text{cyclic perm.} \{ \circ_1 , \circ_2 , \circ_4\} \,  , \{ \bullet_1 , \bullet_2 , \bullet_4 \} }
\end{align}
One can easily see (for example using the \textsc{mathematica} package {\tt LieArt}) that the terms in parenthesis correspond to the weights of the corresponding representation of SO(8), according to the decomposition \eqref{Wexpweights}.

\begin{figure}[H]
    \centering
    \includegraphics[width=0.40\textwidth]{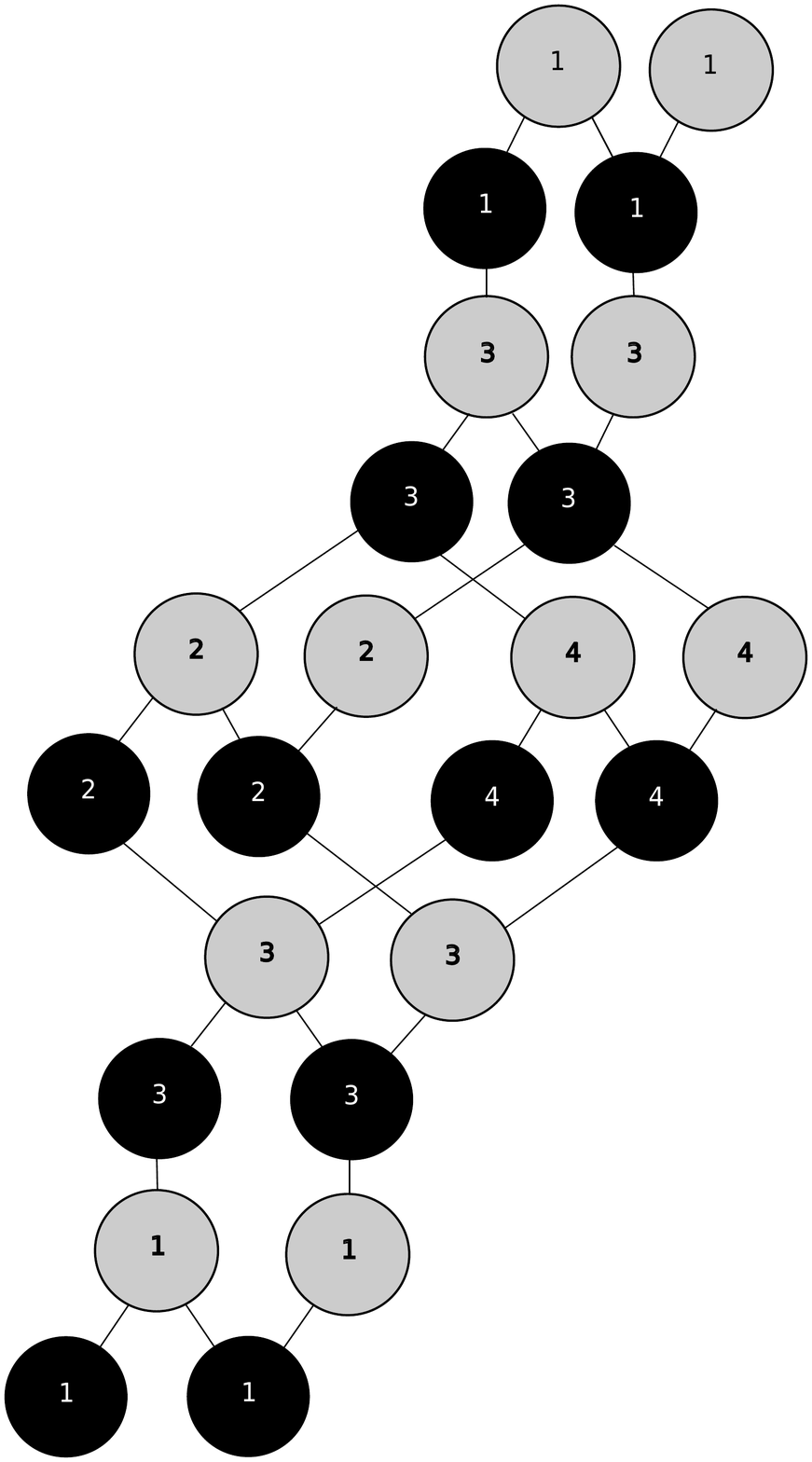}
    \caption{Pyramid arrangement for the $\textbf{35}_s$ of $SO(8)$. Instead of a three dimensional visualization, we have opted for drawing black links where two stones should be touching. Now toric fixed ideals are counted by removing pyramid partitions: stone configurations, such that if a stone is in the configuration, then all the stones immediately above connected via one black link should be in it. The links are non-intersecting.}
    \label{pyramid-SO8-35}
\end{figure}

Finally we would like to consider the remaining fundamental representation of $SO(8)$, given by the $\textbf{28}$. However, for technical reasons, it is easier to compute first the $\textbf{35}_s$, and then use the OPE between representations. The relevant quiver is now
\begin{equation}
\xymatrix@C=8mm{
 & & & & \circ_2  \ar@<-0.5ex>[dd]_{A_2}  \ar@<0.5ex>[dd]^{\tilde{A}_2} &  \\ 
& \circ_1  \ar@<-0.5ex>[dd]_{A_1}  \ar@<0.5ex>[dd]^{\tilde{A}_1} & & \bullet_3 \ar[ll]^{\tilde{\phi}} \ar[rr]_{\tilde{\lambda}} \ar[ur]^{\psi} &  & \circ_4 \ar@<-0.5ex>[dd]_{A_4}  \ar@<0.5ex>[dd]^{\tilde{A}_4} \\
f_{\mathbf{35}_s} \ar@{..>}@<-0.5ex>[ru]_C  \ar@{..>}@<+0.5ex>[ru]^{\tilde{C}}  & & & & \bullet_2 \ar[dl]_{\tilde{\psi}} &  \\
& \bullet_1 \ar[rr]^{\phi} \ar@{..>}@<-0.5ex>[ul]_B  \ar@{..>}@<0.5ex>[ul]^{\tilde{B}} & & \circ_3  \ar@<-0.5ex>[uu]_{A_3}  \ar@<0.5ex>[uu]^{\tilde{A}_3} & & \bullet_4 \ar[ll]_{\lambda}
 }
\end{equation}
where the superpotential
\begin{align} \label{W-SO8-35}
\mathcal{W} =& C \, B \, A_1 - \tilde{C} \, B \, A_1 + \tilde{C} \, \tilde{B} \, A_1 - C \, \tilde{B} \, A_1 + \tilde{\phi} \, A_3 \, \phi \, A_1 - \tilde{\phi} \, \tilde{A}_3 \, \phi \, \tilde{A}_1 
\cr & \
+ \tilde{\psi} \, A_2 \, \psi \, A_1 - \tilde{\psi} \, \tilde{A}_2 \, \psi \, \tilde{A}_3 + \lambda \, A_4 \, \tilde{\lambda} \, A_3 - \lambda \,  \tilde{A}_4 \, \tilde{\lambda} \, \tilde{A}_3 \, ,
\end{align}
is obtained from the analog situation for $SU(3)$, equation \eqref{W-SU3-6}. To build the pyramid arrangement we have to study modules generated by the action of the Jacobian algebra on a single vector $v \in V_f$, with $V_f$ one dimensional. The Jacobian algebra thus generated is naturally graded $\mathscr{J}_{\mathcal{W} } = \bigoplus_{n \ge  0} \, \mathscr{J}_{\mathcal{W} , n}$, and the first few terms are easy to work out
\begin{align}
\mathscr{J}_{\mathcal{W} , 0} & = \{ v \} \, , \cr
\mathscr{J}_{\mathcal{W} , 1} & = \{ \tilde{C} \, v , C \, v \} \, , \cr
\mathscr{J}_{\mathcal{W} , 2} & = \{ \tilde{A}_1 \, \tilde{C} v , A_1 \tilde{C} v = A_1 \, C \, v = \tilde{A}_1 \, C \, v  \} \, , \cr
\mathscr{J}_{\mathcal{W} , 3} & = \{ \phi \, \tilde{A}_1 \, \tilde{C} v , \phi \, \tilde{C} v , A_1 \tilde{C} v = \phi \, A_1 \, C \, v =\phi \, \tilde{A}_1 \, C \, v  \} \, , \cr
\mathscr{J}_{\mathcal{W} , 4} & = \{ A_3 \, \phi \, \tilde{A}_1 \, \tilde{C} v , \tilde{A}_3 \phi \tilde{A}_1 \, \tilde{C} v = A_3 \, \phi \, A_1 \, C \, v = \tilde{A}_3 \, \phi \, \tilde{A}_1 \, C \, v \} \, .
\end{align}
At the next levels we have four independent vectors
\begin{align}
\mathscr{J}_{\mathcal{W} , 5} &= \{ \psi \, A_3 \, \phi \, \tilde{A}_1 \, \tilde{C} \, v , \psi \, \tilde{A}_3 \, \phi \, \tilde{A}_1 \, C \, v ,  \tilde{\lambda} \, A_3 \, \phi \, \tilde{A}_1 \, \tilde{C} \, v , \tilde{\lambda} \, \tilde{A}_3 \, \phi \, \tilde{A}_1 \, C \, v  \} \, , \cr
\mathscr{J}_{\mathcal{W} , 6} &= \{ A_2 \,  \psi \, A_3 \, \phi \, \tilde{A}_1 \, \tilde{C} \, v , \tilde{A}_2 \,  \psi \, A_3 \, \phi \, \tilde{A}_1 \, \tilde{C} \, v , A_4 \,  \tilde{\lambda} \, A_3 \, \phi \, \tilde{A}_1 \, \tilde{C} \, v , \tilde{A}_4 \,  \tilde{\lambda} \, A_3 \, \phi \, \tilde{A}_1 \, \tilde{C} \, v \} \ ,
\end{align}
where we have used the F-term relations obtained from \eqref{W-SO8-35} to show
\begin{align}
A_2 \, \psi \, A_3 \, \phi \, \tilde{A}_1 \, \tilde{C} \, v = \tilde{A}_2 \, \psi \, \tilde{A}_3 \, \phi \, \tilde{A}_1 \, C \, v = A_2 \, \psi \, \tilde{A}_3 \, \phi \tilde{A}_1 \, C \, v \, , \cr
A_4 \, \tilde{\lambda} \, A_3 \, \phi \, \tilde{A}_1 \, \tilde{C} \, v = A_4 \, \tilde{\lambda} \, \tilde{A}_3 \, \phi \, \tilde{A}_1 \, C \, v = \tilde{A}_4 \, \tilde{\lambda} \, \tilde{A}_3 \, \phi \, \tilde{A}_1 \, C \, v \, .
\end{align}
Similar arguments give
\begin{align}
\mathscr{J}_{\mathcal{W} , 7} & = \{ \tilde{\psi} \, \tilde{A}_2 \, \psi \, A_3 \, \phi \, \tilde{A}_1 \, \tilde{C} \, v \ , \ \tilde{\psi} \, A_2 \, \psi \, A_3 \, \phi \, \tilde{A}_1 \, \tilde{C} \, v \} \, , \cr
\mathscr{J}_{\mathcal{W} , 8} & = \{ A_3 \,  \tilde{\psi} \, \tilde{A}_2 \, \psi \, A_3 \, \phi \, \tilde{A}_1 \, \tilde{C} \, v \ , \ \tilde{A}_3 \,  \tilde{\psi} \, \tilde{A}_2 \, \psi \, A_3 \, \phi \, \tilde{A}_1 \, \tilde{C} \, v \} \, , \cr
\mathscr{J}_{\mathcal{W} , 9} & = \{ \tilde{\phi} \, A_3 \,  \tilde{\psi} \, \tilde{A}_2 \, \psi \, A_3 \, \phi \, \tilde{A}_1 \, \tilde{C} \, v \ , \ \tilde{\phi}  \, \tilde{A}_3 \,  \tilde{\psi} \, \tilde{A}_2 \, \psi \, A_3 \, \phi \, \tilde{A}_1 \, \tilde{C} \, v \} \, , \cr
\mathscr{J}_{\mathcal{W} , 10} & = \{ \tilde{A}_1 \,  \tilde{\phi} \, A_3 \,  \tilde{\psi} \, \tilde{A}_2 \, \psi \, A_3 \, \phi \, \tilde{A}_1 \, \tilde{C} \, v \ ,  A_1 \,  \tilde{\phi} \, A_3 \,  \tilde{\psi} \, \tilde{A}_2 \, \psi \, A_3 \, \phi \, \tilde{A}_1 \, \tilde{C} \, v \} \ .
\end{align}
Putting all together we have built the pyramid arrangement of Figure \ref{pyramid-SO8-35}, where to avoid graphical complications, we have ``dismembered" the pyramid and drawn a black link where two stones are supposed to be touching. 
The counting of framed BPS degeneracies proceeds now as usual, albeit being considerably more involved. Toric fixed ideals are labelled by pyramid partitions, configurations of stones which can be removed from Figure \ref{pyramid-SO8-35}, subjected to the condition that if a stone is in the pyramid partition, then any stone immediately above connected by a black link is part of the pyramid partition as well. Each fixed point contributes to the degeneracies with a sign, given by
\begin{equation}
(-1)^{\sum_{i=1}^4 (d_{\circ_i}^2 + d_{\bullet_i}^2 ) - d_{\circ_1} - d_{\bullet_1} - \sum_{i=1}^4  d_{\circ_i} d_{\bullet_i} - d_{\bullet_1} d_{\circ_3}-d_{\bullet_3} d_{\circ_1}-d_{\bullet_2} d_{\circ_3}-d_{\bullet_3} d_{\circ_2}-d_{\bullet_4} d_{\circ_3}-d_{\bullet_3} d_{\circ_4}} \, ,
\end{equation}
where $\mathbf{d} = (d_{\bullet_1} , d_{\bullet_2}, d_{\bullet_3},d_{\bullet_4},d_{\circ_1},d_{\circ_2},d_{\circ_3},d_{\circ_4} )$ denotes the dimension vector of a cyclic module. The framed BPS spectrum consists of four states with $\mathtt{DT} = + 4$ and dimension vectors
\begin{equation}
\{
 (2,0,2,0,2,1,2,1),(2,1,2,0,2,2,2,1),(2,0,2,1,2,1,2,2),(2,1,2,1,2,2,2,2)
 \} \ ,
\end{equation}
one state with $\mathtt{DT} = + 3$ and dimension vector $(2,1,2,1,2,1,2,1)$; $36$ vector multiplet states with $\mathtt{DT}=-2$, and dimension vectors
\begin{equation}
\begin{gathered}
\{
(0,0,0,0,1,0,0,0),(1,0,0,0,2,0,0,0),(1,0,0,0,2,0,1,0),(2,0,0,0,2,0,1,0), \cr
(2,0,1,0,2,0,2,0),(2,0,1,0,2,1,2,0),(2,0,2,0,2,1,2,0),(2,0,1,0,2,0,2,1), \cr
(2,0,2,0,2,0,2,1),(2,0,1,0,2,1,2,1),(2,0,2,0,2,2,2,1),(2,0,2,0,2,1,2,2), \cr
(2,1,2,0,2,2,2,0),(2,1,2,0,2,1,2,1),(2,2,2,0,2,2,2,1),(2,1,2,0,2,2,2,2), \cr
(2,0,2,1,2,1,2,1),(2,0,2,1,2,0,2,2),(2,0,2,2,2,1,2,2),(2,0,2,1,2,2,2,2), \cr
(2,1,1,1,2,1,2,1),(2,1,2,1,2,2,2,1),(2,1,2,1,2,1,2,2),(2,2,2,1,2,2,2,2), \cr
(2,1,2,2,2,2,2,2),(2,1,2,1,2,1,3,1),(2,1,2,1,2,2,3,2),(2,2,2,1,2,2,3,2), \cr
(2,1,2,2,2,2,3,2),(2,2,2,2,2,2,3,2),(1,1,2,1,2,1,2,1),(2,2,3,2,2,2,4,2), \cr
(2,1,2,1,3,1,2,1),(2,2,3,2,3,2,4,2),(2,2,4,2,3,2,4,2),(3,2,4,2,4,2,4,2) 
\} \ ,
\end{gathered}
\end{equation}
and $97$ other hypermultiplets with $\mathtt{DT} = +1$. Recall that in our conventions the charges of the BPS states are given by $e_{\mathbf{35}_s} + \mathbf{e} \cdot \mathbf{d}$. From these degeneracies we can write down the vev as
\be
\langle W_{\zeta , \mathbf{35}_s} \rangle =  \langle W_{\zeta , \mathbf{35}_s} \rangle_{\rm diag} + \langle W_{\zeta , \mathbf{35}_s} \rangle_{\rm off diag}
\ee
with
\begin{small}
\begin{align}
 & \langle W_{\zeta , \mathbf{35}_s} \rangle_{\rm diag} =  \Big[
\frac{1}{X_{\bullet_1}^2 X_{\bullet_2} X_{\bullet_3}^2 X_{\bullet_4} X_{\circ_1}^2 X_{\circ_2} X_{\circ_3}^2 X_{\circ_4}}
+
\frac{1}{X_{\bullet_1} X_{\bullet_2} X_{\bullet_3}^2 X_{\bullet_4} X_{\circ_1} X_{\circ_2} X_{\circ_3}^2 X_{\circ_4}}
+
\frac{1}{X_{\bullet_2} X_{\bullet_3}^2 X_{\bullet_4} X_{\circ_2} X_{\circ_3}^2 X_{\circ_4}}
\nonumber\\[4pt] & \
+
\frac{1}{X_{\bullet_1} X_{\bullet_2} X_{\bullet_3} X_{\bullet_4} X_{\circ_1} X_{\circ_2} X_{\circ_3} X_{\circ_4}}
+
\frac{1}{X_{\bullet_1} X_{\bullet_2} X_{\bullet_3} X_{\circ_1} X_{\circ_2} X_{\circ_3}}
+
\frac{1}{X_{\bullet_1} X_{\bullet_3} X_{\bullet_4} X_{\circ_1} X_{\circ_3} X_{\circ_4}}
\nonumber\\[4pt] & \
+
\frac{1}{X_{\bullet_2} X_{\bullet_3} X_{\bullet_4} X_{\circ_2} X_{\circ_3} X_{\circ_4}}
+
\frac{1}{X_{\bullet_1} X_{\bullet_3} X_{\circ_1} X_{\circ_3}}
+
\frac{1}{X_{\bullet_2} X_{\bullet_3} X_{\circ_2} X_{\circ_3}}
+
\frac{1}{X_{\bullet_3} X_{\bullet_4} X_{\circ_3} X_{\circ_4}}
\nonumber\\[4pt] & \
+
\frac{1}{X_{\bullet_2} X_{\bullet_4} X_{\circ_2} X_{\circ_4}}
+
\frac{1}{X_{\bullet_3} X_{\circ_3}}
+
\frac{1}{X_{\bullet_1} X_{\circ_1}}
+
\frac{1}{X_{\bullet_2} X_{\circ_2}}
+
\frac{1}{X_{\bullet_4} X_{\circ_4}}
+
\frac{X_{\bullet_4} X_{\circ_4}}{X_{\bullet_2} X_{\circ_2}}
+
3
+
\frac{X_{\bullet_2} X_{\circ_2}}{X_{\bullet_4} X_{\circ_4}}
\nonumber\\[4pt] & \
+
X_{\bullet_4} X_{\circ_4}
+
X_{\bullet_2} X_{\circ_2}
+
X_{\bullet_1} X_{\circ_1}
+
X_{\bullet_3} X_{\circ_3}
+
X_{\bullet_2} X_{\bullet_4} X_{\circ_2} X_{\circ_4}
+
X_{\bullet_3} X_{\bullet_4} X_{\circ_3} X_{\circ_4}
\nonumber\\[4pt] & \
+
X_{\bullet_2} X_{\bullet_3} X_{\circ_2} X_{\circ_3}
+
X_{\bullet_1} X_{\bullet_3} X_{\circ_1} X_{\circ_3}
+
X_{\bullet_2} X_{\bullet_3} X_{\bullet_4} X_{\circ_2} X_{\circ_3} X_{\circ_4}
+
X_{\bullet_1} X_{\bullet_3} X_{\bullet_4} X_{\circ_1} X_{\circ_3} X_{\circ_4}
\nonumber\\[4pt] & \
+
X_{\bullet_1} X_{\bullet_2} X_{\bullet_3} X_{\circ_1} X_{\circ_2} X_{\circ_3}
+
X_{\bullet_1} X_{\bullet_2} X_{\bullet_3} X_{\bullet_4} X_{\circ_1} X_{\circ_2} X_{\circ_3} X_{\circ_4}
+
X_{\bullet_2} X_{\bullet_3}^2 X_{\bullet_4} X_{\circ_2} X_{\circ_3}^2 X_{\circ_4}
\nonumber\\[4pt] & \
+
X_{\bullet_1} X_{\bullet_2} X_{\bullet_3}^2 X_{\bullet_4} X_{\circ_1} X_{\circ_2} X_{\circ_3}^2 X_{\circ_4}
+
X_{\bullet_1}^2 X_{\bullet_2} X_{\bullet_3}^2 X_{\bullet_4} X_{\circ_1}^2 X_{\circ_2} X_{\circ_3}^2 X_{\circ_4}
\Big]
\end{align}
\end{small}
and
\begin{small}
\begin{align}
& \langle W_{\zeta , \mathbf{35}_s} \rangle_{\rm off diag}  =  
-\frac{2}{X_{\bullet_1}}
-\frac{2}{X_{\bullet_2}}
-\frac{2}{X_{\bullet_3}}
+\frac{1}{X_{\bullet_1} X_{\bullet_3}}
+\frac{1}{X_{\bullet_2} X_{\bullet_3}}
-\frac{2}{X_{\bullet_4}}
+\frac{4}{X_{\bullet_2} X_{\bullet_4}}
+\frac{1}{X_{\bullet_3} X_{\bullet_4}}
-\frac{2}{X_{\bullet_2} X_{\bullet_3} X_{\bullet_4}}
\nonumber\\[4pt] & \
+\frac{1}{X_{\bullet_1} X_{\bullet_3} X_{\circ_1}}
-2 X_{\circ_1}
+\frac{X_{\circ_1}}{X_{\bullet_1}}
+\frac{1}{X_{\bullet_2} X_{\bullet_3} X_{\circ_2}}
-\frac{2}{X_{\bullet_2} X_{\bullet_4} X_{\circ_2}}
-\frac{2}{X_{\bullet_2} X_{\bullet_3} X_{\bullet_4} X_{\circ_2}}
-2 X_{\circ_2}
\nonumber\\[4pt] & \
+\frac{X_{\circ_2}}{X_{\bullet_2}}
+\frac{4 X_{\circ_2}}{X_{\bullet_4}}
-\frac{2 X_{\circ_2}}{X_{\bullet_2} X_{\bullet_4}}
-\frac{2 X_{\bullet_2} X_{\circ_2}}{X_{\bullet_4}}
+\frac{1}{X_{\bullet_1} X_{\bullet_3} X_{\circ_3}}
+\frac{1}{X_{\bullet_2} X_{\bullet_3} X_{\circ_3}}
+\frac{1}{X_{\bullet_1} X_{\bullet_2} X_{\bullet_3} X_{\circ_3}}
\nonumber\\[4pt] & \
+\frac{1}{X_{\bullet_3} X_{\bullet_4} X_{\circ_3}}
+\frac{1}{X_{\bullet_1} X_{\bullet_3} X_{\bullet_4} X_{\circ_3}}
+\frac{1}{X_{\bullet_2} X_{\bullet_3} X_{\bullet_4} X_{\circ_3}}
+\frac{1}{X_{\bullet_1} X_{\bullet_2} X_{\bullet_3} X_{\bullet_4} X_{\circ_3}}
\nonumber\\[4pt] & \
+\frac{1}{X_{\bullet_1} X_{\bullet_2} X_{\bullet_3} X_{\circ_1} X_{\circ_3}}
+\frac{1}{X_{\bullet_1} X_{\bullet_3} X_{\bullet_4} X_{\circ_1} X_{\circ_3}}
+\frac{1}{X_{\bullet_1} X_{\bullet_2} X_{\bullet_3} X_{\bullet_4} X_{\circ_1} X_{\circ_3}}
+\frac{1}{X_{\bullet_1} X_{\bullet_2} X_{\bullet_3} X_{\circ_2} X_{\circ_3}}
\nonumber\\[4pt] & \
+\frac{1}{X_{\bullet_2} X_{\bullet_3} X_{\bullet_4} X_{\circ_2} X_{\circ_3}}
+\frac{1}{X_{\bullet_1} X_{\bullet_2} X_{\bullet_3} X_{\bullet_4} X_{\circ_2} X_{\circ_3}}
+\frac{1}{X_{\bullet_1} X_{\bullet_2} X_{\bullet_3} X_{\bullet_4} X_{\circ_1} X_{\circ_2} X_{\circ_3}}
-2 X_{\circ_3}
+\frac{X_{\circ_3}}{X_{\bullet_3}}
\nonumber\\[4pt] & \
+X_{\circ_1} X_{\circ_3}
+X_{\bullet_1} X_{\circ_1} X_{\circ_3}
+X_{\bullet_3} X_{\circ_1} X_{\circ_3}
+X_{\circ_2} X_{\circ_3}
+X_{\bullet_2} X_{\circ_2} X_{\circ_3}
+X_{\bullet_3} X_{\circ_2} X_{\circ_3}
\nonumber\\[4pt] & \
+X_{\bullet_3} X_{\circ_1} X_{\circ_2} X_{\circ_3}
+X_{\bullet_1} X_{\bullet_3} X_{\circ_1} X_{\circ_2} X_{\circ_3}
+X_{\bullet_2} X_{\bullet_3} X_{\circ_1} X_{\circ_2} X_{\circ_3}
-\frac{2}{X_{\bullet_2} X_{\bullet_4} X_{\circ_4}}
+\frac{1}{X_{\bullet_3} X_{\bullet_4} X_{\circ_4}}
\nonumber\\[4pt] & \
-\frac{2}{X_{\bullet_2} X_{\bullet_3} X_{\bullet_4} X_{\circ_4}}
+\frac{1}{X_{\bullet_2} X_{\bullet_3}^2 X_{\bullet_4} X_{\circ_2} X_{\circ_4}}
-\frac{2}{X_{\bullet_2} X_{\bullet_3} X_{\bullet_4} X_{\circ_2} X_{\circ_4}}
-\frac{2 X_{\circ_2}}{X_{\bullet_4} X_{\circ_4}}
+\frac{X_{\circ_2}}{X_{\bullet_2} X_{\bullet_4} X_{\circ_4}}
\nonumber\\[4pt] & \
+\frac{1}{X_{\bullet_1}^2 X_{\bullet_2} X_{\bullet_3}^2 X_{\bullet_4} X_{\circ_2} X_{\circ_3}^2 X_{\circ_4}}
-\frac{2}{X_{\bullet_1} X_{\bullet_2} X_{\bullet_3}^2 X_{\bullet_4} X_{\circ_2} X_{\circ_3}^2 X_{\circ_4}}
-\frac{2}{X_{\bullet_1}^2 X_{\bullet_2} X_{\bullet_3}^2 X_{\bullet_4} X_{\circ_1} X_{\circ_2} X_{\circ_3}^2 X_{\circ_4}}
\nonumber\\[4pt] & \
+\frac{1}{X_{\bullet_1} X_{\bullet_3} X_{\bullet_4} X_{\circ_3} X_{\circ_4}}
+\frac{1}{X_{\bullet_2} X_{\bullet_3} X_{\bullet_4} X_{\circ_3} X_{\circ_4}}
+\frac{1}{X_{\bullet_1} X_{\bullet_2} X_{\bullet_3} X_{\bullet_4} X_{\circ_3} X_{\circ_4}}
\nonumber\\[4pt] & \
+\frac{1}{X_{\bullet_1} X_{\bullet_2} X_{\bullet_3} X_{\bullet_4} X_{\circ_1} X_{\circ_3} X_{\circ_4}}
-\frac{2}{X_{\bullet_2} X_{\bullet_3}^2 X_{\bullet_4} X_{\circ_2} X_{\circ_3} X_{\circ_4}}
-\frac{2}{X_{\bullet_1} X_{\bullet_2} X_{\bullet_3}^2 X_{\bullet_4} X_{\circ_2} X_{\circ_3} X_{\circ_4}}
\nonumber\\[4pt] & \
+\frac{1}{X_{\bullet_1} X_{\bullet_2} X_{\bullet_3} X_{\bullet_4} X_{\circ_2} X_{\circ_3} X_{\circ_4}}
+\frac{1}{X_{\bullet_1} X_{\bullet_2} X_{\bullet_3}^2 X_{\bullet_4} X_{\circ_1} X_{\circ_2} X_{\circ_3} X_{\circ_4}}
-2 X_{\circ_4}+\frac{4 X_{\circ_4}}{X_{\bullet_2}}
+\frac{X_{\circ_4}}{X_{\bullet_4}}
\nonumber\\[4pt] & \
-\frac{2 X_{\circ_4}}{X_{\bullet_2} X_{\bullet_4}}
-\frac{2 X_{\bullet_4} X_{\circ_4}}{X_{\bullet_2}}
-\frac{2 X_{\circ_4}}{X_{\bullet_2} X_{\circ_2}}
+\frac{X_{\circ_4}}{X_{\bullet_2} X_{\bullet_4} X_{\circ_2}}
+4 X_{\circ_2} X_{\circ_4}
-\frac{2 X_{\circ_2} X_{\circ_4}}{X_{\bullet_2}}
-2 X_{\bullet_2} X_{\circ_2} X_{\circ_4}
\nonumber\\[4pt] & \
-\frac{2 X_{\circ_2} X_{\circ_4}}{X_{\bullet_4}}
+\frac{X_{\circ_2} X_{\circ_4}}{X_{\bullet_2} X_{\bullet_4}}
+\frac{X_{\bullet_2} X_{\circ_2} X_{\circ_4}}{X_{\bullet_4}}
-2 X_{\bullet_4} X_{\circ_2} X_{\circ_4}
+\frac{X_{\bullet_4} X_{\circ_2} X_{\circ_4}}{X_{\bullet_2}}
+X_{\circ_3} X_{\circ_4}
+X_{\bullet_3} X_{\circ_3} X_{\circ_4}
\nonumber\\[4pt] & \
+X_{\bullet_4} X_{\circ_3} X_{\circ_4}
+X_{\bullet_3} X_{\circ_1} X_{\circ_3} X_{\circ_4}
+X_{\bullet_1} X_{\bullet_3} X_{\circ_1} X_{\circ_3} X_{\circ_4}
+X_{\bullet_3} X_{\bullet_4} X_{\circ_1} X_{\circ_3} X_{\circ_4}
-2 X_{\circ_2} X_{\circ_3} X_{\circ_4}
\nonumber\\[4pt] & \
-2 X_{\bullet_2} X_{\circ_2} X_{\circ_3} X_{\circ_4}
+X_{\bullet_3} X_{\circ_2} X_{\circ_3} X_{\circ_4}
+X_{\bullet_2} X_{\bullet_3} X_{\circ_2} X_{\circ_3} X_{\circ_4}
-2 X_{\bullet_4} X_{\circ_2} X_{\circ_3} X_{\circ_4}
\nonumber\\[4pt] & \
-2 X_{\bullet_2} X_{\bullet_4} X_{\circ_2} X_{\circ_3} X_{\circ_4}
+X_{\bullet_3} X_{\bullet_4} X_{\circ_2} X_{\circ_3} X_{\circ_4}
+X_{\bullet_3} X_{\circ_1} X_{\circ_2} X_{\circ_3} X_{\circ_4}
+X_{\bullet_1} X_{\bullet_3} X_{\circ_1} X_{\circ_2} X_{\circ_3} X_{\circ_4}
\nonumber\\[4pt] & \
+X_{\bullet_2} X_{\bullet_3} X_{\circ_1} X_{\circ_2} X_{\circ_3} X_{\circ_4}
+X_{\bullet_1} X_{\bullet_2} X_{\bullet_3} X_{\circ_1} X_{\circ_2} X_{\circ_3} X_{\circ_4}
+X_{\bullet_3} X_{\bullet_4} X_{\circ_1} X_{\circ_2} X_{\circ_3} X_{\circ_4}
\nonumber\\[4pt] & \
+X_{\bullet_1} X_{\bullet_3} X_{\bullet_4} X_{\circ_1} X_{\circ_2} X_{\circ_3} X_{\circ_4}
+X_{\bullet_2} X_{\bullet_3} X_{\bullet_4} X_{\circ_1} X_{\circ_2} X_{\circ_3} X_{\circ_4}
+X_{\bullet_2} X_{\bullet_4} X_{\circ_2} X_{\circ_3}^2 X_{\circ_4}
\nonumber\\[4pt] & \
-2 X_{\bullet_2} X_{\bullet_3} X_{\bullet_4} X_{\circ_2} X_{\circ_3}^2 X_{\circ_4}
-2 X_{\bullet_2} X_{\bullet_3} X_{\bullet_4} X_{\circ_1} X_{\circ_2} X_{\circ_3}^2 X_{\circ_4}
+X_{\bullet_1} X_{\bullet_2} X_{\bullet_3} X_{\bullet_4} X_{\circ_1} X_{\circ_2} X_{\circ_3}^2 X_{\circ_4}
\nonumber\\[4pt] & \
-2 X_{\bullet_2} X_{\bullet_3}^2 X_{\bullet_4} X_{\circ_1} X_{\circ_2} X_{\circ_3}^2 X_{\circ_4}
+X_{\bullet_2} X_{\bullet_3}^2 X_{\bullet_4} X_{\circ_1}^2 X_{\circ_2} X_{\circ_3}^2 X_{\circ_4}
-2 X_{\bullet_1} X_{\bullet_2} X_{\bullet_3}^2 X_{\bullet_4} X_{\circ_1}^2 X_{\circ_2} X_{\circ_3}^2 X_{\circ_4} \ .
\end{align}
\end{small}

The terms in $ \langle W_{\zeta , \mathbf{35}_s} \rangle_{\rm diag} $ correspond precisely to the weights of the $\mathbf{35}_s$ representation of $SO(8)$ as in the decomposition \eqref{Wexpweights}. The remaining terms are non-perturbative in nature. Finally, our prediction for the remaining fundamental Wilson line $\textbf{28}$ is given by the OPE
\begin{equation}
\langle W_{\zeta , \mathbf{28}} \rangle= \langle W_{\zeta , \mathbf{8}_s} \rangle \langle W_{\zeta , \mathbf{8}_s} \rangle - \langle W_{\zeta , \mathbf{35}_s} \rangle - 1 \, .
\end{equation}
according to the tensor product decomposition $\mathbf{8}_s \otimes \mathbf{8}_s = \mathbf{28} +\mathbf{35}_s+1$.

These quantities pass a very non trivial check. As argued in \cite{BPSlinesCluster} the Wilson line defects correspond to conserved charges of a discrete integrable system associated with the unframed BPS quiver. In this case this is the $Q$-system associated with the $D_4$ Dynkin diagram. More in general for a $D_r$ Dynkin diagram, define the commutative variables $\{ \mathcal{R}_{\alpha , n} \, : \, \alpha \in I_r , n \in \mathbb{Z} \}$. These variables obey the recursion relation
\begin{equation} \label{RrelationsD}
\mathcal{R}_{\alpha , n+1} \, \mathcal{R}_{\alpha , n-1} = \mathcal{R}^2_{\alpha , n} +\prod_{\underset{C_{\alpha \beta} = -1}{ \beta \in I_r} } \mathcal{R}_{\beta , n}  \ , \qquad \mathcal{R}_{0,n} = \mathcal{R}_{r+1 , n} = 1 \ , \qquad  ( \alpha \in I_r \ , n \in \mathbb{Z} ) \ .
\end{equation}
with $I_r = 1 , \dots, r$. This system can be recast within the formalism of cluster algebras by interpreting \eqref{RrelationsD} as exchange relations for the seed  $\left( {\mathbf{x}} , B \right)$ with
\begin{equation}
\mathbf{x} = \left( \mathcal{R}_{1,0} , \dots , \mathcal{R}_{r,0} ; \mathcal{R}_{1,1} , \dots , \mathcal{R}_{r,1} \right) \, , \qquad B = \left( \begin{matrix} 0 & -C \\ C & 0 \end{matrix}  \right)
\end{equation}
where $C$ is the Cartan matrix of $D_r$ (in the conventions such that $B$ is the exchange matrix of the unframed BPS quiver for SO(4) \eqref{BPSquiverSO8}). Define the $Y$-seed variables
\begin{equation} \label{y-variables-SO8}
Y_{i} = \prod_i x_{i}^{B_{ij}} \, .
\end{equation}
These variables are the untwisted coordinates related to the $X_\gamma$ functions which we have been using throughout this paper by a quadratic refinement \cite{Gaiotto:2010be}. It is easy to see that converting the vevs we have computed above in terms of the $Y$-variables simply amounts to neglect all the minus signs. In these variables it is easy to compute the evolution of the $D_4$ $Q$-system which simply amounts to the rational transformation
\begin{equation} \label{RtopSO8}
\tilde{R} \equiv
\begin{cases}
  &Y_{\bullet_1}  \to 1/ Y_{\bullet_1}
     \\
   & Y_{\bullet_2} \to  1/Y_{\bullet_2}
         \\
  & Y_{\bullet_3} \to  1/Y_{\bullet_3}
         \\ 
  & Y_{\bullet_4} \to  1/Y_{\bullet_4}
         \\
   & Y_{\circ_1} \to  \frac{Y_{\bullet_1}^2 (1+Y_{\bullet_3}) Y_{\circ_1}}{(1+Y_{\bullet_1})^2}
          \\
   & Y_{\circ_2} \to  \frac{Y_{\bullet_2}^2 (1+Y_{\bullet_3}) Y_{\circ_2}}{(1+Y_{\bullet_2})^2}
     \\
        & Y_{\circ_3} \to \frac{(1+Y_{\bullet_1}) (1+Y_{\bullet_2}) Y_{\bullet_3}^2 (1+Y_{\bullet_4}) Y_{\circ_3}}{(1+Y_{\bullet_3})^2}
       \\
   & Y_{\circ_4} \to \frac{(1+Y_{\bullet_3}) Y_{\bullet_4}^2 Y_{\circ_4}}{(1+Y_{\bullet_4})^2}
         \end{cases}   \, .
\end{equation}
This transformation coincides with the $1/12$-fractional monodromy of the BPS quiver \eqref{BPSquiverSO8} when composed with the permutation $\sigma = \{ (\circ_1 , \bullet_1) , (\circ_2 , \bullet_2) , (\circ_3 , \bullet_3) , (\circ_4 , \bullet_4) \} $.

To our knowledge there is no systematic way to compute the conserved charges for this $Q$-system. We claim that these coincide with Wilson lines of the supersymmetric field theory which we have just computed. This follows from the results of \cite{BPSlinesCluster}, to which we refer the reader for a more detailed discussion. Indeed now one can directly check that the vevs of the Wilson line operators $W_{\zeta , \mathbf{8}_s} $,  $W_{\zeta , \mathbf{8}_c} $,  $W_{\zeta , \mathbf{8}_v} $ and  $W_{\zeta , \mathbf{28}}$ (as well as $W_{\zeta , \mathbf{35}}$), when expressed in terms of the commuting variables $Y_\gamma$, are invariant under $\sigma \circ \tilde{R}$: they precisely correspond to the fundamental constants of motion of the $Q$-system of $D_4$. This is a highly non-trivial confirmation of our result.

\section{Conclusions}

This paper was devoted to the study of framed BPS degeneracies from an algebraic viewpoint. Framed BPS degeneracies correspond to certain BPS invariants of Donaldson-Thomas type associated with moduli spaces of framed quiver representations. We have introduced a general formalism to compute directly these quantities using localization techniques and detailed several rather explicit examples. In the special case of cyclic or co-cyclic stability conditions we can give a complete classification of the fixed points and determine explicitly the contribution of each fixed point to the framed BPS index. Localization reduces the problem to a purely combinatorial one, counting certain combinatorial arrangements defined in terms of the quiver data. These steps, however technically involved, can be carried out directly for any framed quiver with superpotential. In principle given a framed quiver corresponding to a line defect in any $\cN=2$ model, the problem of computing the framed BPS spectrum can be solved applying our formalism. 

A different perspective, based on the transformation properties of framed quivers under mutations, is detailed in \cite{BPSlinesCluster}. The framed BPS spectra computed with these two different methods agree.

We conclude by collecting here a few open problems which we are currently investigating:
\begin{itemize}
\item It would be interesting to consider more general theories. We have only considered certain asymptotically free theories, but compactifications of the $\mathcal{N}=(0,2)$ theory produce a variety of strongly coupled or non-lagrangian or other exotic theories. In most cases no result for line operators is available. We believe our methods should be very useful to attack this problem.
\item We have focussed on the simpler limit $q \longrightarrow - 1$ of the protected spin character, but one could easily envision an extension of our methods to general values of $q$. As we have already mentioned we expect techniques from quantum cluster algebras \cite{FG} and the theories of motives as used in \cite{KS1,KS2,motives,Cirafici:2011cd}, to play a role here. This would be an important step towards a full fledged categorification of the ring of line operators, together with their OPE relations.
\item From a geometric perspective a line defect in a theory of class $\cS$ can be seen as a link in $S^1 \times \cC$. It is therefore natural to suspect that the BPS degeneracies computed in this paper could be interpreted as some kind of link invariant. Therefore the generalization of our formalism to the full Protected Spin Character should have some description in terms of a version of Khovanov homology for links. It would be very interesting to make these vague statements more precise.
\end{itemize}

\section*{Acknowledgements}

I am especially indebted to Michele Del Zotto for many discussions and collaboration on a closely related project. I thank Dylan Allegretti, Jun Nian, Maxim Kontsevich, Vasily Pestun, Boris Pioline and Harold Williams for discussions on this and related projects. The work of MC was partially supported by FCT/Portugal and IST-ID through EXCL/MAT-GEO/0222/2012 and the program Investigador FCT IF2014, contract IF/01426/2014/CP1214/CT0001. MC is thankful to the Theory Division of CERN for the hospitality and support during the last stages of this project. MC acknowledges support by the Action MP1405 QSPACE from the European Cooperation in Science and Technology (COST). The author also acknowledges the support of IH\'ES during a visit.  The research of M.C. on this project has received funding from the European Research Council (ERC) under the European Union's Horizon 2020 research and innovation programme (QUASIFT grant agreement 677368).

\end{document}